\documentclass[10pt,twoside]{thesis_uchile}

\usepackage{epsfig,natbib,deluxetable}
\usepackage{aastex_hack}
\usepackage{mymacros}
\bibstyle{aa}

\begin{document}

 \title{EL M\'ETODO DE PATRONES ESTANDARIZABLES PARA\\ SUPERNOVAS TIPO II-PLATEAU} 
 \author{FELIPE ANDR\'ES OLIVARES ESTAY}
 \masterscience
 \thesis
 \copyrightnotice
 \degreemonth{Julio}\degreeyear{2008}
\maketitle

\begin{frontmatter}

 \setcounter{page}{1}
\begin{abstract}

Durante el desarrollo de esta tesis estudiamos el M\'etodo de Patrones
Lum\'inicos Estandarizables (SCM) para supernovas Tipo II ``plateau''
haciendo uso de fotometr\'ia $BVRI$ y espectroscop\'ia \'optica. Se
implement\'o un procedimiento anal\'itico para ajustar funciones a las
curvas de luz, de color y de velocidad de expansi\'on. Encontramos que
el color \vi\ de estas supernovas, medido hacia el final de la \'epoca
``plateau'', puede ser utilizado para estimar el enrojecimiento
provocado por el material interestelar de la galaxia anfitriona con
una precisi\'on de $\sigma(A_V)=0.2$~mag. Tras realizar las
correcciones necesarias a la fotometr\'ia se recupera la relaci\'on
luminosidad versus velocidad de expansi\'on, reportada previamente en
la literatura cient\'ifica. Ocupando esta relaci\'on y asumiendo una
ley de extinci\'on est\'andar ($R_V=3.1$) obtenemos diagramas de
Hubble con dispersiones de $\sim$~0.4~mag en las bandas $BVI$. Por
otra parte, si permitimos variaciones en $R_V$ en favor de incertezas
menores obtenemos una dispersi\'on final de 0.25--0.30~mag, lo que
implica que estos objetos pueden entregar distancias tan precisas como
12--14\%. El valor resultante para $R_V$ es de $1.4\pm0.1$, que
sugiere una ley de extinci\'on no-est\'andar en nuestra l\'inea de
visi\'on hacia este tipo de supernovas. Utilizando dos objetos con
distancia Cefeida para calibrar la relaci\'on luminosidad-velocidad
obtenemos una constante de Hubble de $70\pm8$~\dimho, en buen acuerdo
con el valor que obtuvo el HST Key Project.

\end{abstract}

 \begin{acknowledgement}

Quiero agradecer en primer lugar a mi amada mujer, Natalia Armijo,
quien logr\'o mantenerme enfocado y motivado. Su constante apoyo fue
vital para mi constancia. Junto con mi familia jugaron el importante
rol de estar incondicionalmente presentes, respald\'andome. Esas
visitas de fin de semana a la casa de mi familia en Valpara\'iso
fueron una contribuci\'on infalible en favor del relajo y en contra
del estr\'es. Mi hermano Sebasti\'an, mi abuela Mariana y mi madre Ana
Mar\'ia me recibieron siempre con los brazos abiertos. En especial mi
madre, qui\'en fue crucial al momento de empezar mis estudios fuera de
casa. Sin su constante apoyo no hubiese podido sobrellevar la
complicada vida del estudiante de provincia en Santiago. Muchas
gracias tambi\'en a todos mis amigos, de quienes recib\'i mucha ayuda
a la hora de despejar la mente. En cuanto a lo acad\'emico, muchos de
mis conocimientos e ideas impregnadas en esta tesis fueron
responsabilidad de mi profesor gu\'ia, Mario Hamuy, qui\'en
consigui\'o transmitirme todo su entusiasmo, ideas innovadoras y
experiencia. Le agradezco a Mario el haberme llevado por el mejor
camino hacia la obtenci\'on de mi grado. Adem\'as quiero agradecer el
apoyo del Centro Milenio para Estudios de Supernovas a trav\'es del
subsidio P06--045--F financiado por el ``Programa Bicentenario de
Ciencia y Tecnolog\'ia de CONICYT'' y por el ``Programa Iniciativa
Cient\'ifica Milenio de MIDEPLAN'', y el apoyo recibido por parte del
Centro de Astrof\'{\i}sica FONDAP 15010003 y de Fondecyt a trav\'es
del subsidio 1060808.

\end{acknowledgement}

 \tableofcontents
 \listoffigures
 \listoftables

 \newpage
\thispagestyle{empty}
\mbox{}
\begin{dedication}
\vspace*{12cm}
\begin{flushright}
{\em \Large Para mi madre,\\ sin ella nada de esto hubiese sido
  posible.}
\end{flushright}
\end{dedication}
\newpage
\thispagestyle{empty}
\mbox{}

\end{frontmatter}

\chapter{Introduction\label{INTR}}

Supernovae (hereafter SNe) correspond to the explosive, high-energy
final stages of some stars. The mechanical energy released in these
powerful events can reach as much as 10$^{51}$ erg (or 1~foe), and
their peak luminosities can be comparable to the total light of their
host-galaxies.  SNe~can be classified in two types, either ``Core
Collapse'' or ``Thermonuclear'', depending on their explosion
mechanisms.

Core-collapse SNe (CCSNe) are closely associated to star forming
regions in late-type galaxies \citep{And08}. Therefore they have been
attributed to massive stars born with $>$~8~\msun\ that undergo the
collapse of their iron cores after a few million years of evolution
and the subsequent ejection of their envelopes \citep{Bu00}. These SNe
leave a compact object as a remnant, either a neutron star or a black
hole \citep{BZ34,Ar96}.  The core-collapse model received considerable
support with the first detection of neutrinos from the Type~II
SN~1987A \citep{Svo87}, although no compact remnant has been found so
far in the explosion site. Among CCSNe we can observationally
distinguish those with prominent hydrogen lines in their spectra
(dubbed Type~II), those with no H but strong He lines (Type~Ib), and
those lacking H or He lines (Type~Ic) \citep{Min41,Fil97}. Although
all of these objects are thought to share the same explosion
mechanism, their different observational properties are explained in
terms of how much of their H-rich and He-rich envelopes were retained
prior to explosion. When the star explodes with a significant fraction
of its initial H-rich envelope, in theory it should display a H-rich
spectrum and a light curve characterized by a phase of $\sim$~100~days
of nearly constant luminosity followed by a sudden drop of~2--3 mag
\citep{Nad03,Utr07,Ber08}. Nearly~50\% of all CCSNe belong to this
class of Type~II ``Plateau'' SNe (\sneiip).

Thermonuclear SNe are characterized by the lack of hydrogen and helium
in their spectra. Their early-time spectra show strong lines due to
intermediate mass elements \citep[e.g. \ion{Si}{2}, \ion{Ca}{2},
  \ion{Mg}{2};][]{Fil97}. They are found both in elliptical, spiral,
or irregular galaxies. These objects are thought to originate in
low-mass stars that end their lifes as white dwarfs and explode after
a period of mass accretion from a companion star, leaving no compact
remnants behind them \citep{HN00}. Observationally they are referred
as Type~Ia SNe.

Given their large intrinsic luminosities, SNe have long been
considered potential probes for extra-galactic distance determinations
and the measurement of the cosmological parameters that drive the
Universe dynamics.  Among all types of SNe, the Type Ia family is the
one displaying the highest degree of homogeneity \citep{Li01}, both
photometrically and spectroscopically. However, these objects are not
perfect standard candles.  Empirical calibrations have allowed us to
standardize their luminosities to levels of~$\sim$~0.15--0.22 mag and
determine distances to their host-galaxies with an unrivaled precision
of~$\sim$~7\%--10\% \citep{Phi93,Ham96a,Phi99}. This powerful
technique led a decade ago to the construction of Hubble diagrams
between $z=0$--0.5 and measure very precisely the history of the
expansion of the Universe over 5~Gyr of look-back time.  Contrary to
our intuition these observations revealed that the Universe dynamics
is described by an accelerated expansion
\citep{Rie98,Per99,Ast06,WV07}. The discovery of the accelerating
Universe is profoundly connected with theoretical cosmology as it
implies the possible existence of a cosmological constant, a concept
initially introduced by Albert Einstein at the beginning of the
20th century, whose origin still is a mystery.

Although the acceleration of the Universe has been indirectly
confirmed by other independent experiments such as the {\it Wilkinson
Microwave Anisotropy Probe} \citep[WMAP;][]{Spe07,Ben03} and the {\it
Baryon Accoustic Oscillations} \citep[BAO;][]{BG03,SE03}, it is
important to obtain independent confirmation of the Type Ia results.
Although not as bright and uniform as the Type~Ia's, \citet{HP02}
showed that the luminosities of \sneiip\ can be standardized to levels
of 0.4~and 0.3~mag in the $V$- and $I$-bands respectively, thus
converting these objects into potentially useful tools to measure
cosmological parameters.

Even though \sneiip\ are 1--2~mag fainter and much less homogeneous
than SNe~Ia, these objects provide two interesting routes to distance
determinations.  First, the {\it Expanding Photosphere Method}
\citep[EPM,][]{KK74}, a theoretical technique based on atmosphere
models that is independent of the extra-galactic distance scale
\citep{Ea96,DH05}. This method can achieve dispersions of 
0.3~mag in the Hubble diagram, which translates
into a 14\% precision in distance \citep{Sch94,H01,JH08}.
Second, the {\it Standardized Candle Method} (SCM for short), an
empirical technique initially proposed by \citet{HP02} that is based
on an observational correlation between the absolute magnitude of the
SN and the expansion velocity of the photosphere, the {\it
  Luminosity-Expansion Velocity} (\lumvel) relation. This correlation
shows that \sneiip\ with greater luminosities have higher expansion
velocities, which permits one to remove the large ($\sim$~4~mag)
luminosity differences displayed by these objects to levels of only
0.3~mag. So far, the SCM has been applied to 24~low-$z$ ($z<0.05$) SNe
\citep{H03} and more recently by \citet{N06} to 5~high-$z$ ($z<0.29$) SNe.
The latter work was the first attempt to derive cosmological parameters
from \sneiip\ and demonstrated the enormous potential of this class of
objects as cosmological probes.

The upcoming years will witness the deployment of several survey
telescopes, such as the Panoramic Survey Telescope and Rapid Response
System \citep[Pan-STARRS;][]{Hod04}, the Large Synoptic Survey
Telescope \citep[LSST;][]{Tys03}, the Visible and Infrared Survey
Telescope \citep[VISTA;][]{Eme04}, the VLT Survey Telescope
\citep[VST;][]{Cap03}, the Dark Energy Survey \citep[DES;][]{Cas07},
and the Skymapper \citep{Gra06}, all of which offer the promise to
discover SNe by the thousands. Whether we use it or not, we will be
inevitably confronted by enormous amounts of data on \sneiip\ which
will contain valuable cosmological information. In spite of the great
promise shown by the SCM, it still suffers from a variety of problems
which need to be addressed: 1)~the lack of a well-defined maximum in
the light curves has prevented us from defining the phase of each
event; 2)~each SN shows a different color evolution, which has
compromised the use of the photometric data for the determination of
host-galaxy extinction; 3)~the assumption for dereddening employed by
\citet{HP02}, namely that all SNe reach the same asymptotic color
toward the end of the plateau phase, has often led to negative
extinctions so it still needs to be further tested; 4)~the small
sample size used so far, especially the scarcity of SNe in the Hubble
flow, has prevented a proper determination of the intrinsic precision
of the method.

The purpose of this research is to take advantage of the larger and
more distant sample of \sneiip\ available to us today to address the
issues mentioned above, refine the SCM, and assess the feasibility of
using \sneiip\ to measure distances in the Universe, in preparation
for the massive samples of high-$z$ SNe which will be produced in the
years to come.  With this purpose in mind we have developed a robust
mathematical procedure to model the light curves, color curves, and
velocity curves, in order to obtain a more accurate determination of
the relevant parameters required by the SCM (magnitudes, colors and
ejecta velocities).  This work makes use of 37 SNe to construct a
Hubble diagram (HD), evaluate the accuracy of the SCM, and obtain an
independent determination of the Hubble constant. Since our sample has
several objects in common with the recent EPM analysis of
\citet{JH08}, we perform a comparison between SCM and EPM.

We organize this thesis as follows.  In \S~\ref{DATA} we describe all
of the observational material used in this work such as the
telescopes, intruments and surveys involved.  The analysis,
methodology and procedures, such as the $A_G,\ K$ and $A_{host}$
corrections, are explained in detail in \S~\ref{METH}.  The
dereddening analysis, the Hubble diagram, the value of the Hubble
constant, and the distance comparison between SCM and EPM are
addressed in \S~\ref{ANLS}.  The final remarks in \S~\ref{DISC}
provide a discussion about possible variations of the reddening law
for \sneiip along with comparisons between the value of $H_0$ computed
by us and those derived from other methods. The first part of this
section explores the possibility of a non-standard extinction law in
the SN host-galaxies.  We present our conclusions in \S~\ref{CONC}.

\chapter{Observational Material\label{DATA}}

This work makes use of data obtained in the course of four systematic
SN follow-up programs carried out between 1986--2003: 1)~the Cerro
Tololo SN program (1986--1996); 2)~the Cal\'an/Tololo SN program (CT,
1990--1993); 3)~the Optical and Infrared Supernova Survey (SOIRS,
1999--2000); 4)~the Carnegie Type~II Supernova Program (CATS,
2002--2003).  As a result of these efforts photometry and spectroscopy
(some IR but mostly optical) was obtained for nearly 100~SNe of all
types, 51~of which belong to the Type~II class. All of the optical
data have been already reduced and they are being prepared for
publication \citep{H08}.  Next we describe in general terms the data
acquisition and reduction procedures. For more details the reader can
refer to \citet{H08}.

\section{Photometric Data}

The photometry was acquired with telescopes from {\it Cerro Tololo
  Inter-American Observatory} (CTIO), {\it Las Campanas Observatory}
(LCO), the {\it European Southern Observatory} (ESO) in La Silla, and
{\it Steward Observatory} (SO).  A host of different telescopes and
intruments were used to generate this dataset as shown in Table
\ref{TbIns}. In all cases we employed CCD detectors and standard
Johnson-Kron-Cousins-Hamuy $UBVRIZ$ filters \citep{John66,Cou71,H01a}.
\begin{table}[p]
\begin{center}
{\scshape \caption{Telescopes and instruments}\label{TbIns}}
\vspace{3mm}
\begin{tabular}{lcc}
\hline\hline
Telescope    &Instrument &Phot/Spec \\
\hline
CTIO 0.9m    &  CCD      &  P   \\
YALO 1.0m    &  ANDICAM  &  P   \\
YALO 1.0m    &  2DF      &  S   \\
CTIO 1.5m    &  CCD      &  P   \\
CTIO 1.5m    &  CSPEC    &  S   \\
Blanco 4.0m  &  CSPEC    &  S   \\
Blanco 4.0m  &  2DF      &  S   \\
Blanco 4.0m  &  CCD      &  P   \\
Swope 1.0m   &  CCD      &  P   \\
du Pont 2.5m &  WFCCD    &  P/S \\ 
du Pont 2.5m &  MODSPEC  &  S   \\
du Pont 2.5m &  2DF      &  S   \\
du Pont 2.5m &  CCD      &  P   \\
Baade 6.5m   &  LDSS2    &  P/S \\
Baade 6.5m   &  B\&C      &  S   \\
Clay 6.5m    &  LDSS2    &  P/S \\
ESO 1.52m    &  IDS      &  S   \\
Danish 1.54m &  DFOSC    &  P/S \\
ESO 2.2m     &  EFOSC2   &  S   \\
NTT 3.58m    &  EMMI     &  S   \\
ESO 3.6m     &  EFOSC    &  S   \\
Kuiper 61"   &  CCD      &  P   \\
Bok 90"      &  B\&C      &  S   \\
\hline
\end{tabular}\vspace{-5mm}

\tablecomments{Whether the instrument was used for photometry (P),
  spectroscopy (S) or both (P/S) is listed in column~3.}


\normalsize
\end{center}
\end{table}

The images were processed with IRAF\footnote{IRAF is distributed by
  the National Optical Astronomy Observatories, which are operated by
  the Association of Universities for Research in Astyronomy, Inc.,
  under cooperative agreement with the National Science Foundation.}
through bias subtraction and flatfielding. All of them were further
processed through the step of galaxy subtraction using template images
of the host-galaxies. Photometric sequences were established around
each SN based on observations of Landolt and Hamuy standards
\citep{Lan92,H94}. The photometry of all SNe was performed
differentially with respect to the local sequence on the
galaxy-subtracted images.  The transformation of instrumental
magnitudes to the standard system was done by taking into account a
linear color-term and a zero-point.  Although this procedure partially
removes the instrument-to-instrument differences in the SN magnitudes,
it should be kept in mind that significant systematic discrepancies
can still remain owing to the non-stellar nature of the SN spectrum
\citep[e.g.~][]{H90}.

\section{Spectroscopic Data}

The spectroscopic data were also obtained with a great variety of
instruments and telescopes as shown in Table \ref{TbIns}. The
observations consisted of the SN observation immediately followed by
an arc lamp taken at the same position in the sky, and 2--3~flux
standards per night from the list of \citet{H94}.

We always used CCD detectors, in combination with different
gratings/grisms and blocking filters. The reductions were performed
with IRAF and consisted in bias subtraction, flatfielding, 1D~spectrum
extraction and sky subtraction, wavelength and flux calibration. No
attempts were done to remove the telluric lines.

\section{Subsample used for this work}\label{sample}

Of the 51~SNe~II observed in the course of these four surveys, a
subset of 33~objects comply with the requirements of 1)~having light
curves with good temporal coverage; 2)~having sufficient spectroscopic
temporal coverage; 3)~being a member of the plateau class. To this
sample we added four SNe from the literature: SN~1999gi, SN~2004dj,
SN~2004et, and SN~2005cs. Complementary photometry for SN~2003gd
obtained by \citet{vDLF03} and \citet{He05} was also incorporated in
our analysis. Table~\ref{Tb1} lists our final sample of 37~\sneiip.
For each SN this table includes the name of the host-galaxy,
equatorial coordinates, the heliocentric redshift and its source, the
reddening due to our own Galaxy \citep{SFD98} and the survey or
reference for the data. \begin{table}[p]
\begin{center}
\footnotesize
{\scshape \caption{Supernova sample}\label{Tb1}}
\begin{tabular}{lclrcccc}
\hline\hline
SN name &Host Galaxy      &RA(J2000)    &DEC(J2000)    &$z_{host}$\tablenotemark{a} &(s)\tablenotemark{b} &$E(\bv)_{GAL}$  &References  \\
\hline                                                                                          
1991al  &LEDA 140858      &19 42 24.00  &--55 06 23.0  &0.01525  &HP02	&0.051          &1      \\   
1992af  &ESO 340-G038     &20 30 40.20  &--42 18 35.0  &0.01847  &NED	&0.052          &1      \\   
1992am &MCG--01--04--039   &01 25 02.70  &--04 39 01.0  &0.04773  &NED	&0.049          &1      \\   
1992ba  &NGC 2082         &05 41 47.10  &--64 18 01.0  &0.00395  &NED	&0.058          &1      \\   
1993A   &$anonymous$      &07 39 17.30  &--62 03 14.0  &0.02800  &NED	&0.173          &1      \\    
1999br  &NGC 4900	   &13 00 41.80  &+02 29 46.0  &0.00320  &NED	&0.024          &2   \\    
1999ca  &NGC 3120         &10 05 22.90  &--34 12 41.0  &0.00931  &NED   &0.109          &2   \\
1999cr  &ESO 576--G034	   &13 20 18.30  &--20 08 50.0  &0.02020  &NED	&0.098          &2   \\    
1999em  &NGC 1637 	   &04 41 27.04  &--02 51 45.2  &0.00267  &NED	&0.040          &2   \\    
1999gi  &NGC 3184 	   &10 18 17.00  &+41 25 28.0  &0.00198  &NED	&0.017          &3     \\    
0210  &MCG +00--03--054   &01 01 16.80  &--01 05 52.0  &0.05140  &NED   &0.036          &4    \\
2002fa  &NEAT J205221.51  &20 52 21.80  &+02 08 42.0  &0.06000  &NED   &0.099           &4    \\
2002gw  &NGC 922 	   &02 25 02.97  &--24 47 50.6  &0.01028  &NED	&0.020          &4    \\    
2002hj  &NPM1G +04.0097    &02 58 09.30  &+04 41 04.0  &0.02360  &NED   &0.115           &4    \\
2002hx  &PGC 23727 	   &08 27 39.43  &--14 47 15.7  &0.03099  &NED	&0.054          &4    \\    
2003B   &NGC 1097 	   &02 46 13.78  &--30 13 45.1  &0.00424  &NED	&0.027          &4 	  \\    
2003E  &MCG--4--12--004    &04 39 10.88  &--24 10 36.5  &0.01490  &J08 &0.048          &4 	  \\    
2003T   &UGC 4864 	   &09 14 11.06  &+16 44 48.0  &0.02791  &NED	&0.031          &4 	  \\    
2003bl  &NGC 5374 	   &13 57 30.65  &+06 05 36.4  &0.01459  &J08 &0.027          &4       \\    
2003bn  &2MASX J10023529  &10 02 35.51  &--21 10 54.5  &0.01277  &NED	&0.065          &4       \\    
2003ci  &UGC 6212 	   &11 10 23.83  &+04 49 35.9  &0.03037  &NED	&0.060          &4       \\    
2003cn  &IC 849 	   &13 07 37.05  &--00 56 49.9  &0.01811  &J08 &0.021         &4       \\    
2003cx  &NEAT J135706.53  &13 57 06.46  &--17 02 22.6  &0.03700  &NED	&0.094          &4       \\    
2003ef  &NGC 4708 	   &12 49 42.25  &--11 05 29.5  &0.01480  &J08 &0.046         &4       \\    
2003fb  &UGC 11522 	   &20 11 50.33  &+05 45 37.6  &0.01754  &J08 &0.183          &4       \\    
2003gd  &M74 		   &01 36 42.65  &+15 44 20.9  &0.00219  &NED	&0.069          &4, 5, 6  \\    
2003hd &MCG--04--05--010   &01 49 46.31  &--21 54 37.8  &0.03950  &NED	&0.013          &4       \\    
2003hg  &NGC 7771 	   &23 51 24.13  &+20 06 38.3  &0.01427  &NED	&0.074          &4       \\    
2003hk  &NGC 1085 	   &02 46 25.76  &+03 36 32.2  &0.02265  &NED	&0.037          &4       \\    
2003hl  &NGC 772 	   &01 59 21.28  &+19 00 14.5  &0.00825  &NED	&0.073          &4       \\    
2003hn  &NGC 1448 	   &03 44 36.10  &--44 37 49.0  &0.00390  &NED	&0.014          &4       \\    
2003ho  &ESO 235--G58 	   &21 06 30.56  &--48 07 29.9  &0.01438  &NED	&0.039          &4       \\    
2003ip  &UGC 327          &00 33 15.40  &+07 54 18.0  &0.01801  &NED   &0.066           &4       \\
2003iq  &NGC 772 	   &01 59 19.96  &+18 59 42.1  &0.00825  &NED	&0.073          &4       \\    
2004dj  &NGC 2403         &07 37 17.00  &+65 35 58.1  &0.00044  &NED   &0.040           &7        \\
2004et  &NGC 6946         &20 35 25.30  &+60 07 18.0  &0.00016  &NED   &0.342           &8        \\
2005cs  &NGC 5194	   &13 29 53.40  &+47 10 28.0  &0.00154  &NED   &0.035          &9, 10     \\
\hline
\end{tabular}
\vspace{-4mm}

\tablenotetext{a}{~Heliocentric host-galaxy redshifts}
  
\tablenotetext{b}{~Sources of host-galaxy redshifts:
  HP02~=~\citet{HP02}; J08~=~\citet{JH08}; NED~=~NASA/IPAC
  Extragalactic Database}

\tablerefs{(1) Cal\'an/Tololo Supernova Program; (2) SOIRS; (3)
  \citet{L02b}; (4) Carnegie Type II Supernovae Survey (CATS); (5)
  \citet{vDLF03}; (6) \citet{He05}; (7) \citet{V06}; (8) \citet{S06};
  (9) \citet{P06}; (10) \citet{T06}.}

\normalsize
\end{center}
\end{table}

\chapter{Methodology and Procedures\label{METH}}

\section{AKA corrections} \label{cor}

The photon flux measured by an observer is related to the intrinsic
luminosity of the source, its distance, dust extinction along the line
of sight, and the shift of the spectral energy distribution (SED) to
longer wavelengths caused by the expansion of the Universe.  More
specifically, if $L_{\lambda'}$ is the SN rest-frame emergent
luminosity in units of $erg~s^{-1}~${\it \AA}$^{-1}$, the photon flux
per unit wavelength seen by the observer (in units of
$photons~cm^{-2}~s^{-1}~${\it \AA}$^{-1}$), is

\begin{equation}
 n(\lambda) =
 \frac{L_{\lambda'}~(\lambda'/hc)~A_{host}(\lambda')~A_G(\lambda)}{4
   \pi {d_L}^2}
\label{flux_eq}
\end{equation}

\noindent
where $\lambda'$ is the SN rest-frame wavelength, $\lambda$=$\lambda'
(1+z)$ is the observer's wavelength, and $d_L$ is the luminosity
distance to the source.

The flux is modified along its journey to the observer in the
following order: host-galaxy reddening ($A_{host}$), redshift
($K$-term), and Galactic reddening ($A_{G}$) (or AKA, for short).  In
order to be able to extract the distance from the observed fluxes, it
proves necessary to remove Nature's imprint on the observed magnitudes
in reverse order.  To undo Nature's work, we first need to correct the
observed magnitudes for Galactic extinction ($A_G$), which is
equivalent to moving the observer outside the Milky Way. Then we must
move the observer just outside the host-galaxy, for which we must
correct the spectrum for the redshift caused by the expansion of the
Universe ($K$~correction). Finally, we must correct the magnitudes for
host-galaxy extinction ($A_{host}$), which brings the observer to the
SN rest-frame. For the latter step we used as a first approximation
the reddenings determined by \citet{D08}. In a second iteration we
applied our own reddenings (see~\S~\ref{aho}).

In this work the AKA corrections are computed numerically using a
library of 196~\sneiip\ optical spectra. Most of the spectra comes
from our own database of 44~Type~II SNe, 10~of which are not included
in Table~\ref{Tb1} (SN~1987A, SN~1988A, SN~1989L, SN~1990E, SN~1990K,
SN~1993S, SN~1999eg, SN~2000cb, SN~2002gd, and SN~2003ib). These
10~additional objects belong to the Type~II class, although do not
comply with the requirements of \S~\ref{sample} to be included in this
analysis. The database comprises spectra covering the plateau and
nebular phases.  Each spectrum is brought to the SN rest-frame, i.e.,
to redshift zero and null Galactic reddening using the redshifts and
Galactic reddenings listed in Table~\ref{Tb1}. Also the spectra are
corrected using our own host-galaxy reddenings, as determined below
(see~\S~\ref{aho}).

\subsection{$A_G$ corrections}\label{AGcor}

\begin{figure}[p]
\begin{center}
\includegraphics[angle=0,scale=0.8]{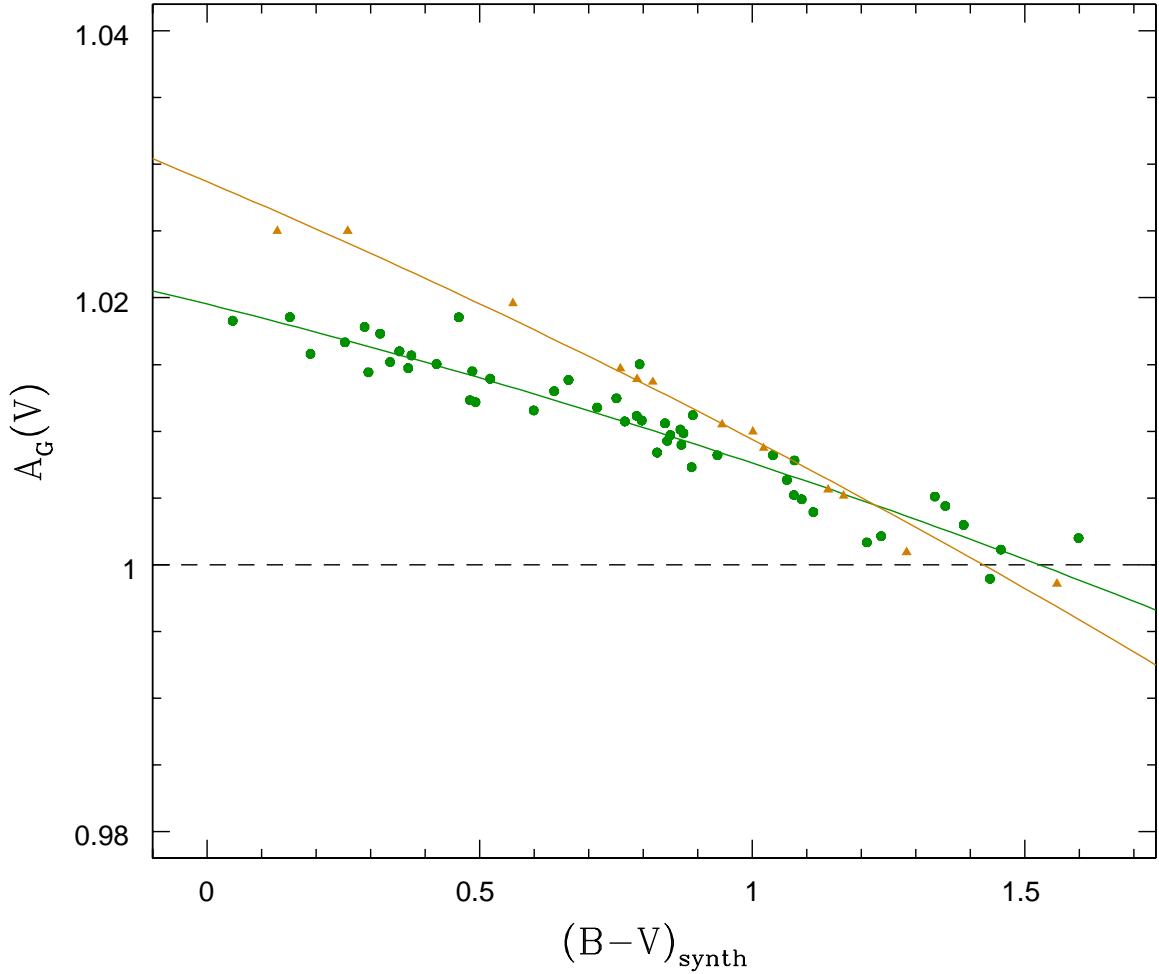}\vspace{-8mm}
\caption[Apparent $A_G(V)$ vs. the synthetic \bv\ color]{The
  dependence of the apparent $A_G(V)$ on the synthetic \bv\ color
  obtained from the library of spectra must be taken into account. The
  dashed line defines the input $A_G^{true}(V)=1$~mag, indicating
  variations up to 0.025~mag. Plateau-phase spectra (filled circles)
  are shown and fitted in green, whereas the data from the nebular
  phase are shown with brown triangles. This diagram shows that the
  $A_G(V)$ versus \bv\ calibration must be treated separately for the
  plateau and nebular phases.
\label{FgAG}}
\end{center}
\end{figure}

We define the synthetic apparent Galactic extinction correction as the
difference between the magnitude of an unextinguished SED and the
magnitude of the extinguished SED, i.e.,

\begin{equation}
 A_G(\overline\lambda) = -2.5~log_{10}~\frac{\int L_{\lambda'} ~
   \lambda' ~ A_{host}(\lambda') ~ A_G(\lambda) ~ S(\lambda) ~
   d\lambda} {\int L_{\lambda'} ~ \lambda' ~ A_{host}(\lambda') ~
   S(\lambda) ~ d\lambda}
\label{AG_eq}
\end{equation}

\noindent
where $S(\lambda)$ is the filter transmission function (see
Appendix~\ref{appA}). We apply this definition to all the spectra of
our library using the $BVRI$ bandpasses.  Since the effective
wavelength $\overline\lambda$ changes as the SN spectrum evolves, we
expect the apparent $A_G$ to be a function of the color of the SED. To
examine this point, Figure~\ref{FgAG} shows $A_G(V)$ against \bv\ for
the specific case of $A_G^{true}(V)$\footnote{$A_G^{true}$ is the
visual absolute extinction at 5500 \r A as defined
by~\citet{Car89}.}~=~1 mag, where we identify with different colors
spectra from the plateau and nebular phases. 
 Although
the apparent $A_G$ does not show large variations, the dependence on
color is evident and must be taken into account. For this purpose we
chose to fit these relations with 3rd order polynomials. With this
calibration we can proceed to interpolate the corresponding value of
the apparent $A_G$ for the specific color of the SN and subtract this
value from the observed magnitudes at every epoch we have photometry
to obtain ($m_{obs} - A_G$).  The RMS of our calibration
---0.0008~mag--- provides an estimate of the uncertainty in the
correction, so we add this number in quadrature to the uncertainty in
the observed magnitudes.

\subsection{$K$~corrections}\label{Kcor}

We use a similar procedure to compute the $K$ corrections. The
synthetic $K$-term is defined as the difference between the magnitude
of a redshifted SED and the magnitude of the zero-redshift SED, i.e.,

\begin{equation}
 K(\overline\lambda) = -2.5~log_{10}~\frac{\int L_{\lambda'} ~
   \lambda' ~ A_{host}(\lambda') ~ S(\lambda) ~ d\lambda} {\int
   L_{\lambda} ~ \lambda ~ A_{host}(\lambda) ~ S(\lambda) ~ d\lambda}
\label{K_eq}
\end{equation}

\begin{figure}[p]
\begin{center}
\includegraphics[angle=0,scale=0.8]{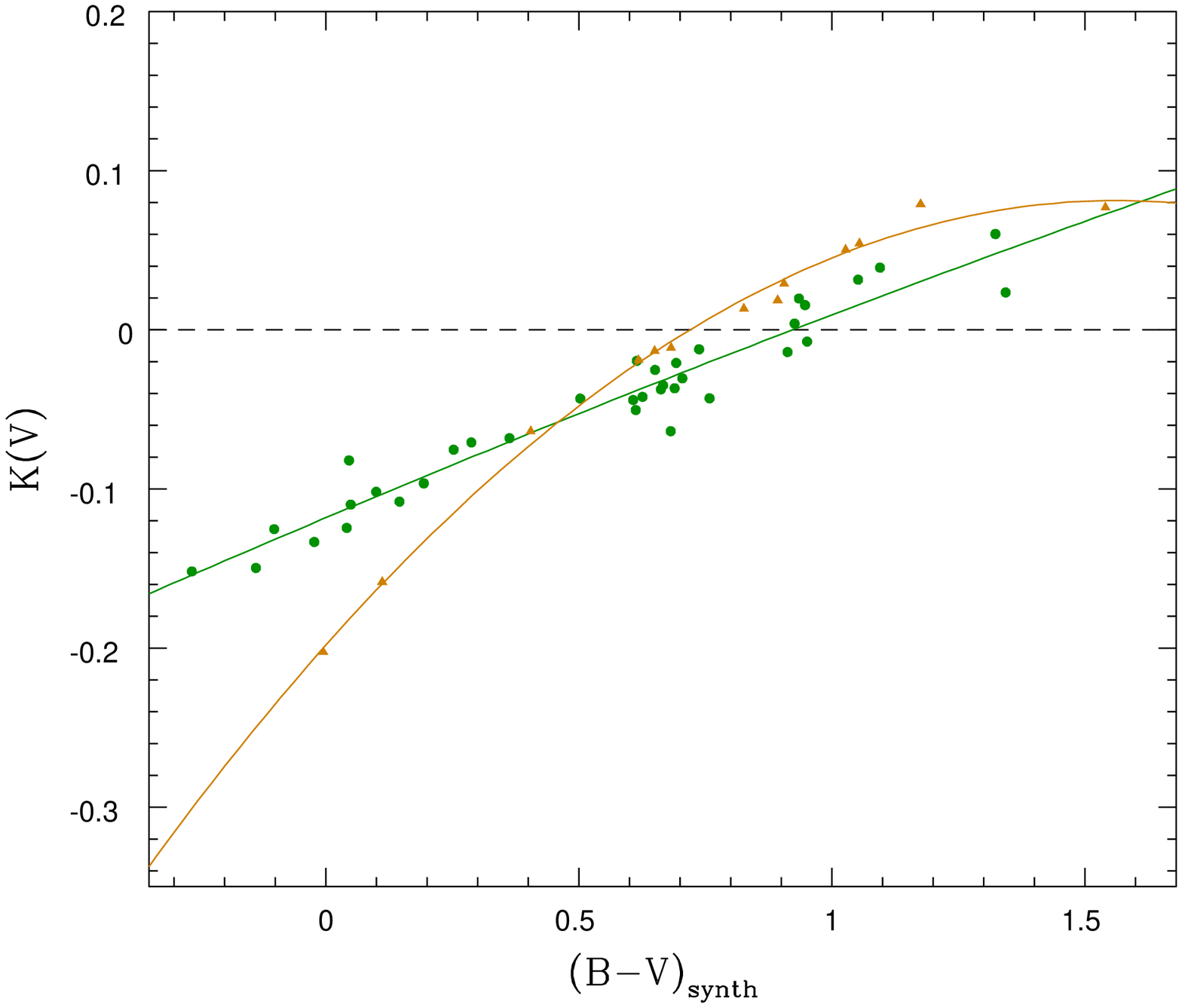}\vspace{-8mm}
\caption[$K(V)$ vs. the synthetic \bv\ color]{The dependence of $K(V)$
  on the synthetic color \bv\ obtained from the library of
  spectra. For a redshift of $z=0.05$ the $K$-terms reach absolute
  values up to 0.2~mag. Plateau-phase spectra (filled circles) are
  shown and fitted in green, whereas data from the nebular phase are
  shown with brown triangles. The diagram shows that the $K(V)$ versus
  \bv\ calibration must be treated separately for the plateau and
  nebular phases.
\label{FgK}}
\end{center}
\end{figure}

\noindent
This definition is equivalent to that given by \citet{Schn83} except
that they use fluxes per unit frequency.  By definition the $K$-term
is a color, therefore we expect it to correlate with the broad-band
colors of the SED.  To demonstrate this point, Figure~\ref{FgK} shows
$K$ versus \bv\ for the specific case of the $V$-band for $z$=0.05.
 This dependence is very useful as it permits us to
interpolate $K$~values for the specific color of the SN and correct
the photometry at the plateau or nebular phases separately.  The
result of this correction is the quantity ($m_{obs} - A_G - K$). The
RMS of the points around the polynomial fits are of the order of
$\sim$~0.001--0.01~mag, and these are quadratically added to the error
of ($m_{obs} - A_G$) in order to account for the uncertainties in the
$K$ correction.

\subsection{$A_{host}$ corrections}

Finally, we must repeat the previous process for the $A_{host}$
correction. The following equation

\begin{equation}
 A_{host}(\overline\lambda) = -2.5~log_{10}~\frac{\int L_{\lambda} ~
   \lambda ~ A_{host}(\lambda) ~ S(\lambda) ~ d\lambda} {\int
   L_{\lambda} ~ \lambda ~ S(\lambda) ~ d\lambda}
\label{Ah_eq}
\end{equation}

\noindent
defines the apparent $A_{host}$ and gives the difference between the
magnitude of the supernova right outside the host-galaxy ($m_{obs} -
A_G - K$) and the magnitude free of host-galaxy dust ($m_{obs} - A_G -
K - A_{host}$).  \begin{figure}[p]
\begin{center}
\includegraphics[angle=0,scale=0.8]{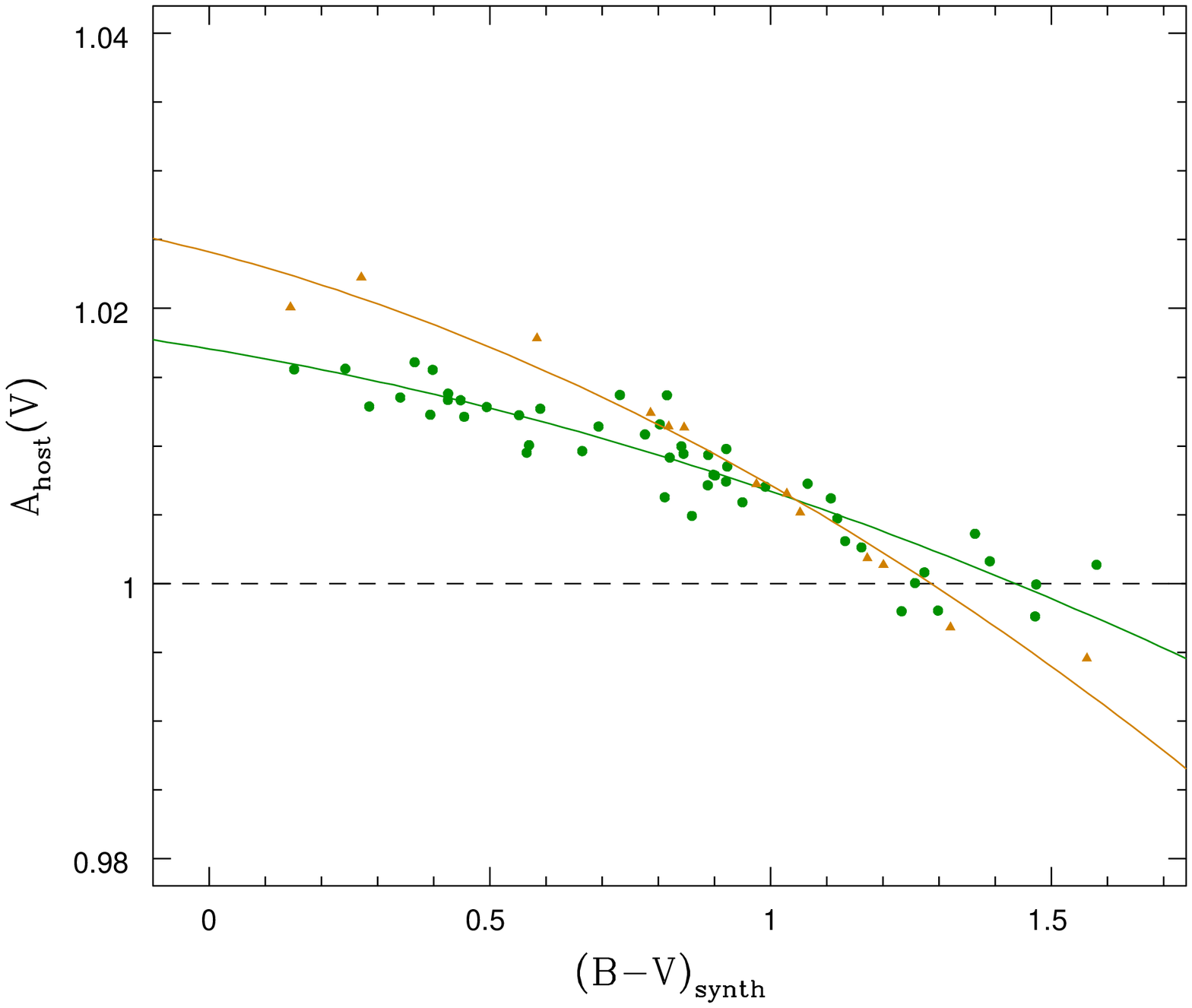}\vspace{-8mm}
\caption[Apparent $A_G(V)$ vs. the synthetic \bv\ color]{The
  dependence of the apparent $A_{host}(V)$ on the synthetic \bv\ color
  obtained from the library of spectra. The dashed line defines the
  input $A_{host}^{true}(V)=1$, indicating variations up to 0.023~mag.
  Plateau-phase spectra (filled circles) are shown and fitted in
  green, whereas data from the nebular phase are shown in brown
  triangles.  This diagram shows that the $A_{host}(V)$ versus
  \bv\ calibration must be treated separately for the plateau and
  nebular phases.
\label{FgAh}}
\end{center}
\end{figure}
 Figure~\ref{FgAh} shows this
correction for all the plateau and nebular spectra of our library
separately for the specific case
$A_{host}^{true}(V)$\footnote{$A_{host}^{true}$ is the visual absolute
  extinction at 5500 \r A as defined by~\citet{Car89}.}~=~1~mag, which
confirms the previous results, namely that the apparent $A_{host}$ is
a function of the broad-band colors.  We fit these relations with a
3$^{\mbox{\scriptsize rd}}$~order polynomial and apply these
calibrations to the SN magnitudes, making sure to include the RMS of
the fits ---varying between~0.003--0.01--- in the error budget.  At
this point, we obtain the magnitude of the SN corrected for Galactic
extinction, redshift and host-galaxy extinction for all epochs
(colors) of the SN.

\section{Fits to light, color and velocity curves}

In the first incarnation of the SCM \citet{HP02} measured all the
relevant quantities (magnitudes, colors and velocities) at fixed
epochs with respect to the time of explosion. In most cases, however,
it proves hard to constrain this time, thus hampering the task to
compare data obtained for different SNe. It would be ideal to have a
conspicuous feature in the light curves but, unlike other SN types,
\sneiip\ generally do not show an obvious maximum during the plateau
phase.

One way around this is to use the end of the plateau as an estimate of
the time origin for each event. Although simple in theory, in practice
it is not easy to measure this time owing to the coarser sampling of
the light curves at this phase.  Thus, our first aim is to implement a
robust light curve fitting procedure in order to obtain a reliable
time origin to be used as a uniform reference epoch to measure
magnitudes, colors, and expansion velocities. In the remainder of this
section we proceed to implement fitting methods to measure reliable
colors and expansion velocities.
\begin{figure}[p]
\begin{center}
\includegraphics[angle=0,scale=0.8]{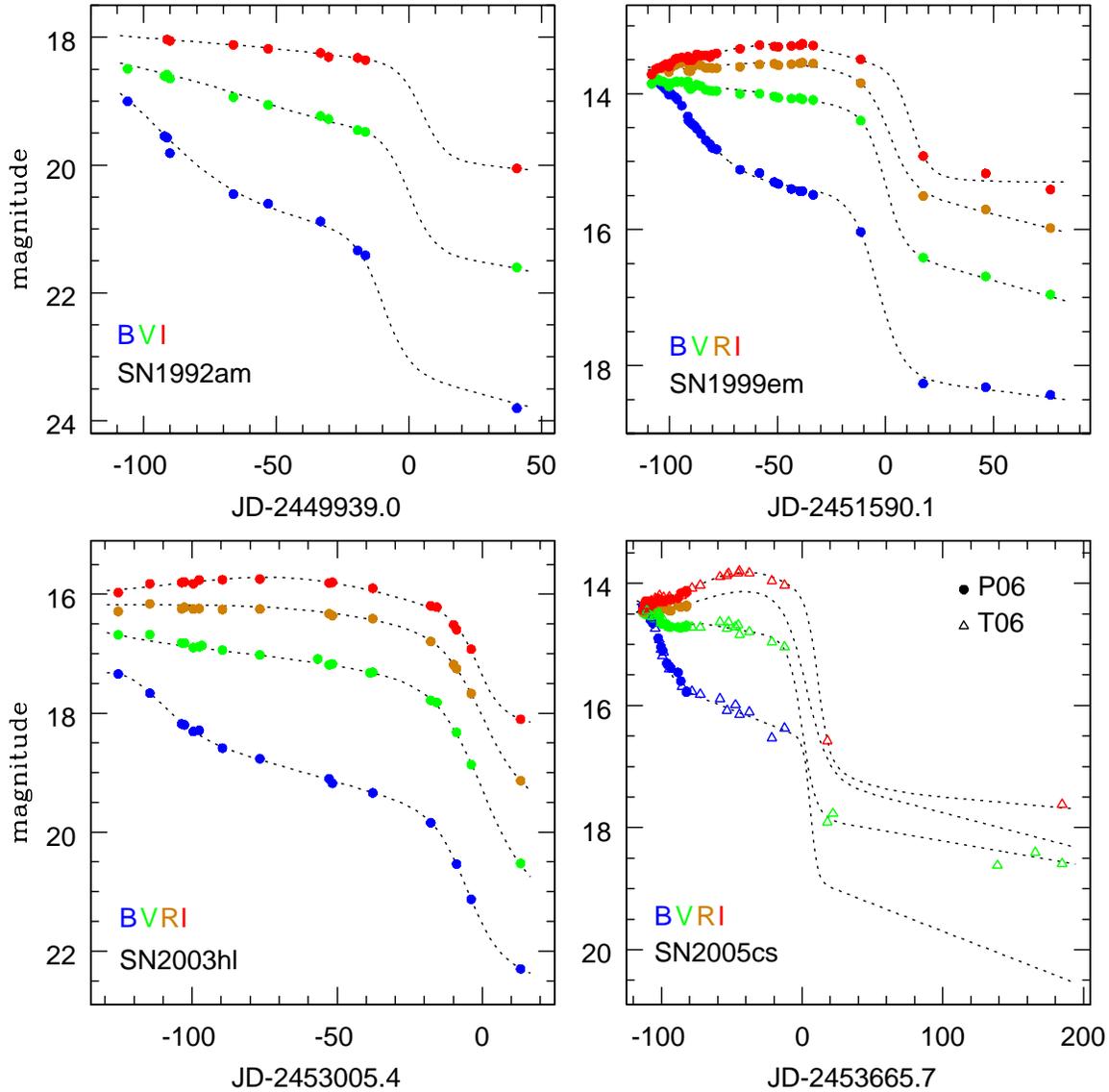}
\caption[$BVRI$ light curves of four SNe]{Four representative light
  curves of the SN sample. Data points are colored according to the
  filters. $BVRI$ magnitudes are respectively shown in blue, green,
  brown and red. The dotted line corresponds to the analytic fit (see
  Fig.~\ref{FgFunc}). In the $B$ and $R$ light curve of SN~2005cs we
  can appreciate the quality of the procedure, since even without data
  points the code manages to achieve reasonable fits. The light curves
  of this SN were complemented with photometry from the literature
  \citep[P06;][]{P06}\citep[T06;][]{T06}.
\label{FgLCs}}
\end{center}
\end{figure}
\begin{figure}[p]
\begin{center}
\includegraphics[angle=0,scale=0.8]{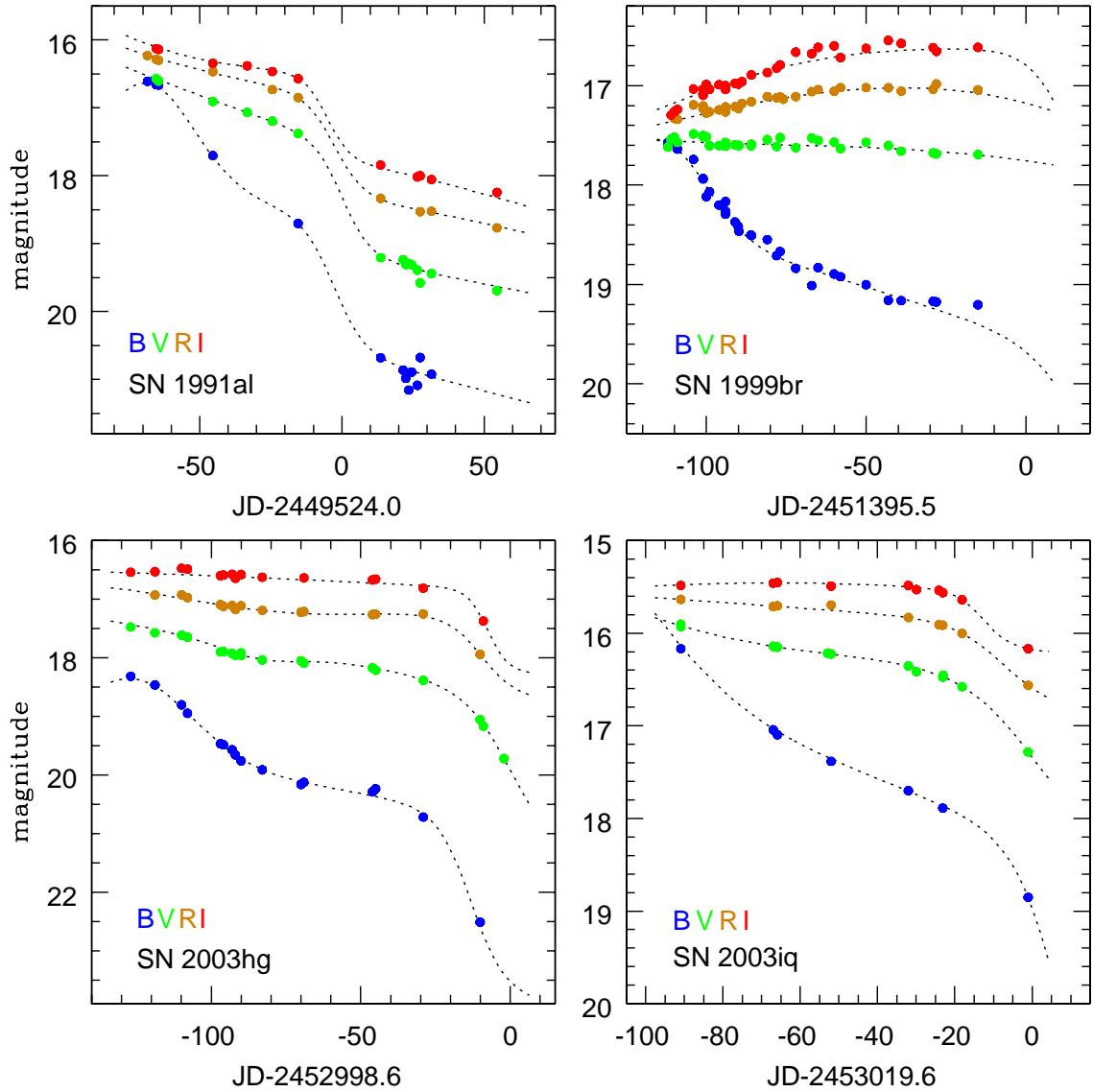}
\caption[$BVRI$ light curves of four other SNe]{Four representative
  light curves of the SN sample. Data points are colored in the same
  way as in Fig.~\ref{FgLCs}. The dotted line corresponds to the
  analytic fit (see Fig.~\ref{FgFunc}).
\label{FgLCs2}}
\end{center}
\end{figure}
\begin{figure}[p]
\begin{center}
\vspace{-1cm}
\plottwo{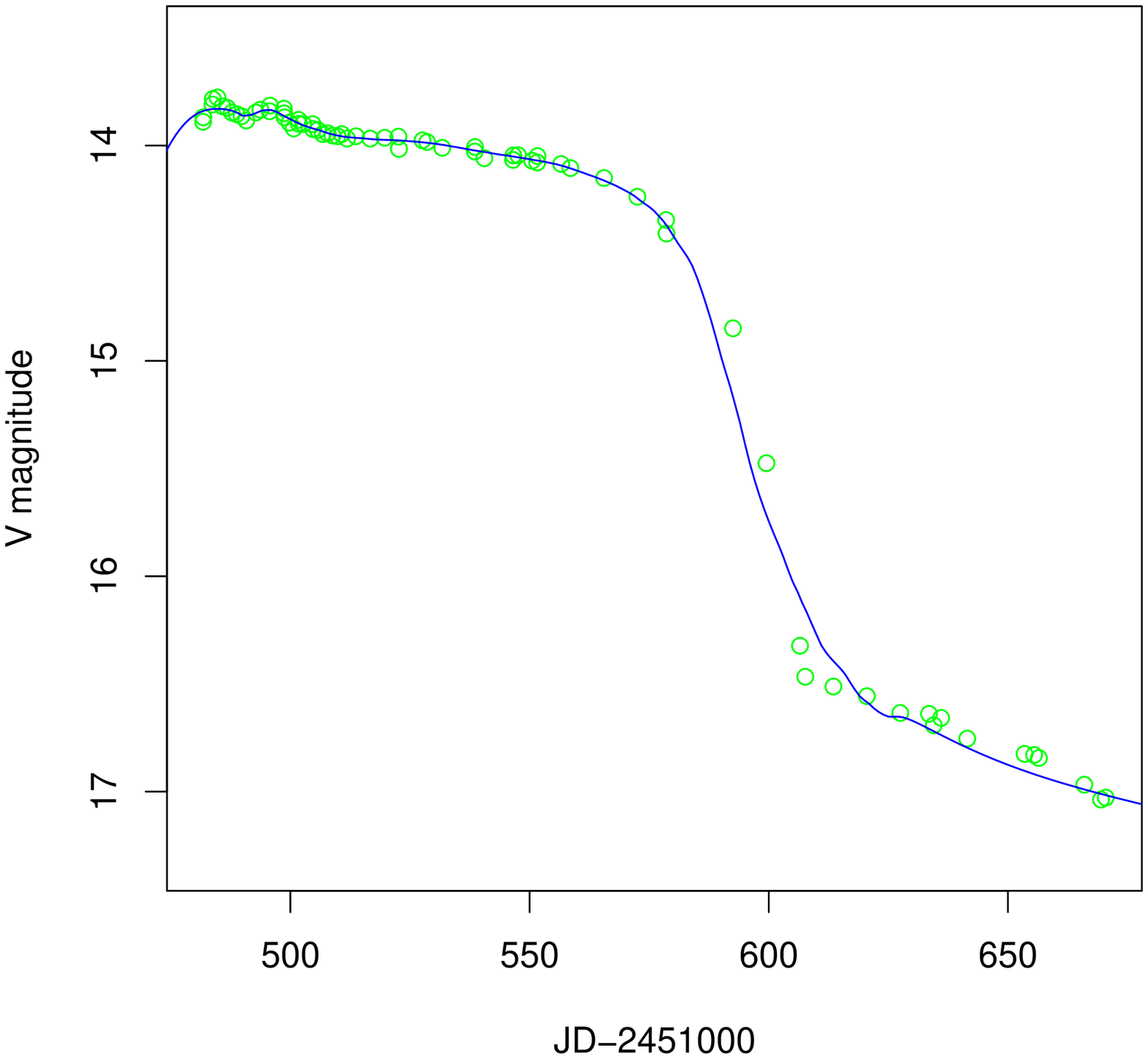}{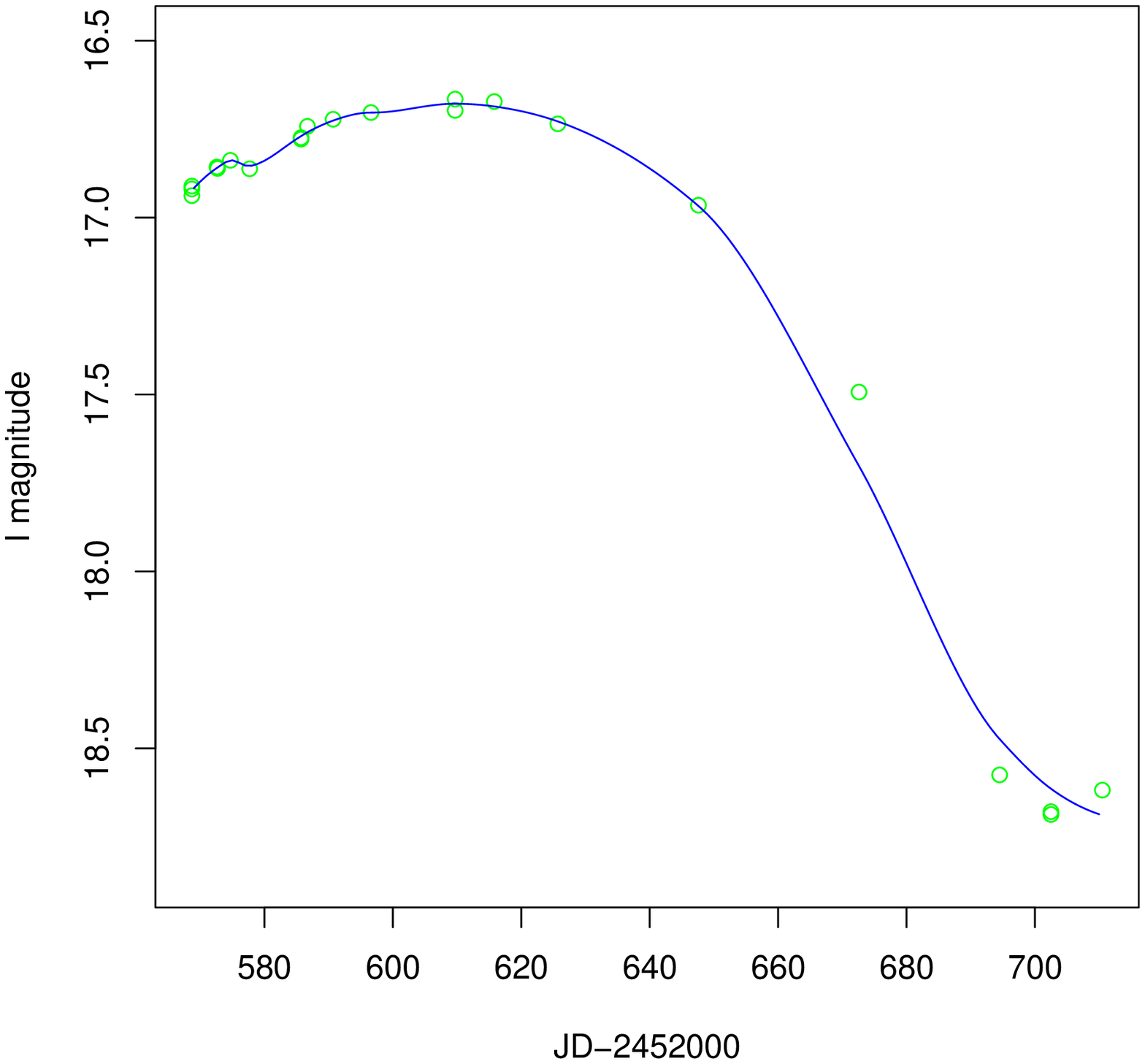}
\vspace{-0.5cm}
\caption[{\tt Loess} fits to light curves]{{\tt Loess} fits to the
  $V$~light curve of SN~1999em (left panel) and to the $I$~light curve
  of SN~2002gw (right panel). Although the fits do a good job on small
  scales, the transition evidently is not well modelled.
\label{FgR1}}
\columnwidth=2.22\columnwidth
\vspace{1cm}
\plottwo{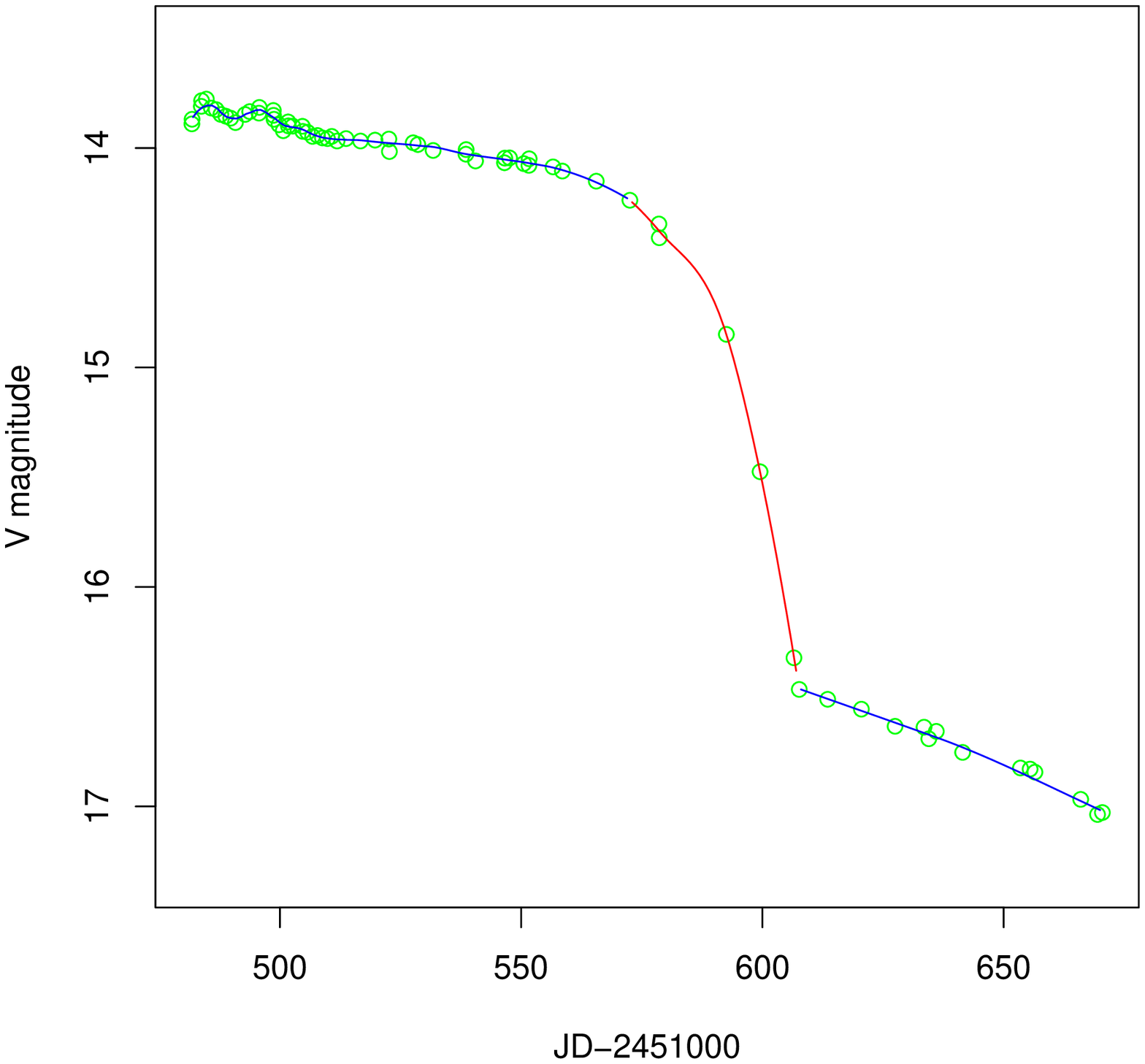}{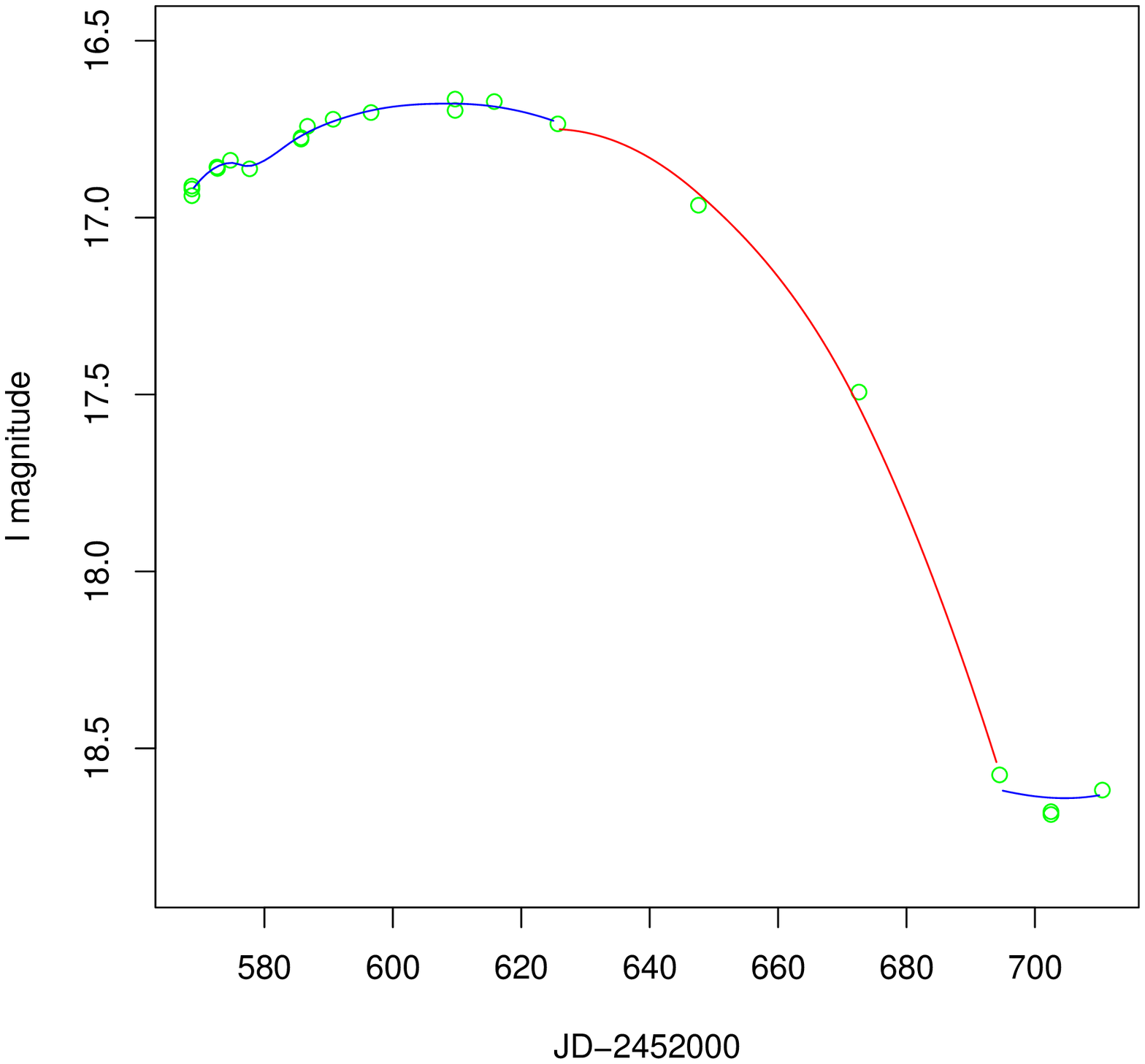}
\vspace{-0.5cm}
\caption[Individual {\tt Loess} fits to the three light curve
  phases]{{\tt Loess} fits to the $V$~light curve of SN~1999em (left
  panel) and to the $I$~light curve of SN~2002gw (right panel). In
  this case the algorithm was applied separately to each phase
  (plateau, transition, and tail).  Despite the satisfactory accuracy
  of the regression, we could not find an easy way to force the
  continuity at the two interfaces.
\label{FgR2}}
\vspace{-0.5cm}
\end{center}
\end{figure}

\subsection{$BVRI$ light curve fits}\label{LCF}

Figures~\ref{FgLCs} and~\ref{FgLCs2} show $BVRI$ light curves of eight
well-observed \sneiip.  These SNe are representative of the whole
sample. As can be seen, there are three distinguishable phases in the
light curves:

\begin{description}

\item[---] The {\it Plateau phase} in which the SN shows an almost
  constant luminosity during the first $\sim$~100~days of its
  evolution. This phase corresponds to the optically thick period in
  which a hydrogen recombination wave recedes in mass, gradually
  releasing the internal energy of the star \citep{Nad03,Utr07,Ber08}.

\item[---] The {\it Linear} or {\it Radioactive Tail}, a linear decay
  in magnitude (exponential in flux) starting about 100~days after the
  explosion. This phase corresponds to the optically thin period
  powered by the $^{56}$Co~$\rightarrow$~$^{56}$Fe radioactive decay
  \citep{WW80}.

\item[---] A {\it Transition phase} of $\sim$~30~days between the
  plateau and linear phases.

\end{description}

Both the plateau and linear phases are trivial to model if taken
separetely, but the abrupt transition makes the fitting task much more
challenging, especially with coarsely sampled light curves. The first
attempt consisted in using the {\it Local Polynomial Regression
Fitting}, a technique developed by \citet{Cl92} within the framework
of the R~environment for statistical computing.  The name of the
method is self-explanatory, since it performs a polynomial regression
over small local intervals along the domain using a routine called
{\tt loess}.  Figure~\ref{FgR1} shows the resulting fits when all the
data are fitted simultaneously by {\tt loess}. The small scale
features of the plateau are nicely reproduced, but {\tt loess} is
unable to model the transition satisfactorily.  Another attempt was
done by fitting separately the three phases of the light curves. The
results, shown in Figure~\ref{FgR2}, satisfy our needs, but the lack
of continuity at the two interfaces led us to look for alternative
approaches.

\begin{figure}[p]
\begin{center}
\includegraphics[angle=0,scale=0.8]{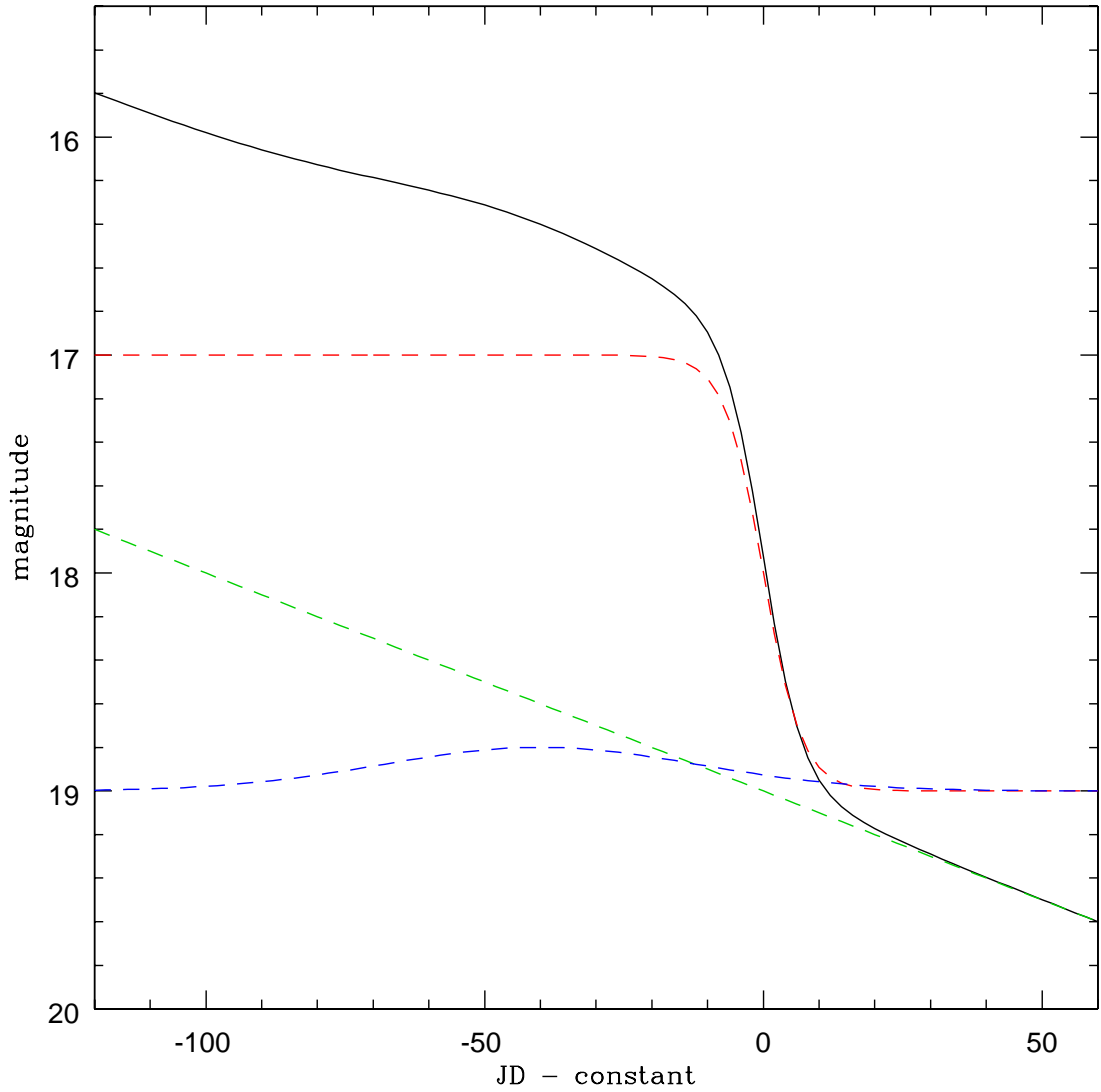}
\caption[Decomposition of the analytic function used to fit the light
  curves]{The analytic function used to fit the light curves is shown
  with the black continuous line. The dashed lines represent the
  decomposition of the main function in its three addends: the
  Fermi-Dirac in red, a straight line in green, and a Gaussian
  function in blue.
\label{FgFunc}}
\end{center}
\end{figure}

After experimenting with several fitting approaches we concluded that
the best fits could be achieved with analytic functions.  After
examining several options we ended up using the arithmetic sum of the
three functions shown in Figure~\ref{FgFunc}:

\begin{description}

\item[$\triangleright$] A Fermi-Dirac function (red dashed line in
  Fig.~\ref{FgFunc}) which provides a very good description of the
  transition between the plateau and radioactive phases.
\begin{equation}
f_{FD}(t)=\frac{-a_0}{1+e^{\frac{t-t_{PT}}{w_0}}}
\end{equation}\label{eqF}
\begin{description}
\item
$a_0$ : represents the height of the step in units of magnitude.
\item
$t_{PT}$ : corresponds to the middle of the transition phase and is a
  natural candidate to define the origin of the time axis.
\item
$w_0$ : quantifies the width of the transition phase. At
  $t=t_{PT}-3w_0$ the height of the step has been reduced by~4.7\%,
  and it decreases down to~95.3\% at $t=t_{PT}+3w_0$.
\end{description}

\item[$\triangleright$] A straight line (green dashed line in
  Fig.~\ref{FgFunc}) which accounts for the slope due to the
  radiactive decay.
\begin{equation}
l(t)=p_0\, (t-t_{PT}) + m_0
\end{equation}
\begin{description}
\item
$p_0$ : corresponds to the slope of radioactive tail and an
  approximate slope for the plateau in units of magnitudes per day.
\item
$m_0$ : corresponds to the zero point in magnitude at $t=t_{PT}$.
\end{description}

\item[$\triangleright$] A Gaussian function (blue dashed line in
  Fig.~\ref{FgFunc}) which serves mainly for fitting the $B$~light
  bump curve and the $I$~light curve curvature during the plateau
  phase.
\begin{equation}
g(t)=-P\,e^{-\left(\frac{t-Q}{R}\right)^2}
\end{equation}
The Gaussian function is also useful for reproducing the small scale
features that can appear in the $V$-band plateau.
\begin{description}
\item
$P$ : height of the Gaussian peak in units of magnitudes.
\item
$Q$ : center of the Gaussian function in days.
\item
$R$ : width of the Gaussian function.
\end{description}
\end{description}

The resulting analytic function we use to model the light curves, is
the sum of the three functions detailed above
\begin{eqnarray}
\mathcal{F}(t)&=&f_{FD}(t)+l(t)+g(t)\nonumber\\
              &=&\frac{-a_0}{1+e^{\frac{t-t_{PT}}{w_0}}}+p_0\,
              (t-t_{PT})+m_0-P\,e^{-\left(\frac{t-Q}{R}\right)^2}
\label{fnc_eq}
\end{eqnarray}
which has 8~free parameters. It is fitted to the individual light
curves using a $\chi^2$ minimizing procedure. In order to find the
minimum $\chi^2$ we use the {\it Downhill Simplex Method} (see
Appendix~\ref{appB}), which although is not very efficient in terms of
the number of iterations required, provides robust solutions. Examples
of the analytic fits are shown as dotted lines in Figures~\ref{FgLCs}
and~\ref{FgLCs2}.  Although we cannot model most of the small scale
features in the plateau, the fitting does a really good job modelling
the transition. Furthermore, the analytic function gives us important
parameters that characterize the light curve shape, particularly
$t_{PT}$ which provides a time origin. In the plots in
Fig.~\ref{FgLCs} the time axis is chosen to coincide with the value of
this parameter obtained from the $V$~light curve.  A critical quantity
in the analysis that follows is $t_{PT}$ and its uncertainty, both of
which determine the uncertainties in all the relevant SCM
quantities. Normally, when the tail phase has been observed the
$\chi^2$ minimizing routine has no difficulties finding $t_{PT}$ and
delivers a credible error.  On the other hand, when the light curve
does not have any late-time data points the routine underestimates
$\sigma(t_{PT})$, in which case we need to provide a more realistic
estimate of this uncertainty. The criteria to fix $\sigma(t_{PT})$
depend on what fraction of the transition phase was sampled. Two
useful parameters are introduced to describe such sampling:

\begin{description}

\item
$t_f$ : the day of the last data point for a given SN.

\item
$t_{end}$ : the approximate day of the end of the $V$-band plateau.
We estimate this time around each observed point, by calculating the magnitude
difference, $\Delta m$, between the previous and the following point. 
We start from the earliest epochs onward in time, until $\Delta m$ exceeds 
0.7 mag. This value is an input parameter for the fitting routine 
(see Appendix \ref{appB}). 

\end{description}

\noindent
The criteria to estimate $t_{PT}$ and $\sigma(t_{PT})$ are the
following,

\begin{enumerate}

\item 
When the transition can be clearly seen, but if we are uncertain that
the last data point (at $t=t_f$) belongs to the tail phase, we set a
conservative estimate of $\sigma(t_{PT})$=5~days.  An example of such
a case is SN~2003hl (Fig.~\ref{FgLCs}).

\item 
When the transition can be clearly seen, but we are sure that the last
data point (at $t=t_f$) does not belong to the tail phase (see
SN~2003hg and SN~2003iq in Fig.~\ref{FgLCs2}), the routine usually
fails to converge to a reasonable value because there are not enough
constraints on the transition phase. We identify the failing cases
when $t_{PT}>t_{end}+45$ because $t_{PT}$ is never greater than
$t_{end}+25$. In these cases, we set $t_{PT}=t_{end}+15$ since the
typical value for the width of the transition is 30~days. For this
same reason the value of $\sigma(t_{PT})$ is set to 15~days, a very
conservative value as it encompasses the full possible range of
$t_{PT}$ values.

\item 
When the plateau was extensively observed for more than 100~days but
could not detect any hints of the transition (see SN~1999br in
Fig.~\ref{FgLCs2}), we arbitrarily set $t_{PT}=t_f+15$, because the
plateau phase usually lasts $\sim$100~days. In the two cases where we
face this situation, $\sigma(t_{PT})$ is set to 20~days, which
generously covers the possibility of a longer plateau.

\end{enumerate}

Regardless of the sampling of the light curves, we assign a minimum of
$\sigma(t_{PT})=2$.

The light curve fits also have a healthy benefit: they allow us to
interpolate magnitudes at epochs where only one of the magnitudes was
obtained and the second magnitude, necessary to construct a color, is
missing. Without a color we would be unable to calculate the AKA
corrections, so that the interpolation feature is extremely beneficial
as it permits one to correct magnitudes at all epochs.

\subsection{Color curves fits}\label{CC}

\begin{figure}[p]
\begin{flushright}
\hspace{-1.5cm}
\includegraphics[angle=0,scale=0.8]{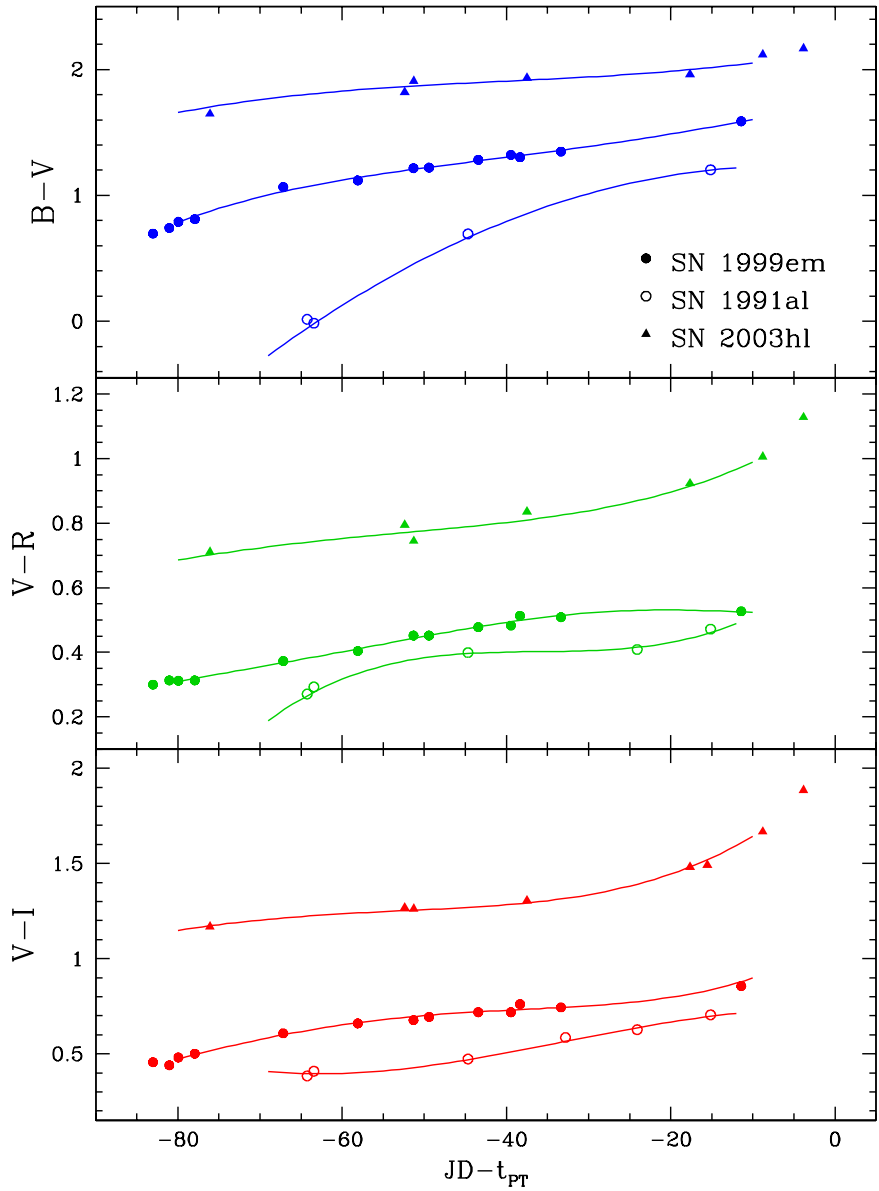}
\caption[(\bv), (\vr), and (\vi) color curves of three SNe]{(\bv),
  (\vr), and (\vi) color curves of SN~1999em (filled circles),
  SN~1991al (open circles) and SN~2003hl (filled triangles) corrected
  for $A_G$ and $K$-terms.  This comparison demonstrates that each SN
  displays a different color evolution, which prevents us to determine
  color excesses from a simple color offset after correcting for $A_G$
  and $K$-terms.
\label{FgColr}}
\end{flushright}
\end{figure}

Figure~\ref{FgColr} shows the (\bv), (\vr), and (\vi) colors of three
proto-typical Type~II-P SNe corrected for $A_G$ and $K$-terms.
 The time origin (the $x$-axis) corresponds to
$t_{PT}$, i.e. the middle of the transition phase. In each case we
employ all the data points between day~--100 and ~--10 to fit a
Legendre polynomial shown with solid lines in Fig.~\ref{FgColr}. The
degree of the polynomial was chosen on a case-by-case basis and varied
between 3rd and 6th order. It is evident that, during the plateau
phase, the photosphere gets redder with time owing to the decrease of
the surface temperature as the SN expands. In theory the photospheric
temperature should approach and never get below the temperature of
hydrogen recombination around 5,000~K. Based on this physical
argument,
\citet{HP02} argued that all \sneiip\ should reach the same intrinsic
colors toward the end of the plateau phase and, therefore, they
proposed that the color excesses measured at this phase could be
attributed to dust reddening in the SN host-galaxy and be exploited to
measure $A_{host}$. \citet{HP02} performed their analysis with a
simple naked-eye estimate of the asymptotic colors. Here we improve
significantly this situation through the formal color curve fitting
procedure describe above.

Armed with the polynomial fits we proceeded to interpolate colors on a
continuous one-day spaced grid between day~--80 and~--10 for analyzing
colors at multiple epochs (see~\S~\ref{aho}). In a handful of cases
the data did not encompass the whole grid and we had to extrapolate
colors, but never by more than 3~days from the nearest data point.

\subsection{\ion{Fe}{2}~based expansion velocity curves}\label{exp}

The third ingredient for SCM is the velocity of the SN ejecta. It is
well known that different spectroscopic lines yield different
expansion velocities. The \ion{Fe}{2}~$\lambda5169$ line is thought to
closely match the SN photospheric velocity and has been usually
employed for SCM and EPM.  Here we use that line as a proxy for the
velocity of the SN ejecta.  Due to the expansion of the envelope, the
spectral lines show a P-Cygni profile with an emission centered at the
rest-frame wavelength $\lambda_0$ and an absorption shifted bluewards
by $\Delta\lambda$. From a measurement of $\Delta\lambda$ we can
compute an expansion velocity

\begin{equation}
 \upsilon_{exp}=c\times\frac{\Delta\lambda}{\lambda_0}
\label{Vexp}
\end{equation}

\begin{figure}[p]
\hspace{-8mm}\includegraphics[angle=0,scale=0.8]{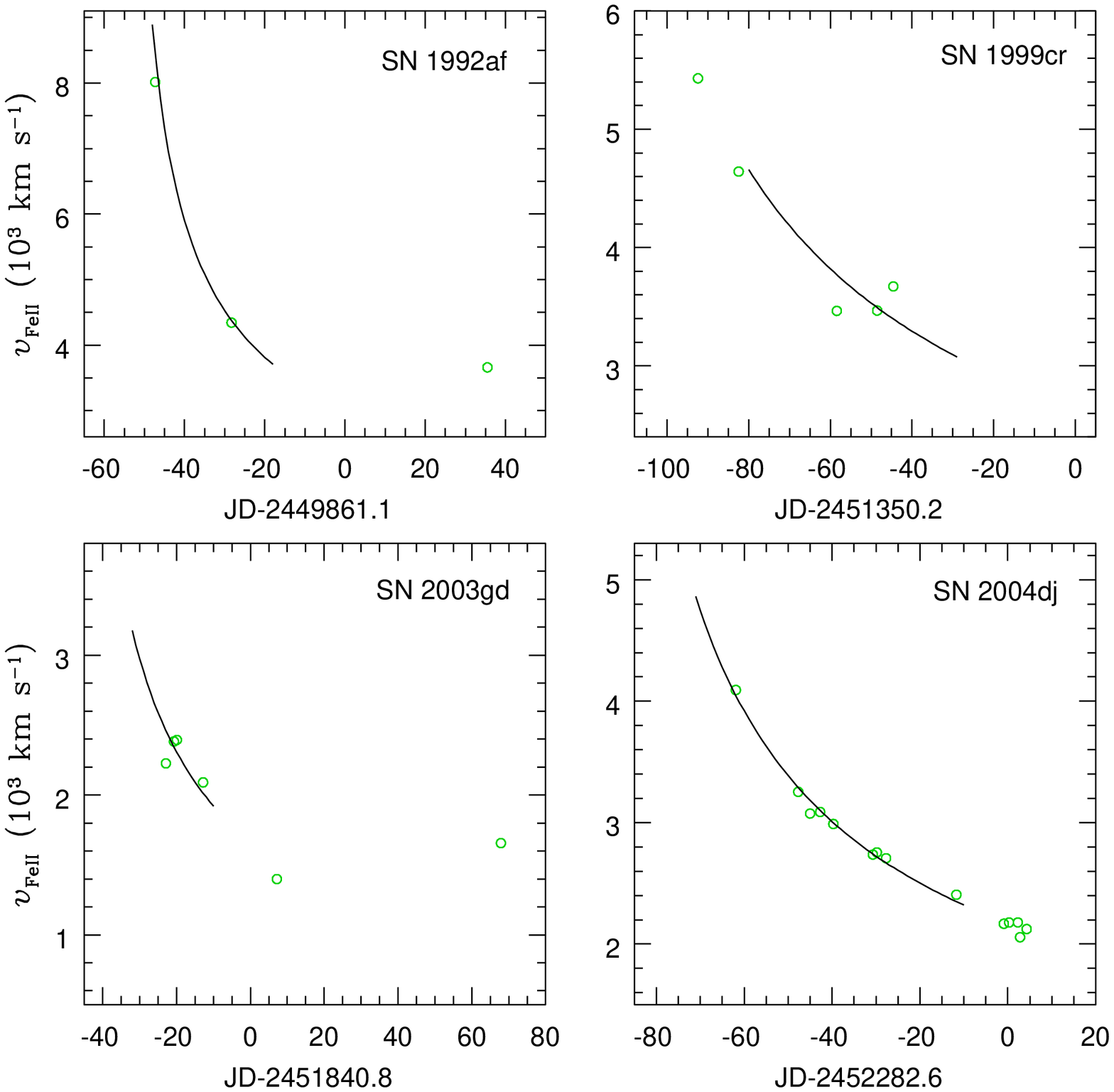}
\caption[\ion{Fe}{2}~velocity curves of four SNe]{Four expansion
  velocity curves representative of the SN sample measured by means of
  the \ion{Fe}{2}~$\lambda$5169 line profile. The solid lines
  correspond to a power law (eq. \ref{vel_eq}). For SN~1992af (upper
  left panel) we fit only two velocities (see \S~\ref{exp} for more
  details).  The upper right panel shows a 15~days extrapolation for
  SN~1999cr beyond the last data point. The lower left and right
  panels show a 10~days extrapolation for SN~2003gd and SN~2004dj
  prior to the first data point.
\label{FgVels}}
\end{figure}

\noindent
In this study we use the expansion velocities measured by
\citet{JH08}. Figure~\ref{FgVels} shows \ion{Fe}{2} velocities as a
function of SN phase, for four different SNe selected for their wide
range of sampling characteristics. 
 In all cases the
SNe show a systematic decrease of their velocities with time. Two
physical arguments support this observational fact: 1)~all the shells
of the SN undergo an homologous expansion, i.e. the outer shells move
faster than the inner shells, and 2)~the photosphere recedes in mass
allowing us to observe deeper and slower layers of the SN as time
passes. As shown by Fig.~\ref{FgVels} the \ion{Fe}{2}~$\lambda5169$
expansion velocity curve during the plateau phase can be properly
modeled with a power law of the form

\begin{equation}
 \upsilon_{exp}(t) = A \times (t-t_0)^\alpha
\label{vel_eq}
\end{equation}

\noindent
where $A$, $t_0$ and $\alpha$ are three free fitting parameters
without obvious physical meaning.

In general we fit for three free parameters ($A,t_0,\alpha$), but when
only two velocity measurements are available (e.g. SN~1992af in the
upper left panel of Fig.~\ref{FgVels}) we fix the $\alpha$ exponent
to~--0.5, which corresponds to a typical value for our sample. As
shown with solid lines in Figure~\ref{FgVels} the fits are quite
satisfactory. Here we choose to restrict the power-law fits to the
plateau phase, since the power-law behavior is not observed for
expansion velocities beyond the transition phase.  We use the same
one-day continous grid as the color curves in order to interpolate
velocities at different epochs between~--80~to~--10~days.  Given the
good quality of the fits and the shallow slope at late epochs, we
allow extrapolations of up to 15~days past the nearest data point (see
SN~1999cr in Fig.~\ref{FgVels}). At the left boundary we reduce the
extrapolations to 10~days prior to the first point, because the power
law gets steeper at early times (see SN~2003gd and SN~2004dj in
Fig.~\ref{FgVels}). When only two velocity measurements are available
we reduce the extrapolations by 5~days.

\section{Host Extinction Determination}\label{aho}

While the determination of Galactic reddening is straighforward
---thanks to the IR dust maps of \citet{SFD98}---, it is much more
challenging to ascertain the extinction due to host-galaxy dust. To
address this issue here we assume that, owing to the hydrogen
recombination nature of their photospheres, all \sneiip\ should evolve
from a hot initial stage to one of similar photospheric
temperature. If two SNe have similar spectra but suffer different
amounts of extinction, all color indices of one object should be
redder than the corresponding color indices of the other object. This
is the hypothesis we want to test in this section.

Although simple in theory there are a couple of practical difficulties
to perform this test. First, it is not always possible to contrain the
time of explosion and line up color curves from different objects.
One way around this is to use the transition time $t_{PT}$ defined in
\S~\ref{LCF}. The second problem is illustrated in Figure
\ref{FgColr}: the color curve shapes can vary significantly from SN to
SN, preventing one to measure a single color offset between two
SNe. Our approach to get around this is to pick a single fiducial
epoch in the color evolution and assess the performance of such color
as reddening estimator. Our polynomial fits to the color curves are
very convenient for this purpose as they allow us to interpolate
reliable colors on a day-to-day basis over a wide range of epochs and
explore which epoch is the one that gives the best results.

If all SNe share the same intrinsic temperature at some epoch we
expect the subset of dereddened SNe to have nearly identical colors
($C_0$) and the remaining objects should show color excesses,
$E(C)=C-C_0$, in direct proportion to their extinctions. A useful
diagnostic to check our underlying assumption is the color-color
plot. Unreddened SNe should occupy a small region in this plane. If we
further assume the same extinction law in the SN host-galaxies, the
subset of extinguished objects should describe a straight line
originating from such region.  The figures of merit in this test are
1)~the color dispersion displayed by the unreddened SNe, 2)~the slope
described by the reddened SNe (which is determined by the extinction
law), and 3)~the dispersion relative to the straight line (the smaller
the better).

\begin{figure}[p]
\begin{center}
\includegraphics[angle=0,scale=0.4]{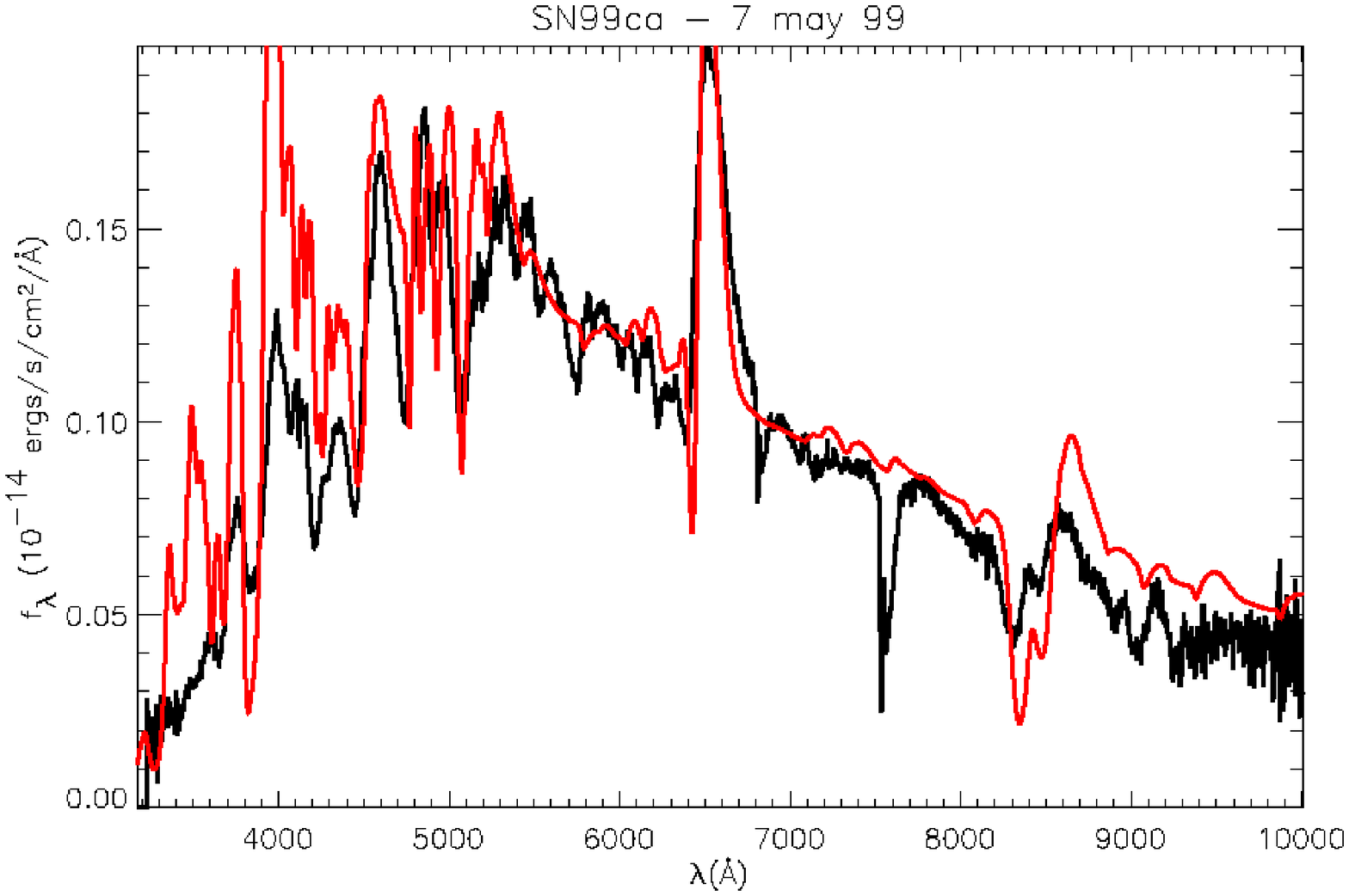}\\
\includegraphics[angle=0,scale=0.4]{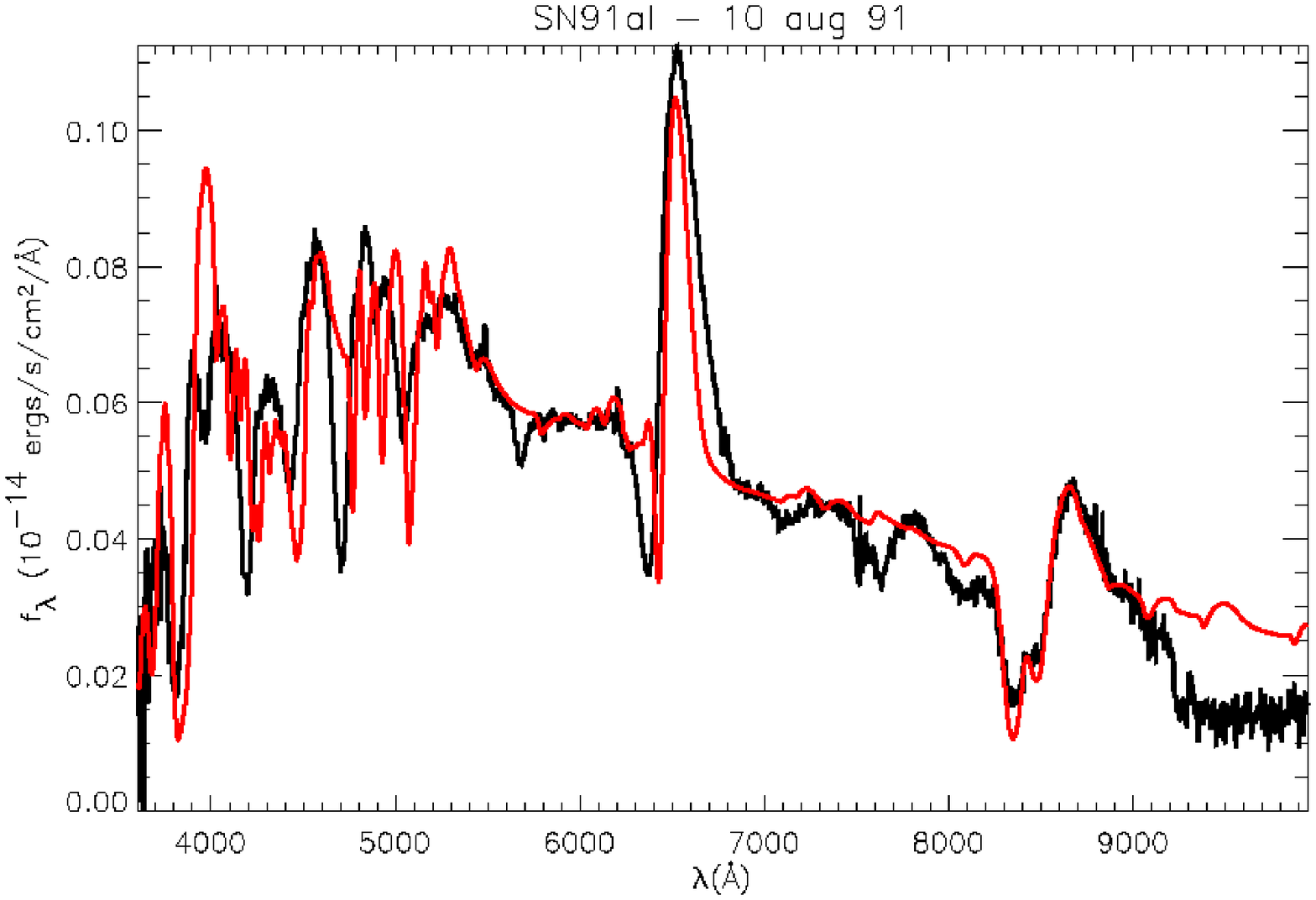}\\
\includegraphics[angle=0,scale=0.4]{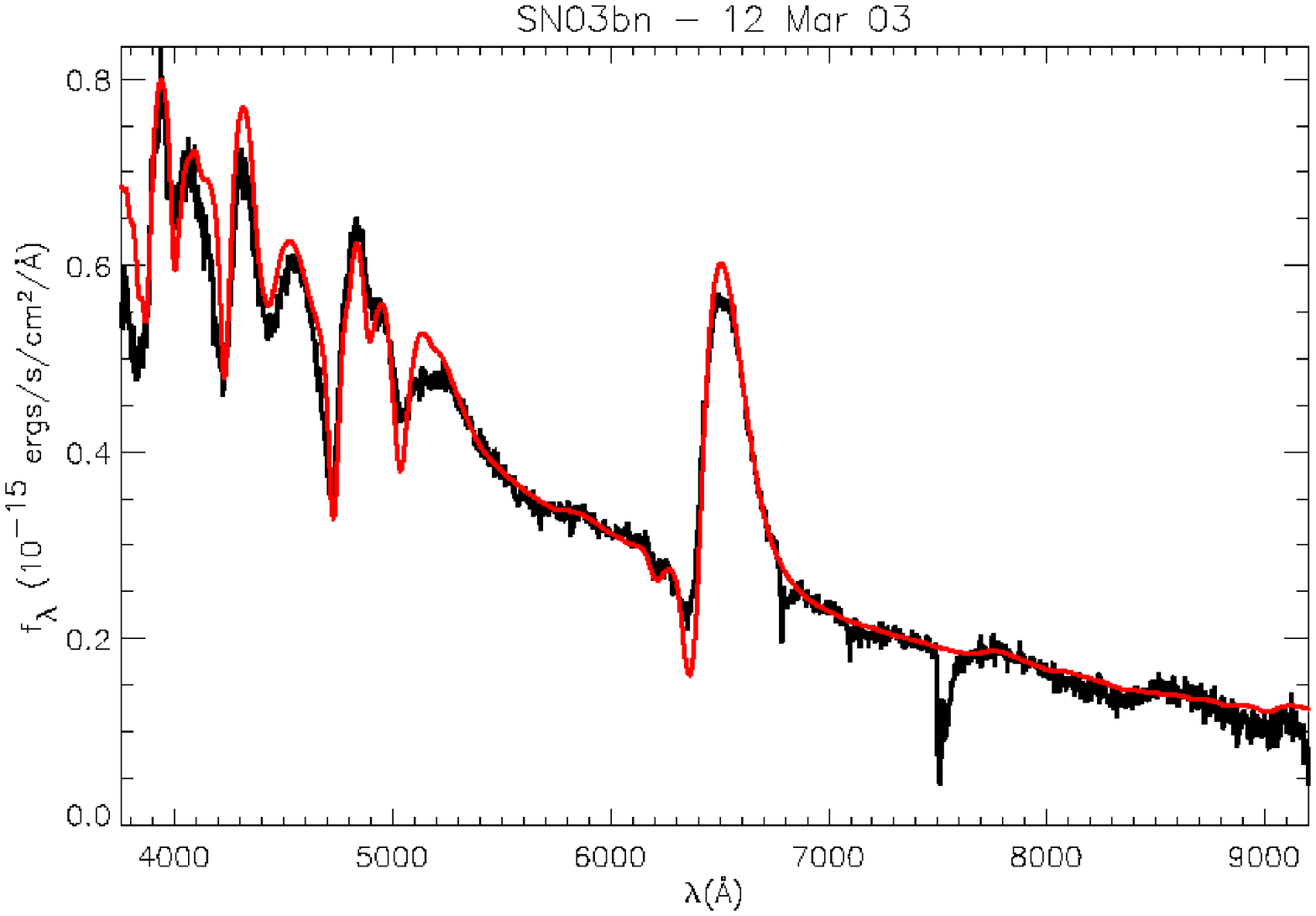}
\end{center}
\caption[Atmosphere models by \citet{D08} fitted to three SN
  spectra]{Type~II-P SN atmosphere models by \citet{D08} (red line)
  fitted to our spectra (black line). The top panel shows an example
  of an unsatisfactory fit to one of our late-time spectra. The middle
  panel shows a better fit to an early-time spectrum of SN~1991al. A
  much better fitting is achieved for an early spectrum of SN~2003bl
  as shown in the bottom panel.
\label{FgSpec}}
\end{figure}

We have identified four objects in our sample (SN~2003B, SN~2003bl,
SN~2003bn, SN~2003cn) consistent with zero reddening. Such objects
were selected for having: 1)~no significant \nad\ interstellar lines
in their spectra at the redshifts of their host-galaxies, and
2)~dust-free early-time spectra.  For the latter we used extinction
values determined by \citet{D08} from fits of Type~II-P SN atmosphere
models to our early-time spectra.  Such models use the SN spectral
lines to constrain the photospheric temperature and the continuum to
restrict the amount of extinction. As shown in Figure~\ref{FgSpec},
the atmosphere models successfully reproduce the early-time SN
spectra. 
 \begin{figure}[p]
\begin{center}
\includegraphics[angle=0,scale=0.8]{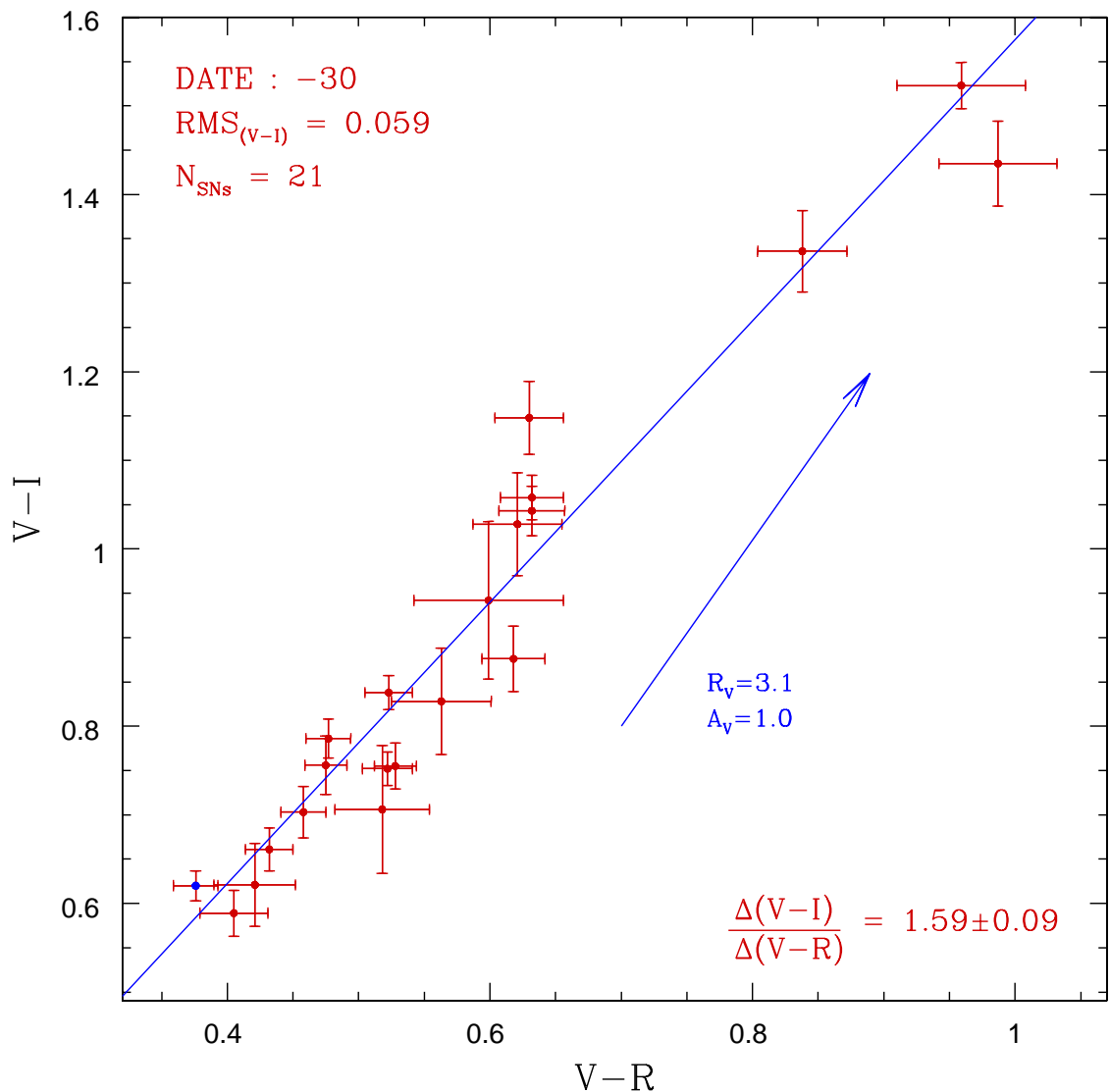}
\caption[(\vi) vs. (\vr) color diagram]{(\vi) versus (\vr) diagram for
  21~\sneiip\ having $VRI$ photometry corrected for $A_G$ and
  $K$-terms. The blue line is a least-squares fit to the data, with a
  slope of $1.59\pm0.09$. The blue arrow has a slope of~2.12 and
  corresponds to the reddening vector for a standard Galactic
  extinction law ($R_V=3.1$). With a blue dot is shown the one SN of
  this subsample consistent with zero extinction having
  $R$-photometry.
\label{FgVIVR}}
\end{center}
\end{figure}
 \begin{figure}[p]
\begin{center}
\includegraphics[angle=0,scale=0.8]{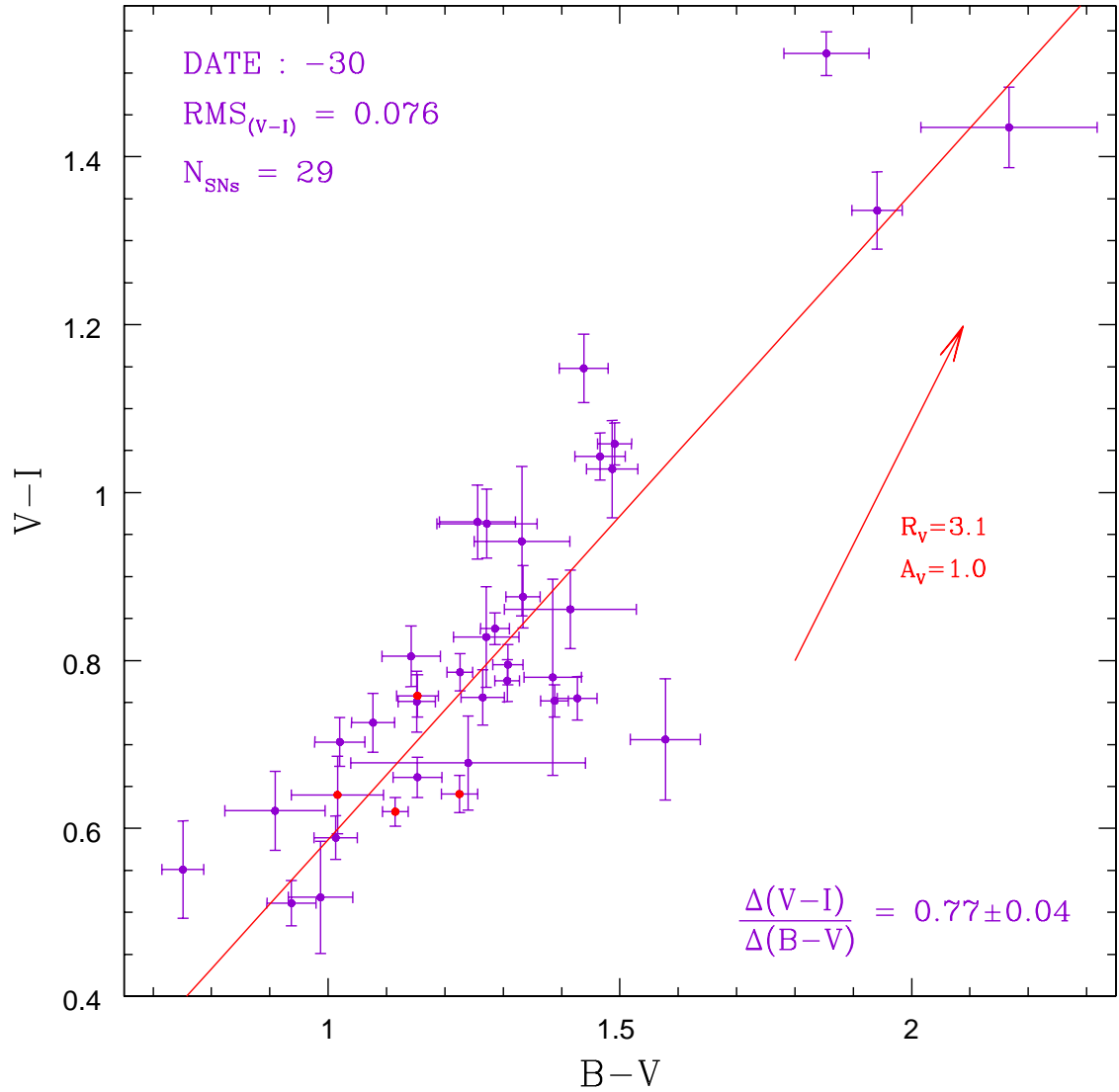}
\caption[(\vi) vs. (\bv) color diagram]{(\vi) versus (\bv) diagram for
  29~\sneiip\ having $BVI$ photometry corrected for $A_G$ and
  $K$-terms. The red line is a least-squares fit to the data, with a
  slope of $0.77\pm0.04$. The red arrow has a slope of~1.38 and
  corresponds to the reddening vector for a standard Galactic
  extinction law ($R_V=3.1$). With red dots are shown the four SNe
  consistent with zero host-galaxy extinction.
\label{FgVIBV}}
\end{center}
\end{figure}

We investigated two color-color plots (\vi\ versus \vr, and
\vi\ versus \bv) over a wide range (from day~--50 to~--15) of epochs
after correcting the photometry for Galactic extinction and $K$-terms.
The best results obtained from our scrutiny is the (\vi) versus (\vr)
diagram constructed from day~--30 and shown in Figure~\ref{FgVIVR}.
At this epoch ---approximately the end of the plateau--- we obtain the
linear behavior expected for a sample with the same intrinsic color
but different degrees of extinction. Shown with a blue dot is the one
SN consistent with zero extinction which is, remarkably, one of the
bluest objects in this diagram; the other three unextinguished SNe do
not have $R$-photometry. A least-squares fit to the data yields a
slope of $1.59\pm0.09$, which is close but not exactly equal to the
$E(V-I)/E(V-R)=2.12$ ratio expected for a Galactic extinction law
($R_V=3.1$), shown as a vector in Figure~\ref{FgVIVR}.  This suggests
a somewhat different extinction law in the SN hosts compared to the
Galaxy.  The dispersion of 0.059~in \vi\ is a promising result as it
translates into an uncertainty of $A_{host}(V)\,=\,0.15$~mag, which
corresponds to the limiting precision of this method. The reduced
$\chi^2$ of 1.55 implies that the dispersion can be accounted almost
solely by our error bars and that any instrinsic color dispersion in
our sample is $\leq 0.06$. The bottom line is that both the (\vr) and
(\vi) colors fulfill the minimum requirements as reddening indicators.

The best results from the (\bv) versus (\vi) analysis were obtained
from day~--30, which are shown in Figure~\ref{FgVIBV}. The four SNe
consistent with zero extinction, shown with red dots, average colors
$(B-V)_0=1.147\pm0.053$ and $(V-I)_0=0.656\pm0.053$. Note that there
are five SNe in this diagram which are slightly bluer than the
unreddened sample. The \vi\ color dispersion of 0.076 is greater than
that obtained in Fg.~\ref{FgVIVR} and is most likely due to the
\bv\ color, since the $B$-band is more sensitive to the metallicity of
the SN, owing to several absorption lines that lie in this spectral
region. Therefore we believe that the greater dispersion in this
diagram could be due to the different metallicities of our SN
sample. A least-squares fit to the data yields a slope of
$0.77\pm0.04$. This slope is quite different than the
$E(V-I)/E(B-V)=1.38$ ratio expected for the Galactic extinction law
(shown as a vector in Figure~\ref{FgVIBV}), in agreement with the
suggestion made in the previous paragraph from the (\vi) versus (\vr)
diagram.

We conclude from our exploration that, while the \bv\ color is
problematic, both the (\vr) and (\vi) colors offer a promising route
for dereddening purposes. In what follows we will employ solely the
\vi\ color since only a small subset of our objects possess $R$ 
photometry. Although the evidence points to a non-Galactic reddening
law, for now we will assume a standard reddening law (later on we will
relax this assumption; see section~\ref{HDRV}). Using our library of
SNe~II spectra we computed synthetically the appropriate conversion
factor between $E(V-I)$ and $A_V$ for Type~II SNe and a standard
reddening law ($R_V=3.1$), which yielded:

\begin{equation}
 \beta_V=\frac{A_V}{E(V-I)}=2.518
\label{beta_eq}
\end{equation}

\noindent
Assuming an intrinsic (\vi)$_0=0.656\pm0.053$ the host-galaxy
extinction can be computed, with its corresponding uncertainty, from:

\begin{eqnarray}
 A_V(V-I)&=&2.518\times[(V-I)-0.656]\label{Ext_eq} \\
 \sigma(A_V)&=&2.518\times\sqrt{\sigma_{(V-I)}+0.053^2+0.059^2}\nonumber
\end{eqnarray}

\noindent \begin{table}[p]
\begin{center}
\small {\scshape \caption{Host-galaxy extinctions for all
    37~SNe}\label{Tb2}}
\vspace{3mm}
\begin{tabular}{l|cc|c|r}
\hline\hline
SN name      &$A_V$(spec)\tablenotemark{a}  &{\it subclass}  &$A_V$(\nad)\tablenotemark{b} &$A_V$(\vi)\tablenotemark{c}  \\
\hline       
1991al        &0.31(16)     &silver   &0.31(06)     &--0.17(21) \\
1992af        &1.24(31)     &coal     &0.17(15)     &--0.37(21) \\
1992am        &\nodata      &         &0.00(43)     &  0.52(23) \\
1992ba        &0.43(16)     &silver   &0.00(03)     &  0.30(21) \\
1993A         &0.00(31)     &bronze   &0.00(54)     &  0.06(25) \\
1999br        &0.25(16)     &silver   &0.00(04)     &  0.94(25) \\
1999ca        &0.12(31)     &coal     &0.34(05)     &  0.25(21) \\
1999cr        &0.47(31)     &coal     &0.69(21)     &  0.12(21) \\
1999em        &0.31(16)     &gold     &1.01(05)     &  0.24(21) \\
1999gi        &0.56(16)     &silver   &0.50(08)     &  1.02(21) \\
0210          &0.31(31)     &bronze   &0.00(23)     &  0.31(36) \\
2002fa        &\nodata      &         &0.00(14)     &--0.35(26) \\
2002gw        &0.40(19)     &silver   &0.00(02)     &  0.18(22) \\
2002hj        &0.16(31)     &bronze   &0.00(06)     &  0.24(22) \\
2002hx        &0.16(25)     &coal     &0.00(16)     &  0.38(22) \\
2003B\mark    &0.00(25)     &silver   &0.12(05)     &--0.09(21) \\ 
2003E         &1.09(31)     &coal     &0.71(07)     &  0.78(23) \\
2003T         &0.53(31)     &coal     &0.19(19)     &  0.35(21) \\
2003bl\mark   &0.00(16)     &gold     &0.11(10)     &  0.26(21) \\
2003bn\mark   &0.09(16)     &silver   &0.00(03)     &--0.04(21) \\ 
2003ci        &0.43(31)     &coal     &0.00(23)     &  0.78(23) \\ 
2003cn\mark   &0.00(25)     &gold     &0.00(09)     &--0.04(23) \\
2003cx        &0.65(25)     &coal     &\nodata      &--0.27(25) \\ 
2003ef        &1.24(25)     &gold     &1.40(12)     &  0.98(21) \\
2003fb        &0.37(31)     &coal     &0.54(23)     &  1.24(23) \\
2003gd        &0.40(31)     &coal     &0.00(04)     &  0.33(21) \\
2003hd        &0.90(31)     &coal     &0.74(27)     &  0.01(21) \\ 
2003hg        &\nodata      &         &2.29(12)     &  1.97(24) \\ 
2003hk        &0.65(31)     &coal     &1.69(20)     &  0.44(25) \\ 
2003hl        &1.24(25)     &gold     &1.84(09)     &  1.72(23) \\ 
2003hn        &0.59(25)     &coal     &0.64(08)     &  0.46(21) \\ 
2003ho        &1.24(31)     &bronze   &1.28(10)     &  2.19(21) \\ 
2003ip        &0.40(31)     &coal     &0.42(08)     &  0.56(22) \\
2003iq        &0.37(16)     &silver   &0.91(04)     &  0.25(22) \\ 
2004dj        &0.50(25)     &coal     &0.26(06)     &--0.09(23) \\
2004et        &0.00(25)     &coal     &1.17(02)     &  0.13(27) \\
2005cs        &\nodata      &         &\nodata      &  0.72(30) \\
\hline
\end{tabular}
\vspace{-4mm}

\tablecomments{The third column lists the subclass defined upon the
  criteria exposed in \S~\ref{dtc}.}

\tablenotetext{a}{\,\citet{D08} with $R_V=3.1$.}
\tablenotetext{b}{\,equivalent width of \nad\ measured by us and
  converted to $A_V$ with a law of \citet{B90} using $R_V=3.1$.}
\tablenotetext{c}{\,this research.}  \tablenotetext{*}{\,SNe picked to
  determine the intrinsic color.}

\end{center}
\end{table}
 where \vi\ corresponds to the color of a
given SN at day~--30 (corrected for $K$-terms and foreground
extinction) and $\sigma_{(V-I)}$ combines the instrumental errors in
the $V$ and $I$ magnitudes, the RMS of the $A_G$ and $K$ terms (see
\S~\ref{AGcor} and \S~\ref{Kcor} respectively), and the uncertainty in
$t_{PT}$ (\S~\ref{LCF}). The uncertainty in the intrinsic (\vi)$_0$
color was also included in the error of $A_V$ along with the (\vi) RMS
in the (\vi) versus (\vr) diagram. The host-galaxy reddening values
obtained from this technique are listed in column~5 of Table~\ref{Tb2}
(with the uncertainties given in parenthesis for the whole sample of
37~SNe). There are eight SNe with (\vi) colors bluer than (\vi)$_0$
which stand out in this table for their negative reddenings. Although
not physically meaningful, these negative values are statistically
consistent with zero or moderate reddenings. In fact, seven out of
these eight objects differ by $<1.1\sigma$ from zero reddening. Only
SN~1992af differs by $1.8\sigma$ from $A_{host}=0$. In the following
section we compare this method to other dereddening techniques.

\clearpage

\chapter{Analysis\label{ANLS}}

\section{Comparing dereddening techniques} \label{dtc}

At our request, \cite{D08} has kindly performed fits of
\sneiip\ atmosphere models to our library of spectra. In these fits
the spectral lines are used to constrain the photospheric temperature
and the corresponding continuum is employed to estimate the
extinction. A crucial condition for this technique to work is the
spectrophotometric quality of the spectra.

\begin{figure}[p]
\begin{center}
\vspace{-2.3cm}
\includegraphics[angle=0,scale=0.8]{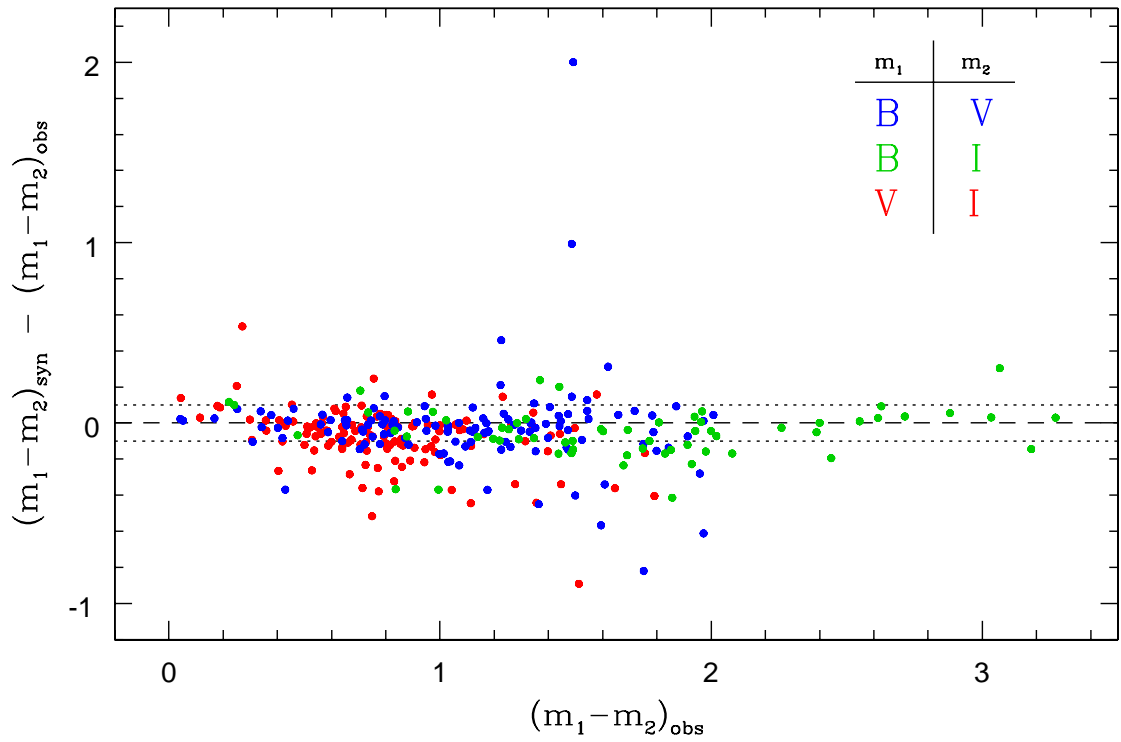}\vspace{-2cm}
\caption[Color differences between spectra and photometry]{Synthetic
  colors minus observed colors versus observed colors for the
  150~library spectra. The 64~cases (18\% of all cases) with colors
  outside the $\sigma=0.1$ contours are selected for a ``mangling''
  correction, whose purpose is to correct the flux calibration of a
  spectrum and make it consistent with the observed photometry.
\label{FgMang}}
\end{center}
\end{figure}

In general our observations were obtained with the slit oriented along
the paralactic angle and the relative shape of the spectra should be
accurate. However, this was not always possible and sometimes the
spectra were contaminated from light of the host-galaxy or the slit
could not be rotated. For all these reasons we first checked the flux
calibration of each spectrum by synthesizing magnitudes and comparing
them to the observed magnitudes, duly interpolated to the time of the
spectroscopy. In general we found good agreement between the observed
and synthetic colors (Figure~\ref{FgMang}), thus confirming our
confidence in the flux calibration. 
 In the 64~cases
(18\% of all cases) where we found significant differences between
synthetic and observed colors ($\geq$~0.1 for any color) we applied a
low-order polynomial correction to the spectrum. Basically, this
``mangling'' correction used the observed photometry to change the
slope of a spectrum. After checking the flux calibration (and mangling
it, if needed) the next step was to correct for Galactic absorption
and de-redshift the spectra. This database was then used by \cite{D08}
for the atmosphere model fits. Examples of the fits are shown in
Fig.~\ref{FgSpec}.

The resulting spectroscopic reddenings are summarized in column~2 of
Table~\ref{Tb2}.  As pointed out by \cite{D08}, the spectrum fitting
technique works much better with early-time spectra than with
late-time observations. At late times the photosphere has receded in
mass exposing inner and more metal-rich layers, which translates into
an over-abundance of heavy-elements absorption lines. Therefore the
fitting to the continuum ---practically a temperature fitting--- is
hampered by the presence of strong absorption lines at late times.
According to this we divided our sample in four quality subcategories
based on the epoch and the flux quality of the spectra used in the
reddening determination:

\begin{description}
\item[\hspace{2cm}$\diamond$\ ] {\it gold :} early-time spectra,
  without mangling correction
\item[\hspace{2cm}$\diamond$\ ] {\it silver :} early-time spectra,
  with mangling correction
\item[\hspace{2cm}$\diamond$\ ] {\it bronze :} late-time spectra,
  without mangling correction
\item[\hspace{2cm}$\diamond$\ ] {\it coal :} late-time spectra, with
  mangling correction
\end{description}

\noindent
Note that the main criterion is whether the spectrum is early or late
and the second criterion corresponds to the flux calibration quality.
We trust more the spectra that do not require any corrections as they
reflect that the observations were better performed, so we consider
the unmangled spectra as higher-quality than the mangled ones. This
sub-classification is given for each SN in column~3 of
Table~\ref{Tb2}. As the reader can notice, the error is directly
related to this sub-classification.  \citet{D08} assigns an error of
$\sigma_{E(B-V)}=0.05$ when using early-time spectra, and
$\sigma_{E(B-V)}=0.1$ when using late-time spectra. We refine this
argument ramping up from 0.05 to 0.10, depending on the number of
spectra employed for each extinction determination.  These values are
given in Table~\ref{Tb2} in units of $A_V=3.1\times E(B-V)$.

\begin{figure}[p]
\begin{center}
\includegraphics[angle=0,scale=0.8]{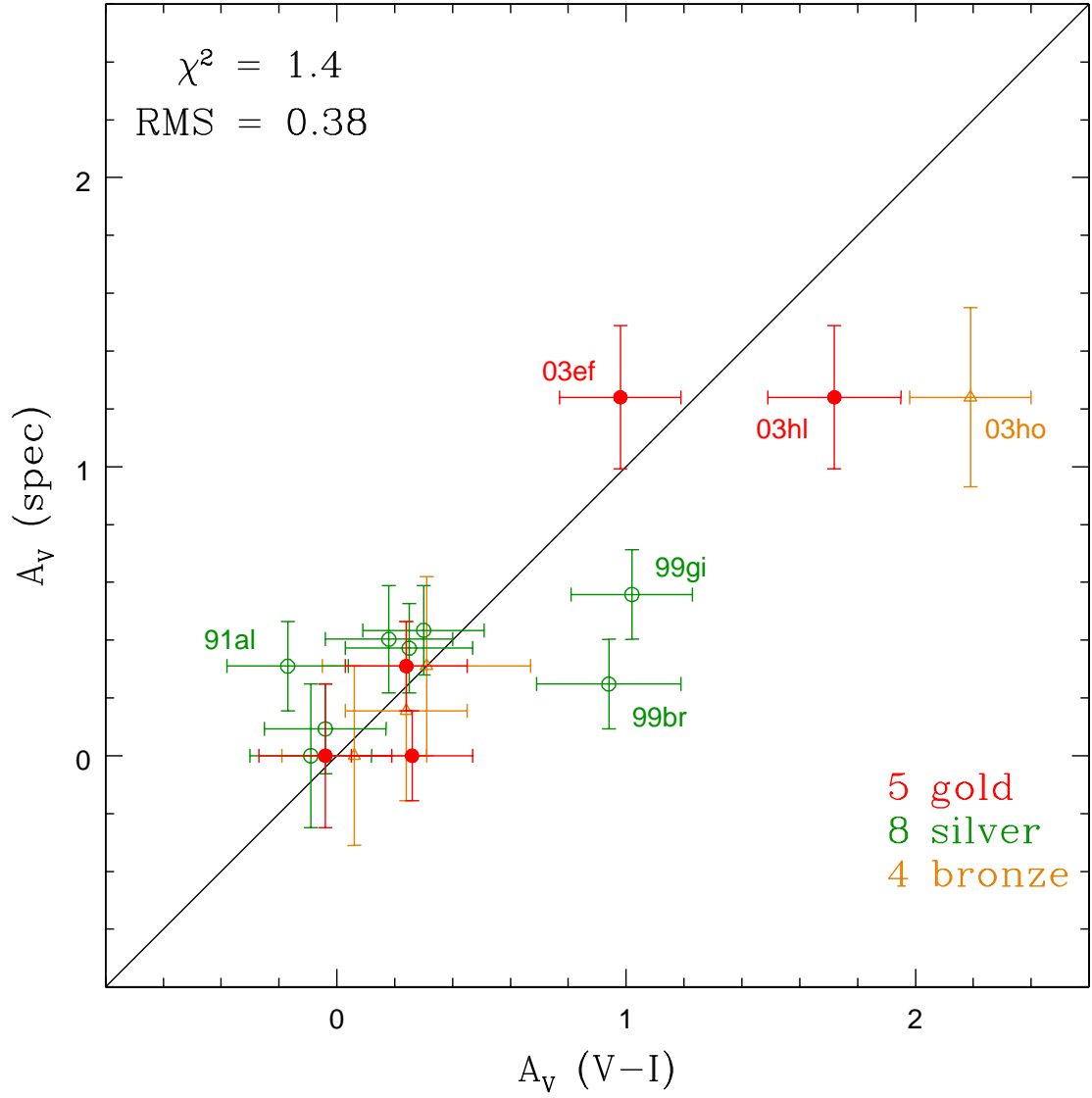}
\caption[Best spectroscopic extinctions against (\vi) based
  extinctions]{Comparison between two dereddening techniques: the
  spectrum fitting method \citep[$y$-axis,][]{D08} and our technique
  ($x$-axis, \S~\ref{aho}).  The figure shows three subclasses defined
  by the quality of the spectroscopic data used by \citet{D08}: the
  {\it gold}+{\it silver}+{\it bronze} sample (see \S~\ref{dtc}).  The
  solid black line is the one-to-one relation.
\label{FgRd1}}
\end{center}
\end{figure}
\begin{figure}[p]
\begin{center}
\includegraphics[angle=0,scale=0.8]{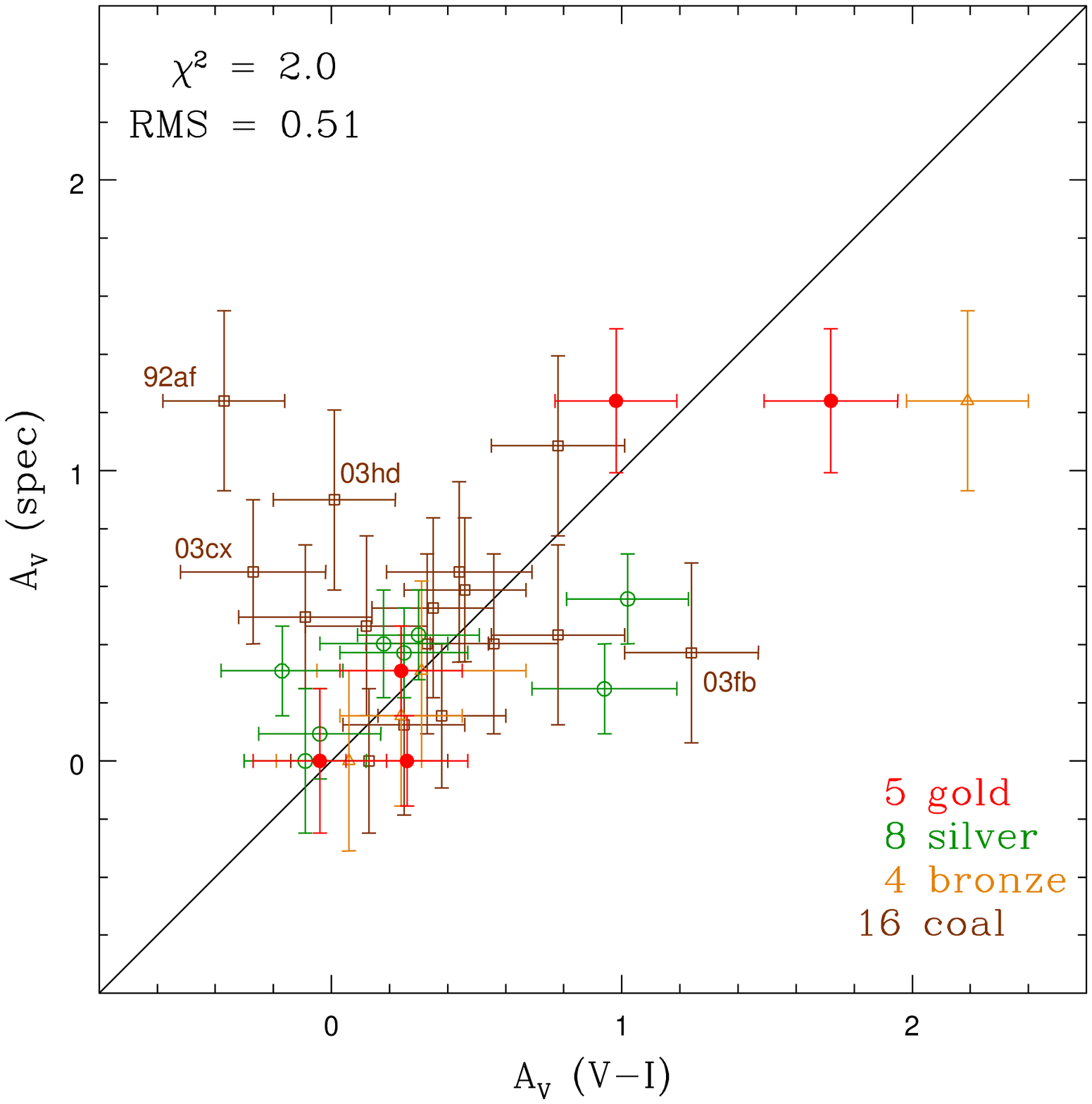}
\caption[All spectroscopic extinctions against (\vi) based
  extinctions]{Comparison between two dereddening techniques: the
  spectrum fitting method \citep[$y$-axis,][]{D08} and our technique
  ($x$-axis, \S~\ref{aho}).  The figure shows the same three
  subclasses as Fig.~\ref{FgRd1} plus the worst quality {\it coal}
  sample (see \S~\ref{dtc}).  The solid black line is the one-to-one
  relation.
\label{FgRd2}}
\end{center}
\end{figure}

Figure~\ref{FgRd1} shows a comparison between the spectroscopic
reddenings $A_V$(spec) and our color reddenings $A_V(V-I)$ derived in
\S~\ref{aho}, for the 17~SNe belonging to the top three subclasses
({\it gold}+{\it silver}+{\it bronze}). A good agreement is displayed
between both techniques with a dispersion of 0.38~mag. The resulting
$\chi^2=1.4$ suggests that this dispersion is consistent with the
combined errors between both techniques. The exceptions are two
objects: SN~1999br and SN~2003ho.  The first object (SN~1999br) only
has plateau photometry making hard the determination of
$t_{PT}$. Although the error in $t_{PT}$ is quite large (20~days),
this uncertainty does not have a great impact on the error of the \vi\
color due to the flatness of the color curve at these epochs (+0.0024
per day). Another cause for the disagreement is the extreme properties
(low luminosity and velocity) of this SN which might also have a \vi\
color instrinsically redder than that of the bulk of the SNe.  The
second discrepant object (SN 2003ho) belongs to the {\it bronze} group
so it is possible that the difference could be due to the use of a
late-time of the spectrum in the determination of the spectroscopic
reddening.

Figure~\ref{FgRd2} shows the same comparison, but this time we include
the 16~lowest-quality {\it coal} SNe. This inclusion clearly
deteriorates the good agreement seen in Fig.~\ref{FgRd1} from
$\sigma=0.38$~mag to $\sigma=0.51$~mag ($\chi^2=2.0$), namely due to
SN~1992af, SN~2003cx, SN~2003hd, and SN~2003fb. This confirms the
warning by \citet{D08}, namely, that his spectroscopic technique works
much better with early-time spectra. The large value of $\chi^2$ also
suggests that the errors in the spectroscopic reddenings derived from
late-time spectra are underestimated.

\begin{figure}[p]
\begin{center}
\includegraphics[angle=0,scale=0.8]{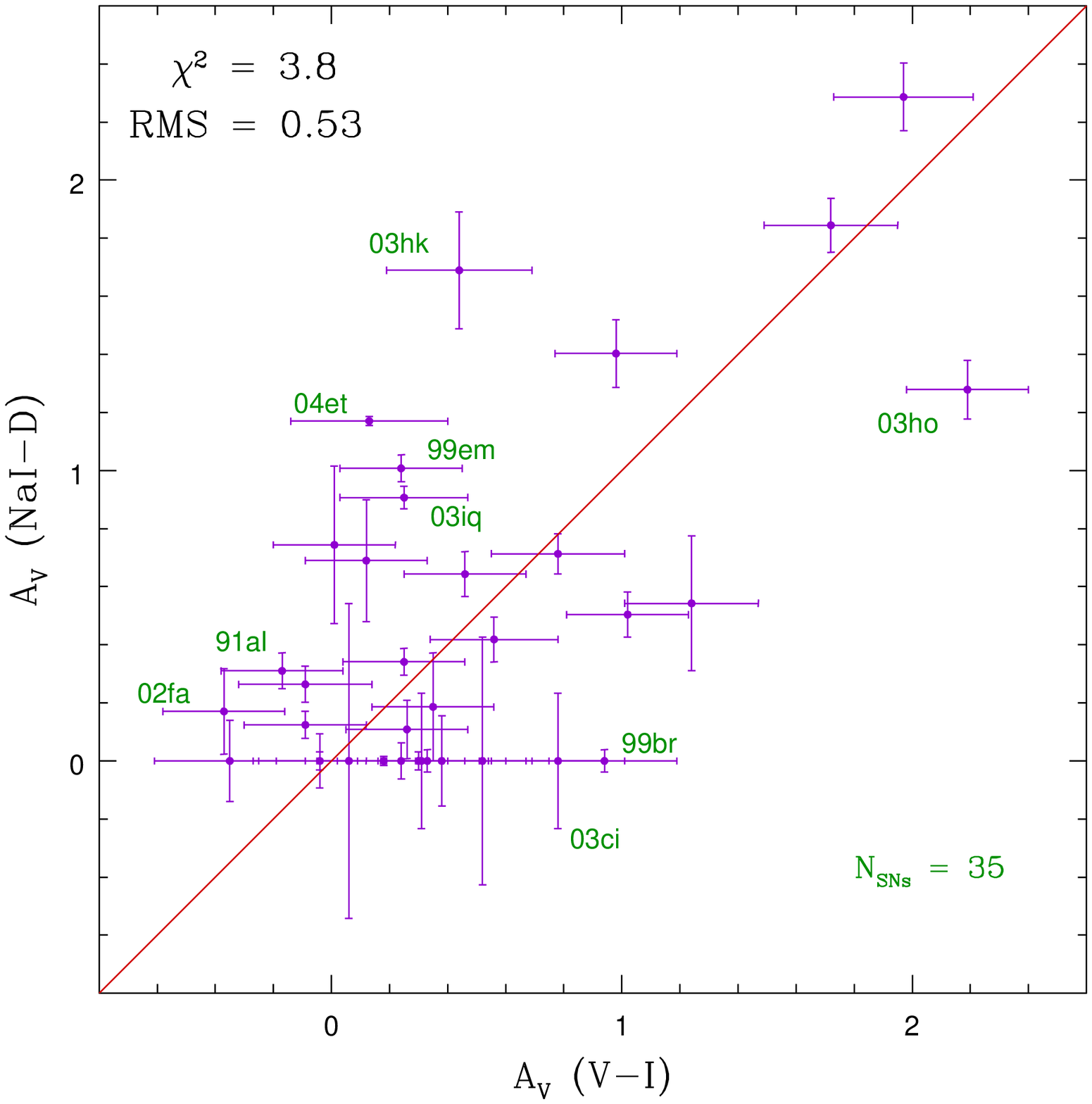}
\caption[\nad\ interstellar line extinctions against (\vi) based
  extinctions]{Comparison between two dereddening techniques: the
  \nad\ interstellar line ($y$-axis) and our technique ($x$-axis,
  \S~\ref{aho}). The equivalent width of the \nad\ interstellar line
  are transformed into visual extinctions according to the calibration
  given by \citet{B90}.
\label{FgNad} }
\end{center}
\end{figure}

Another way to estimate host-galaxy reddening is from the
\nad\ $\lambda\lambda$5893,5896 interstellar absorption doublet
observed in the SN spectrum at the host-galaxy redshift.  Whenever the
line was detected we measured its equivalent width; in those cases
where we did not detect the \nad\ line we assigned it a null value. In
all cases we estimated the uncertainty in the equivalent width ($EW$)
based on the signal-to-noise of the continuum around this line. We
converted these measurements into visual extinctions $A_V$(\nad) using
the calibration of \citet{B90}:

\begin{equation}
 E(B-V)\,\simeq\,0.25\,\times\,EW(\mbox{\nad}),
\end{equation}

\noindent
and tabulate our results in column~4 of Table~\ref{Tb2}. The
comparison between $A_V$(\nad) and our technique, shown in
Figure~\ref{FgNad}, exhibits a dispersion of 0.53~mag ($\chi^2=3.8$),
which is much higher than the $\sigma$=0.38~mag obtained from the
previous comparison. The disagreement can be attributed to the fact
that the absorption line traces the gas content along the line of
sight, but does not necessarily probe dust \citep{MZ97}. Furthermore,
reddenings derived from the EW of the \nad\ lines measured from
low-dispersion spectra simply cannot be expected to be precise which
is exactly our case. The basic problem is that the D lines produced by
a typical interstellar cloud are saturated \citep{Hob74}. The only way
that one can hope to derive the reddening from the D lines is via very
high-dispersion spectra that resolve the lines and allow the column
density to be derived.

\section{The Luminosity-Expansion Velocity relation}

Armed with the dereddening method based on late-time colors we can now
revisit the \lumvel\ relation originally discovered by \citet{HP02}
which is at the core of the SCM.  For this purpose we applied AKA
corrections (\S~\ref{cor}) to our photometry, we used our analytic
fits (\S~\ref{LCF}) to interpolate $BVI$ magnitudes on day~--30, and
we employed the CMB redshifts\footnote{the CMB redshifts were computed
  by adding the heliocentric redshifts given in Table~\ref{Tb1} and
  the projection of the velocity of the Sun relative to the CMB in the
  direction of the SN host galaxy. For the latter we adopted a
  velocity of 371~\kmpsec\ in the direction $(l,b)=(264.1,48.3)$ given
  by \citet{Fix96}.}  in Table~\ref{TbHDs} to convert apparent
magnitudes to absolute values (assuming $H_0=70$~\dimho). The
expansion velocities were determined from the minimum of the
\ion{Fe}{2} $\lambda$5169 P-Cygni line profiles. We performed a
power-law fit to interpolate a velocity contemporaneous to the
photometry (day~--30) as described in section~\ref{LCF}.  From our
original sample of 37~\sneiip\ we were able to use 30~SNe to build
this relation, since five of them do not have \ion{Fe}{2} velocities
at day~--30, and two others have extremely low redshifts
($cz_{CMB}<300$~\kmpsec).

\begin{figure}[p]
\begin{center}
\includegraphics[angle=0,scale=0.8]{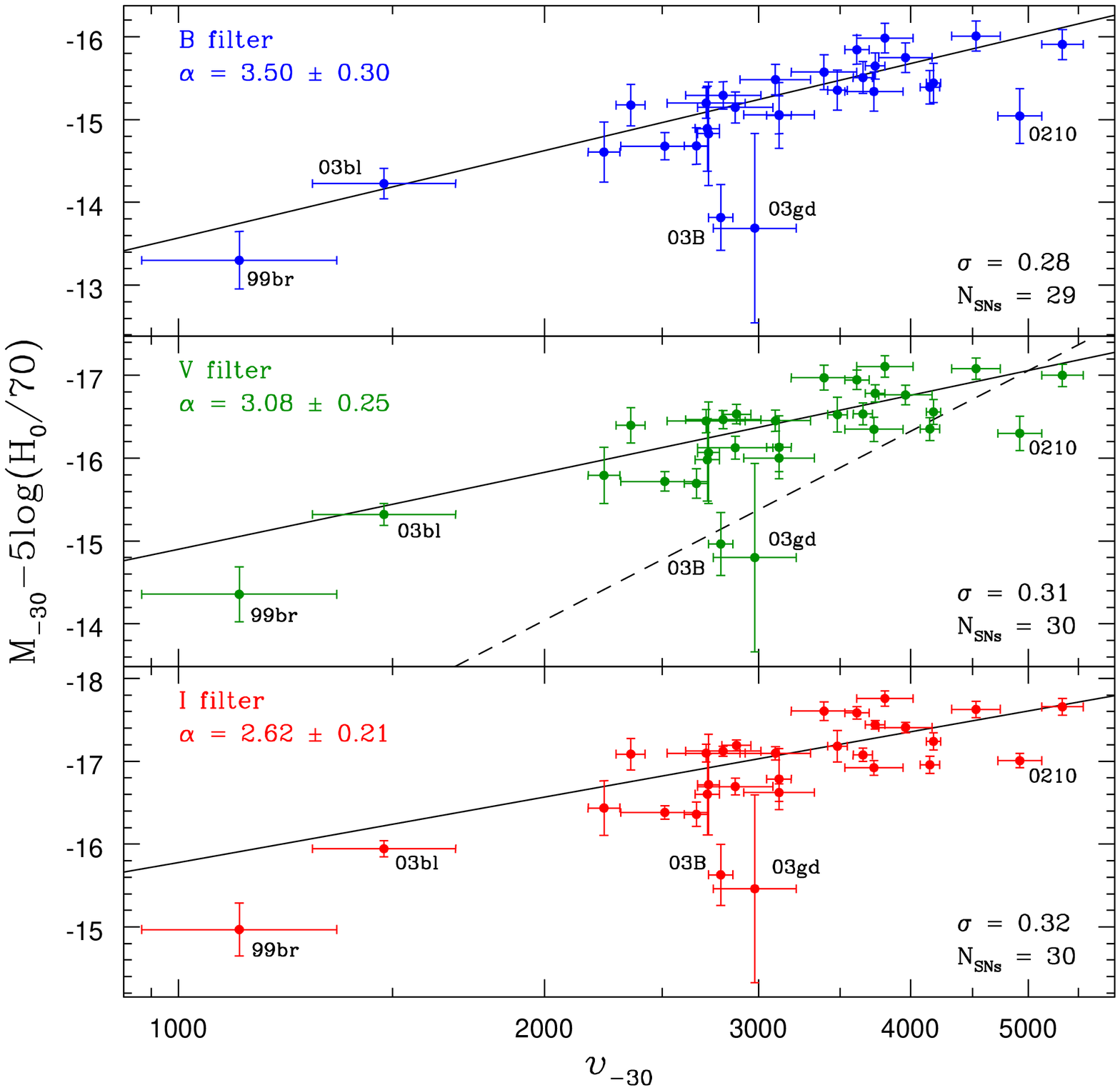}
\vspace{-5mm}
\caption[Luminosity vs. expansion velocity relation for $BVI$
  photometry]{$BVI$ band absolute magnitudes ($y$-axis) against the
  \ion{Fe}{2} expansion velocity ($x$-axis) for 29--30~SNe.  We use
  the magnitudes corrected for Galactic extinction, $K$-terms, and
  host-galaxy extinction together with a value of the Hubble constant
  of 70~\dimho. The dashed line in the middle panel represents the
  \lumvel\ relation for the $V$-band found by \citet{HP02} obtained
  from magnitudes and velocities measured at day~50 past the explosion
  (approximately day~--60 in our own time scale).
\label{FgLvelV}}
\end{center}
\end{figure}

Figure~\ref{FgLvelV} shows the resulting \lumvel\ relation (absolute
magnitude versus expansion velocity) for all $BVI$ bands. Evidently we
recover the result of \citet{HP02}, namely that the most luminous SNe
have greater expansion velocities.  In their case the data were
modeled with a linear function.  Our sample suggests that the relation
may be quadratic, but we need more SNe at low expansion velocities to
confirm this suspicion.  Linear least-squares fits to our $BVI$ data
yields the following solutions

\begin{eqnarray}
M_{abs}(B)&=&3.50(\pm 0.30)~\log{(\upsilon_{\mbox{\scriptsize
      FeII}}/5000)} - 16.01(\pm 0.20)\\
M_{abs}(V)&=&3.08(\pm 0.25)~\log{(\upsilon_{\mbox{\scriptsize
      FeII}}/5000)} - 17.06(\pm 0.14)\\
M_{abs}(I)&=&2.62(\pm 0.21)~\log{(\upsilon_{\mbox{\scriptsize
      FeII}}/5000)} - 17.61(\pm 0.10)
\end{eqnarray}

\noindent
which are shown with solid lines in Figure~\ref{FgLvelV}.
 The relation found by \citet{HP02} for the $V$-band
is shown with the dashed line in the middle panel of the same
figure. Given that the study of \citet{HP02} was performed using data
at day~50 after the explosion (around day~--60 in our time scale), it
is not unexpected that their \lumvel\ relation is shifted to higher
expansion velocities. Some of the difference in slope is explained by
the inclusion of SN~2003bl in our sample, which flattens the
correlation. This relation exhibits a dispersion of 0.3~mag, similar
to that reported before. This low scatter is very encouraging as it
implies that the expansion velocities can be used to predict the SN
luminosities, to standardize them, and to derive distances.

\clearpage
\section{Hubble diagrams}\label{HDgr}

The \lumvel\ relation shown in the previous section implies that a
spectroscopic measurement of the expansion velocity of a SN~II-P can
be used to compute a corrected luminosity which should be
approximately the same for all SNe. We examine the reality of this
property of \sneiip\ in the magnitude-redshift Hubble diagram. For
this purpose we employ \ion{Fe}{2} velocities (in units of \kmpsec),
$BVI$ apparent magnitudes corrected for $K$-terms, Galactic reddening,
and host-galaxy reddening determined from \vi\ colors, and host-galaxy
redshifts in the CMB frame. Table~\ref{TbHDs} lists these values for
the 37~SNe of our sample, of which 35 meet the requirement of being in
the Hubble flow ($cz_{CMB}>300$~\kmpsec). A perfect distance indicator
would describe a straight line in the Hubble diagram, so the figure of
merit to assess the performance of this method is the dispersion from
the fit. Along this section we use dispersions weighted by the errors
to evaluate the precision of the Hubble diagrams. \begin{table}[p]
\begin{center}
\footnotesize
{\scshape \caption{Magnitudes, expansion velocities, and \vi\ colors
    for day~--30}\label{TbHDs}}
\vspace{5mm}
\begin{tabular}{lrccccc}
\hline\hline
SN name &$cz_{CMB}$\mark &$B$     &$V$           &$I$        &$\upsilon_{\mbox{\scriptsize FeII}}$ &\vi \\
\hline                                                                          
1991al    & 4480  &17.944(036)   &16.917(029)   &16.333(020)    &5328(202)      &0.589(022)  \\
1992af    & 5359  &18.024(038)   &17.096(029)   &16.706(053)    &4529(212)      &0.511(027)  \\
1992am    &14007  &20.443(117)   &19.053(038)   &18.218(033)    &\nodata        &0.830(046)  \\
1992ba    & 1245  &16.962(040)   &15.657(031)   &14.886(022)    &2237(068)      &0.776(021)  \\
1993A     & 8907  &20.636(127)   &19.309(089)   &18.657(057)    &\nodata        &0.678(047)  \\
1999br    & 1285  &19.010(022)   &17.582(013)   &16.574(008)    &1127(205)      &1.028(011)  \\
1999ca    & 3108  &17.901(071)   &16.489(055)   &15.735(034)    &\nodata        &0.755(025)  \\
1999cr    & 6363  &19.435(067)   &18.418(051)   &17.723(035)    &3095(207)      &0.703(025)  \\
1999em    &  670  &15.331(034)   &13.998(023)   &13.245(017)    &2727(056)      &0.752(019)  \\
1999gi    &  831  &16.554(032)   &15.060(019)   &14.011(022)    &2725(061)      &1.058(025)  \\
0210      &15082  &21.955(044)   &20.572(028)   &19.733(035)    &4923(206)      &0.780(110)  \\
2002fa    &17847  &21.222(072)   &20.243(052)   &19.709(088)    &4176(061)      &0.518(065)  \\
2002gw    & 2878  &18.575(020)   &17.491(029)   &16.752(014)    &2669(063)      &0.726(035)  \\
2002hj    & 6869  &19.705(084)   &18.582(061)   &17.939(042)    &3657(065)      &0.751(031)  \\
2002hx    & 9573  &20.328(056)   &19.164(035)   &18.360(028)    &3960(203)      &0.805(031)  \\
2003B     & 1105  &17.078(022)   &15.966(018)   &15.341(015)    &2795(067)      &0.620(017)  \\
2003E     & 4380  &19.659(046)   &18.368(026)   &17.472(024)    &2869(203)      &0.965(024)  \\
2003T     & 8662  &20.541(028)   &19.227(020)   &18.423(022)    &2803(203)      &0.795(024)  \\
2003bl    & 4652  &20.158(042)   &18.961(020)   &18.230(019)    &1475(202)      &0.758(025)  \\
2003bn    & 4173  &18.638(045)   &17.403(034)   &16.769(024)    &2719(202)      &0.641(020)  \\
2003ci    & 9468  &\nodata       &19.634(035)   &18.648(029)    &2876(077)      &0.963(034)  \\
2003cn    & 5753  &19.852(024)   &18.827(015)   &18.182(014)    &2510(207)      &0.640(017)  \\
2003cx    &11282  &20.417(100)   &19.513(050)   &19.050(067)    &3734(205)      &0.551(058)  \\
2003ef    & 4504  &19.304(032)   &17.837(024)   &16.792(017)    &\nodata        &1.043(015)  \\
2003fb    & 4996  &20.521(122)   &19.085(092)   &17.940(056)    &3120(206)      &1.148(040)  \\
2003gd    &  359  &15.208(045)   &13.965(036)   &13.167(023)    &2976(230)      &0.786(019)  \\
2003hd    &11595  &20.462(031)   &19.320(014)   &18.658(018)    &3741(069)      &0.661(022)  \\
2003hg    & 3921  &20.247(091)   &18.067(040)   &16.603(025)    &3398(211)      &1.435(027)  \\
2003hk    & 6568  &19.478(036)   &18.201(024)   &17.380(023)    &3618(087)      &0.828(026)  \\
2003hl    & 2198  &19.124(062)   &17.219(038)   &15.806(027)    &2354(063)      &1.336(027)  \\
2003hn    & 1102  &16.416(020)   &15.153(011)   &14.307(015)    &3121(074)      &0.836(018)  \\
2003ho    & 4134  &20.779(084)   &18.946(026)   &17.419(016)    &4152(080)      &1.523(022)  \\
2003ip    & 5050  &18.894(046)   &17.549(035)   &16.664(025)    &3813(204)      &0.876(021)  \\
2003iq    & 2198  &17.396(049)   &16.124(037)   &15.362(024)    &3482(063)      &0.756(018)  \\
2004dj    &  180  &12.973(074)   &12.046(035)   &11.416(033)    &2725(202)      &0.621(043)  \\
2004et    &--133  &13.683(220)   &12.099(171)   &11.378(104)    &2901(202)      &0.707(070)  \\
2005cs    &  635  &16.099(062)   &14.749(051)   &13.824(055)    &\nodata        &0.957(067)  \\
\hline
\end{tabular}

\tablenotetext{*}{~The {\it Velocity Calculator} tool at the NASA/IPAC
  Extragalactic Database webpage computes the CMB redshift ($cz_{CMB}$
  in \kmpsec) given the coordinates and the heliocentric redshift of
  an object. The error of this quantity is dominated by the error in
  the determination of the Local Group velocity, which is
  187~\kmpsec.}

\tablecomments{All the data in this table has been interpolated to
  day~--30 with errors accounting for the interpolation, the error of
  $t_{PT}$, and the instrumental error. $BVI$ corresponds to apparent
  magnitudes corrected for Galactic absorption, $K$-terms, but
  uncorrected for host-galaxy extinction. The \vi\ colors do not
  exactly match the $V$ minus $I$ magnitude difference, since light
  and color curves are interpolated independently.}

\normalsize
\end{center}
\end{table}

\subsection{Using $A_V$(\,\vi) and $A_V$(spec)}\label{HDAcs}
\begin{figure}[p]
\begin{center}
\includegraphics[angle=0,scale=0.8]{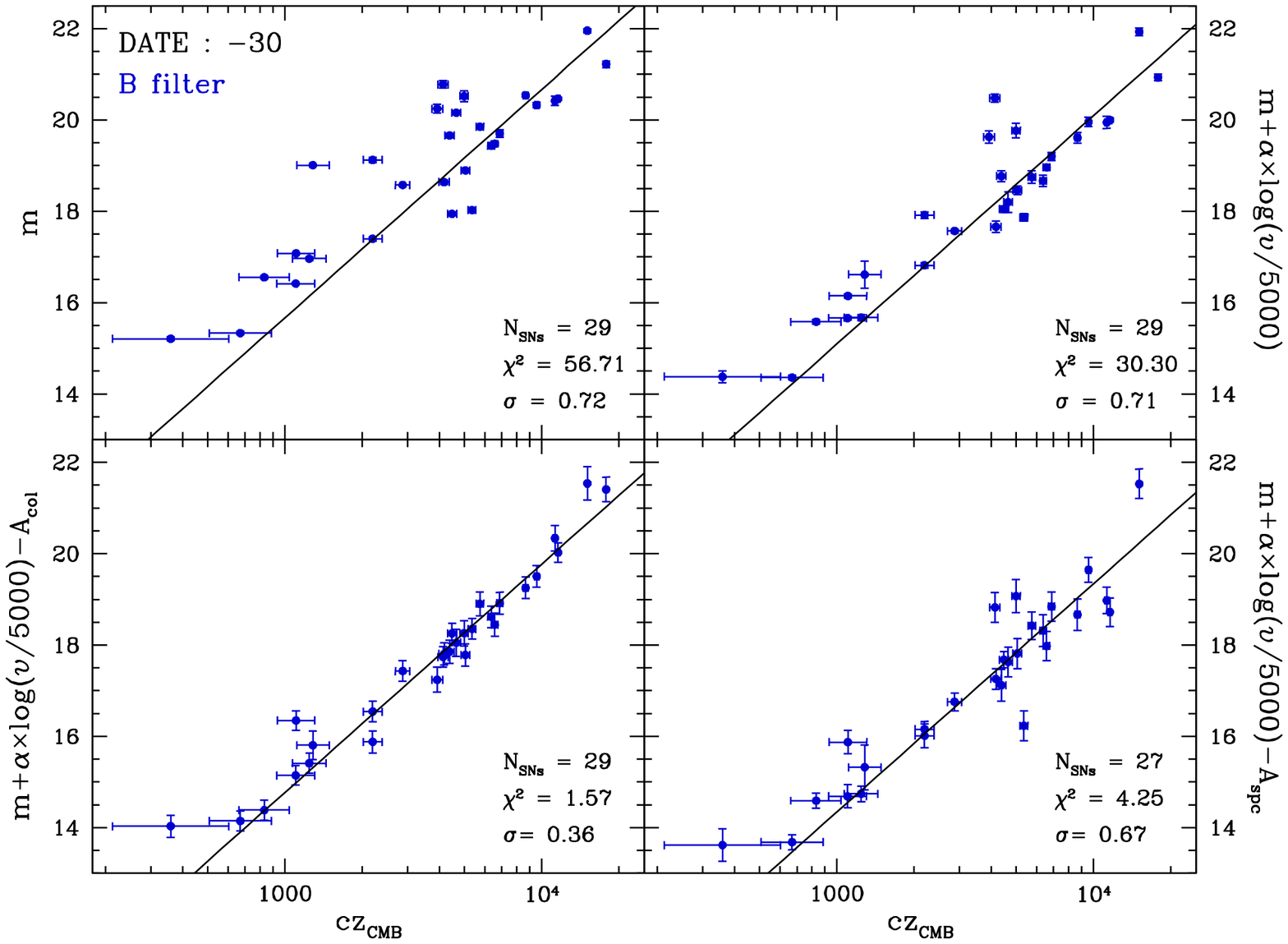}
\vspace{-5mm}
\caption[$B$-band Hubble diagrams]{$B$-band Hubble diagrams. The top
  panel shows magnitudes interpolated to day~$-30$ previously
  corrected for Galactic extinction and $K$-terms; the top right panel
  shows magnitude additionally corrected for expansion velocities; the
  bottom left panel includes further corrections for color-based
  host-galaxy extinction (using the \vi\ color calibration given in
  section~\ref{aho}); the bottom right panel replaces the color-based
  extinction for the spectroscopic reddenings of \citet{D08}.
\label{HDB}}
\end{center}
\end{figure}

The top left panel of Figure~\ref{HDB} shows a Hubble diagram
constructed from $B$-band magnitudes interpolated to day~--30
previously corrected for Galactic extinctions and $K$-terms. The top
right panel shows magnitudes additionally corrected for the
\lumvel\ relation, the bottom left panel includes further corrections
for host-galaxy extinction (using the \vi\ color calibration given in
section~\ref{aho}). In each case we perform a linear least-squares fit
of the form, 
\begin{figure}[p]
\begin{center}
\includegraphics[angle=0,scale=0.8]{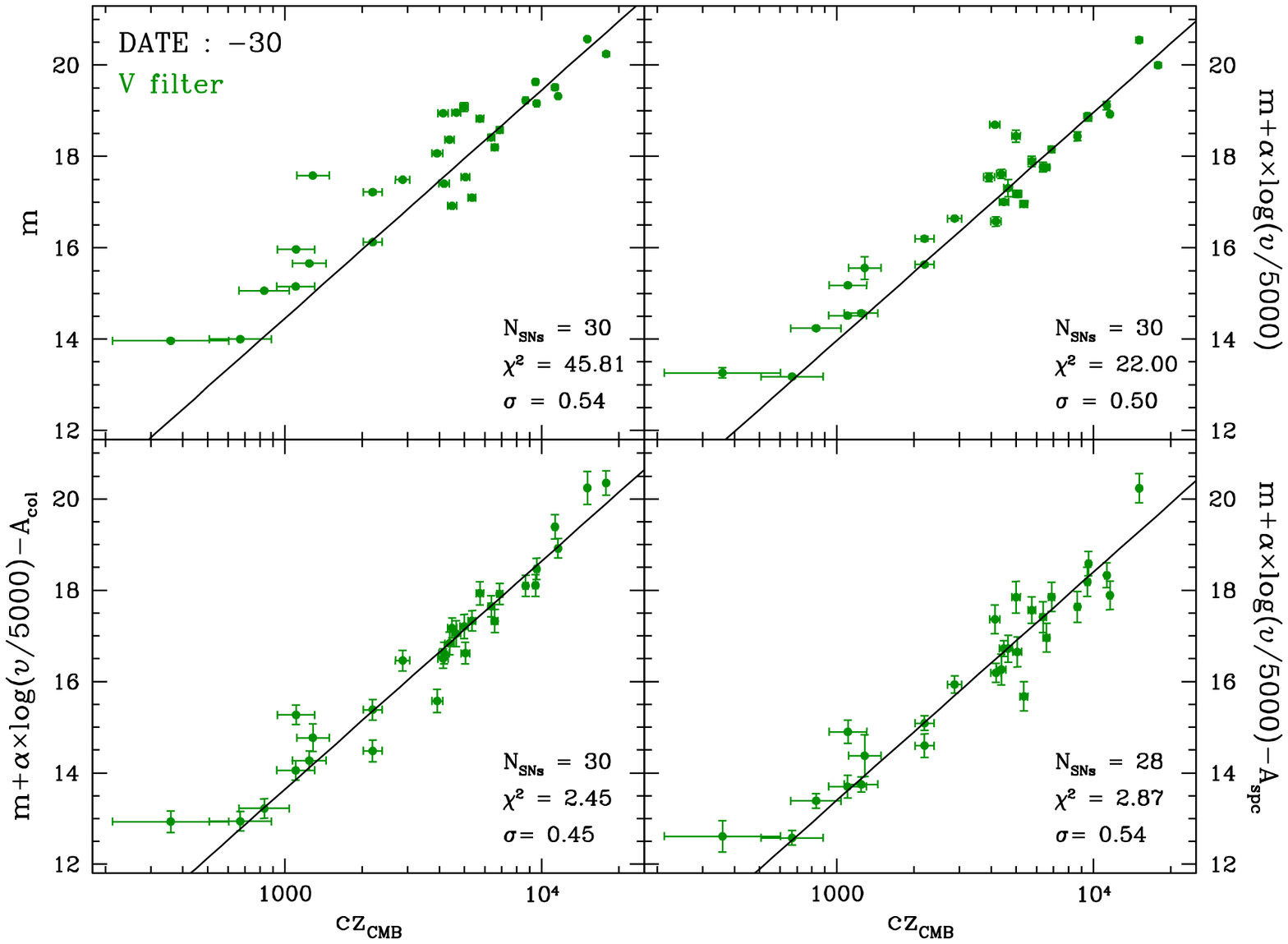}
\vspace{-5mm}
\caption[$V$-band Hubble diagrams]{$V$-band Hubble diagrams. The top
  panel shows magnitudes interpolated to day~--30 previously corrected
  for Galactic extinction and $K$-terms; the top right panel shows
  magnitude additionally corrected for expansion velocities; the
  bottom left panel includes further corrections for color-based
  host-galaxy extinction (using the \vi\ color calibration given in
  section~\ref{aho}); the bottom right panel replaces the color-based
  extinction for the spectroscopic reddenings of \citet{D08}.
\label{HDV}}
\end{center}
\end{figure}
\begin{figure}[p]
\begin{center}
\includegraphics[angle=0,scale=0.8]{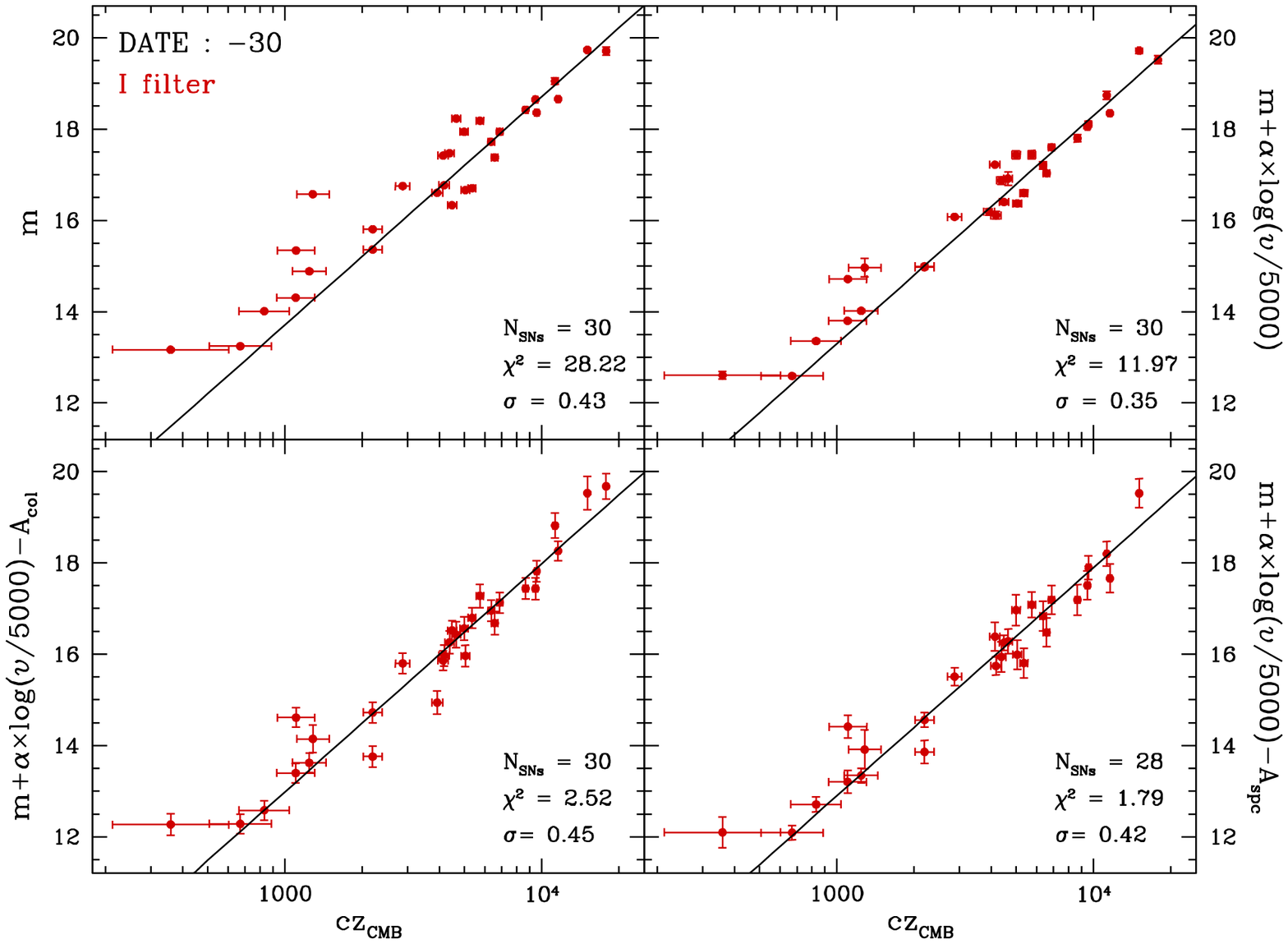}
\vspace{-5mm}
\caption[$I$-band Hubble diagrams]{$I$-band Hubble diagrams. The top
  panel shows magnitudes interpolated to day~$-30$ previously
  corrected for Galactic extinction and $K$-terms; the top right panel
  shows magnitude additionally corrected for expansion velocities; the
  bottom left panel includes further corrections for color-based
  host-galaxy extinction (using the \vi\ color calibration given in
  section~\ref{aho}); the bottom right panel replaces the color-based
  extinction for the spectroscopic reddenings of \citet{D08}.
\label{HDI}}
\end{center}
\end{figure}

\begin{equation}
 m~+~\alpha\,\log{(\upsilon_{\mbox{\scriptsize FeII}}/5000)}~-~A_{host}
 ~=~5\,\log{cz}~+~zp
\label{HDeq1}
\end{equation}

\noindent
where $m$ is the apparent magnitude corrected for $K$-terms and
Galactic absorption, $\upsilon_{FeII}$ is the expansion velocity,
$A_{host}$ is the host-galaxy absorption, and $z$ is the CMB
redshift. The only fitting parameters are $\alpha$ and $zp$; in the
top left panel we set $\alpha=0$ and we only fit for the zero point.

The dramatic decrease in the dispersion, from 0.72 to 0.36~mag clearly
demonstrates the benefitial effects of adding the velocity and
host-galaxy extinction terms. An inspection of the $V$-band diagram
(Fig.~\ref{HDV}) shows a large scatter of 0.54~mag in the top left
panel. When we correct for expansion velocities, the scatter drops to
0.50~mag. This is certainly not unexpected given the \lumvel\ relation
reported in the previous section. It is encouraging to notice that the
dispersion drops from 0.50 to 0.45~mag when we include our host-galaxy
extinction corrections. This indicates the usefulness of our
color-based dereddening technique. The reduced $\chi^2$ value of 2.45
implies that most of the scatter is accounted by the observational
errors.  We performed the same analysis using other epochs and we
found that day~--30 yielded the lowest dispersion. This is also the
day for which we reach the maximum number of SNe in our HDs,
i.e. moving backwards or forwards in time means losing SNe data
(velocity or magnitudes out of the observation range).

If we turn our attention to the $I$-band (Figure~\ref{HDI}) the final
dispersion is 0.45~mag, identical to that found in the $V$-band. It
seems that the dispersion could have some dependence on wavelength,
since it decreases from 0.45 in $VI$ to 0.36~mag in $B$. However, it
may be due to a sampling effect, because the $VI$ diagrams have one SN
more than the $B$ diagram. The scatter of $\sim$~0.4 mag in $BVI$ is
comparable but somewhat larger than the $\sim$~0.34 dispersion
obtained in previous SCM studies \citep{HP02,H03}. It is important to
notice that the dispersions are independent of the intrinsic (\vi)$_0$
color calculated in section~\ref{aho}.

In the bottom right panels of Figs.~\ref{HDB}, \ref{HDV},
and~\ref{HDI} we examine the SCM using the spectroscopic reddenings
determined by \cite{D08} for a set of 28~\sneiip\ (Table~\ref{Tb2})
instead of our color-based extinctions. The resulting dispersions in
$BVI$ are (0.67, 0.54, 0.42), which compare to (0.36, 0.45, 0.45) when
using color-based extinctions. All the fitting parameters derived from
both dereddening techniques are compiled in Table~\ref{TbPar}.

\subsection{Leaving $R_V$ as a free parameter}\label{HDRV}

The previous analysis suggests that the dispersions are somewhat
larger than the observational errors. One possible source of scatter
could be the reddening law. As shown in \S~\ref{aho} and
Fig.~\ref{FgVIVR} we have evidence pointing toward a somewhat
different extinction law in the SN hosts compared to the Galaxy.  Here
we take this idea a step further and we analyze the Hubble diagram
leaving $R_V$ as a free parameter. To accomplish this we model the
data with the following expression

\begin{equation}
 m + \alpha\,\log{(\upsilon_{\mbox{\scriptsize FeII}}/5000)} - \beta
 (\mbox{\vi})~=~5\,\log{cz} + zp
\label{HDeq2}
\end{equation}

\noindent Note that we are replacing $A_{host}$ in
equation~\ref{HDeq1} with the term $\beta$(\vi).  The $\beta$ factor
is a new free parameter to be marginalized along with $\alpha$ and
$zp$, and \vi\ is the color on day~--30 corrected for $K$-terms,
Galactic extinction but {\it uncorrected} for host-galaxy dust. Once
$\beta$ is known we can used it to solve for host-galaxy extinction
from $A(\lambda)=\beta\,\Bigl[\,$(\vi)$-0.656\,\Bigr]$, where 0.656 is
the intrinsic \vi~color of \sneiip\ (see equations~\ref{beta_eq} and
\ref{Ext_eq} in \S~\ref{aho}).

\begin{figure}[p]
\begin{center}
\includegraphics[angle=0,scale=0.8]{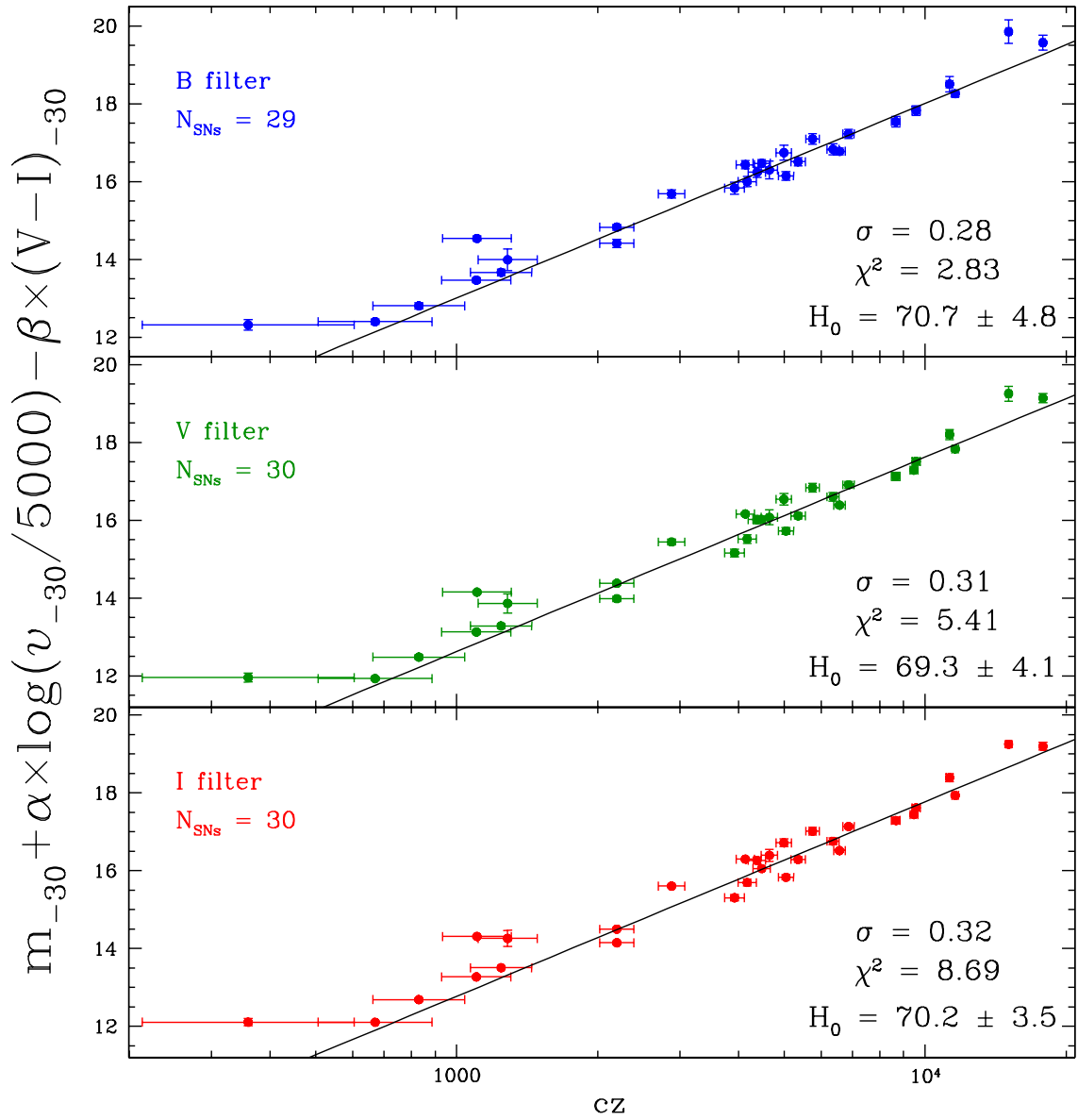}
\caption[$BVI$ Hubble diagrams leaving $R_V$ as a free
  parameter.]{$BVI$ Hubble diagrams leaving $R_V$ as a free parameter.
\label{HDBVI}}
\end{center}
\end{figure}

Figure~\ref{HDBVI} shows the $BVI$ Hubble diagrams for our set of
30~\sneiip. We get dispersions of (0.28,~0.31,~0.32) in $BVI$
respectively, which compare to (0.36,~0.45,~0.45) when $R_V$ is kept
fixed. The increase in $\chi^2$ is due to the fact that we are not
using the intrinsic color, which gives the major contribution to the
errors. Although we expect a reduction in the scatter due to the
inclusion of an additional parameter, the large drop in the dispersion
is remarkable.
The last three lines of Table~\ref{TbPar} shows the parameters we
obtain by minimizing the dispersion in the HD for each band. When we
restrict the sample to objects with $cz_{CMB}>3000$~\kmpsec (leaving
aside SNe with potentially greater peculiar velocities), we end up
with 19~SNe in the $B$-band and 20~SNe in the $VI$ bands. The
resulting HDs in $BVI$ show dispersions of (0.25,~0.28,~0.30),
respectively. Not surprisingly these are lower than the
(0.28,~0.31,~0.32) dispersions obtained from the whole sample, and the
Hubble constant shows only a mild increase of 3\%.

\begin{figure}[p]
\begin{center}
\includegraphics[angle=0,scale=0.8]{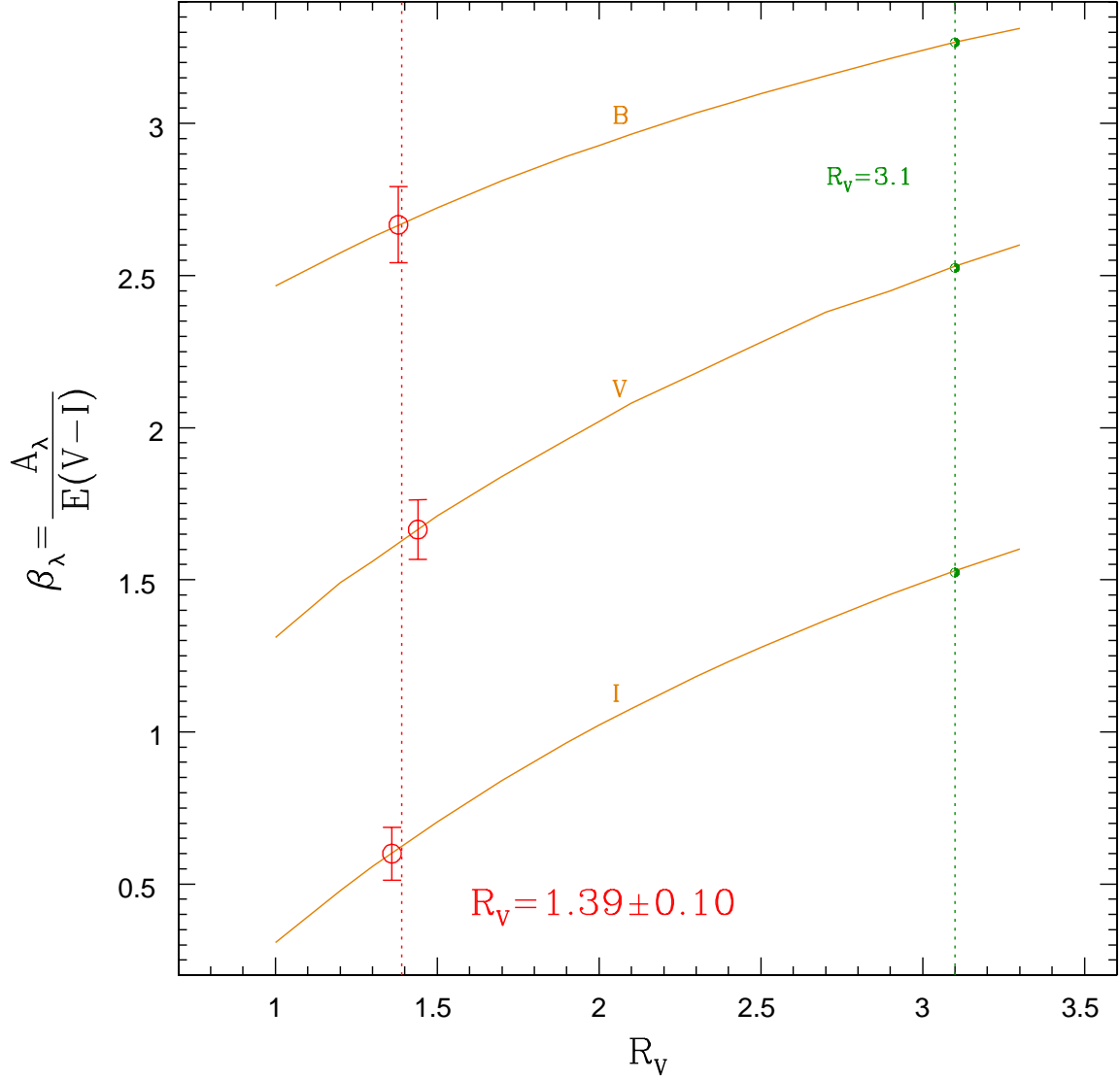}
\caption[$R_V$ versus $\beta$ from minimizing the Hubble
  diagrams]{$R_V$ versus $\beta$. The solid lines are computed from
  our library of SNe~II spectra for each of the $BVI$ bands. The
  $\beta$ parameters derived from the HDs in Figure~\ref{HDBVI} are
  shown with red open circles for each of the $BVI$ bands. The red
  dotted line shows the $R_V$ weighted mean for the $BVI$ values of
  $\beta$.  The values for the standard Galactic extinction law are
  also shown with green dots.
\label{FgRv}}
\end{center}
\end{figure}

By definition the term $\beta$(\vi) in equation~\ref{HDeq2}
corresponds to the extinction in a broad-band magnitude (with central
wavelenth $\lambda$).  Thus, $\beta$ is the ratio $A(\lambda)/E(V-I)$
(see eq.~\ref{beta_eq} in \S~\ref{aho}) and is related to the shape of
the extinction law. For each of the $BVI$ bands we used our library of
SNe~II spectra to compute synthetically the value of $\beta_\lambda$
as a function of $R_V=A_V/E(B-V)$ (see Figure~\ref{FgRv}). This
allowed us to convert the $\beta_\lambda$ values resulting from our
fits into $R_V$ values. Our fits yield $\beta_\lambda=(2.67\pm0.13,
1.67\pm0.10, 0.60\pm0.09)$ for the $BVI$ bands, which translate into
$R_V=(1.38^{+0.27}_{-0.24}, 1.44^{+0.13}_{-0.14},
1.36^{+0.11}_{-0.12})$.  These values are remarkably consistent to
each other and significantly lower than the $R_V=3.1$ value of the
standard reddening law.  Independent evidence for a low $R_V$ law has
been already reported from studies of SNe~Ia (see
\S~\ref{DISC}).

The dispersion of 0.28--0.32~mag in the HDs translates into an
accuracy of 13--14\% in the determination of distances. Although not
as good as the $\sim$~7\%--10\% precision of SNe~Ia
\citep{Phi93,Ham96a,Phi99}, this is a very encouraging result which
demonstrates that \sneiip\ have great potential to determine
extra-galactic distances, and therefore, cosmological parameters.

\clearpage

\begin{table}[p]
\begin{center}
{\scshape \caption{Fitting parameters from the Hubble diagrams}\label{TbPar}}
\vspace{4mm}
\begin{tabular}{l|c|ccc}
\hline\hline
Reddening & & & & \\
estimator             &filter  &{\large $\alpha$} &{\large $\beta$}   &{\large $zp$}        \\
\hline
(\vi)\tablenotemark{a} &$B$  &~3.27$\,\pm\,$0.37  &\nodata      &--0.24$\,\pm\,$0.08   \\
                       &$V$  &~3.01$\,\pm\,$0.36  &\nodata      &--1.36$\,\pm\,$0.08   \\
                       &$I$  &~2.99$\,\pm\,$0.36  &\nodata      &--2.02$\,\pm\,$0.08   \\
\hline                                                          
spectral                  &$B$  &~4.77$\,\pm\,$0.47  &\nodata      &--0.60$\,\pm\,$0.11   \\
analisis\tablenotemark{b} &$V$  &~4.22$\,\pm\,$0.45  &\nodata      &--1.60$\,\pm\,$0.10   \\
                          &$I$  &~3.67$\,\pm\,$0.43  &\nodata      &--2.11$\,\pm\,$0.10   \\
\hline  
(\vi) +                         &$B$  &~3.50$\,\pm\,$0.30  &~2.67$\,\pm\,$0.13  &--1.99$\,\pm\,$0.11   \\
variable $R_V$\tablenotemark{c} &$V$  &~3.08$\,\pm\,$0.25  &~1.67$\,\pm\,$0.10  &--2.38$\,\pm\,$0.09   \\
                                &$I$  &~2.62$\,\pm\,$0.21  &~0.60$\,\pm\,$0.09  &--2.23$\,\pm\,$0.07   \\
\hline\end{tabular}
\vspace{-4mm}

\tablenotetext{a}{\,Reddenings from \S~\ref{aho}}
\tablenotetext{b}{\,Reddenings from \citet{D08}}
\tablenotetext{c}{\,Leaving $R_V$ as a free parameter}

\tablecomments{These are the results of fitting
  $$m+\alpha\,\log{(\upsilon_{\mbox{\scriptsize
        FeII}}/5000)}-A_{host}=5\,\log{cz}+zp$$ and
  $$m+\alpha\,\log{(\upsilon_{\mbox{\scriptsize
        FeII}}/5000)}-\beta(\mbox{\vi})=5\,\log{cz}+zp$$ to the SN
  data in Tables~\ref{Tb2} and~\ref{TbHDs}.}


\end{center}
\end{table}

\clearpage
\section{The Hubble constant}\label{H0}

The Hubble constant is a parameter of central importance in cosmology
which can be determined from our Hubble diagrams. This can be
accomplished as long as we can convert apparent magnitudes into
distances. This calibration was done with two objects for which we
found Cepheid distances in the literature: SN~1999em
\citep[$\mu=30.34\pm0.19$;][]{L02a} and SN~2004dj
\citep[$\mu=27.48\pm0.24$;][]{F01}. For each calibrating SN we can
solve for $H_0$ using the following expression

\begin{equation}
 \log\,H_0~=~0.2\times\Bigl[\,m+\alpha\,\log{(\upsilon_{\mbox{\scriptsize
         FeII}}/5000)}-\beta(\mbox{\vi})-\mu-zp+25\,\Bigr]
\end{equation}

\noindent
In this equation $m$ is the apparent magnitude corrected for Galactic
absorption and $K$-terms, $\upsilon_{FeII}$ is the expansion velocity,
and \vi\ is the color of the calibrating SN, all measured on day~--30
and given in Table~\ref{TbHDs}; $\alpha$, $\beta$, and $zp$ are the
fitting parameters given in the bottom three lines of
Table~\ref{TbPar}.

\begin{figure}[p]
\begin{center}
\includegraphics[angle=0,scale=0.8]{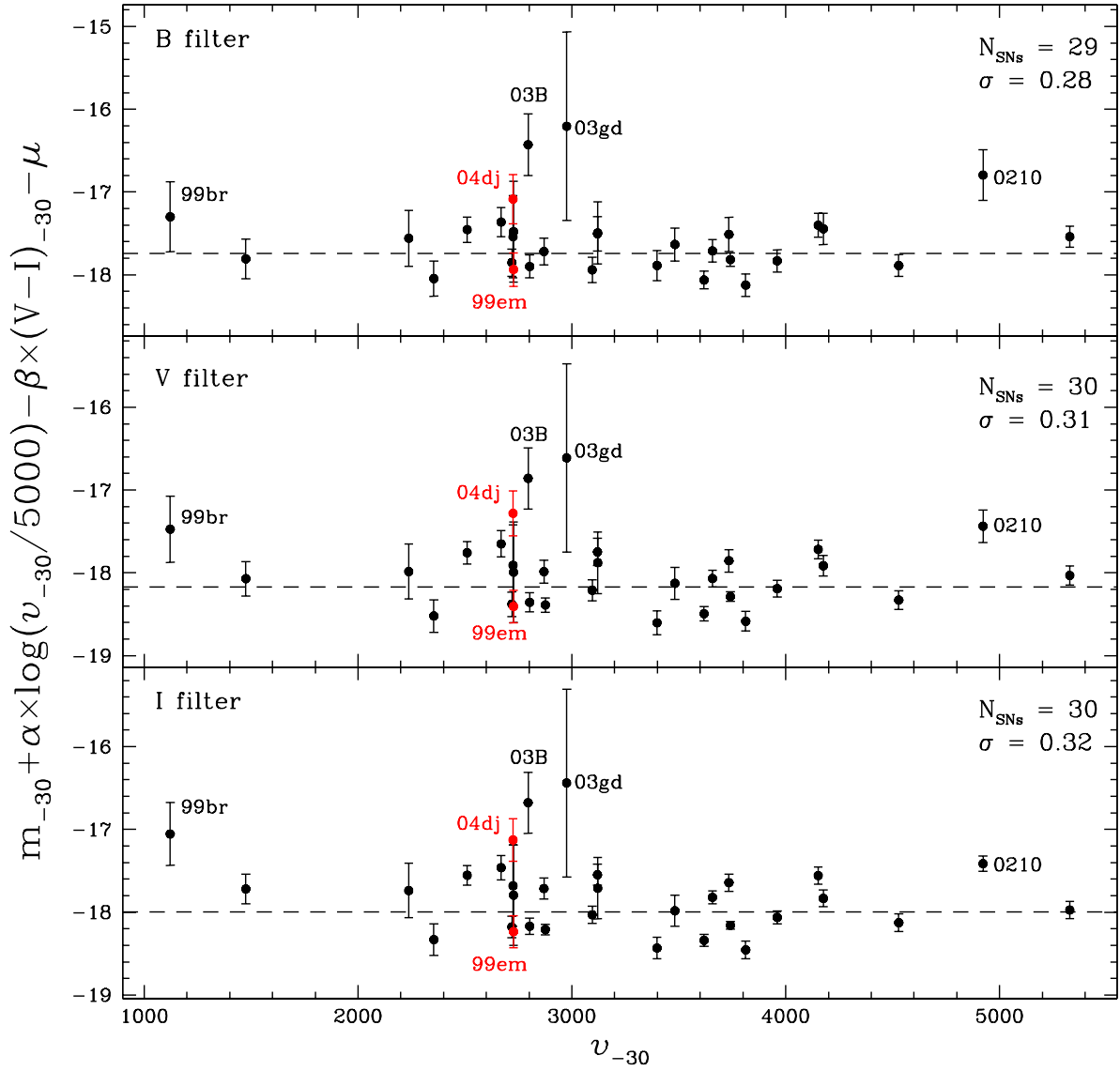}
\caption[Residuals in the $BVI$ corrected absolute magnitudes against
  the \ion{Fe}{2} expansion velocity]{$BVI$ corrected absolute
  magnitudes ($y$-axis) against the \ion{Fe}{2} expansion velocity
  ($x$-axis) for 29--30~SNe (black dots). We use the $H_0$ values of
  Table~\ref{TbHo} and the CMB redshifts to compute the distance
  moduli. The dashed lines show the mean corrected absolute magnitude
  for the black dots. The red dots are the two calibrating SNe, whose
  corrected absolute magnitudes were calculated using their Cepheid
  distances (see Table~\ref{TbHo}). Note that, as expected, the two
  calibrating SNe fall on each side of the corrected absolute
  magnitude distributions.
\label{Mcor}}
\end{center}
\end{figure}

Table~\ref{TbHo} summarizes our calculations for the two calibrators.
This table shows that the corrected $BVI$ absolute magnitudes of
SN~1999em and SN~2004dj differ from each other by 0.86--1.13~mag.
This is somewhat greater than the scatter of 0.3~mag in the
\lumvel\ relation, but statistically plausible. This can be seen in
Figure~\ref{Mcor} where we plot with red dots the corrected absolute
magnitudes of the two calibrating SNe, on top of the whole sample of
SNe employed in the HDs (black dots). This difference leads to $H_0$
values in the range 62--105~\dimho. A weighted average of the values
of Table~\ref{TbHo} yields a Hubble constant

\begin{equation}
 H_0\ =\ 70\ \pm\ 8 \quad\mbox{\dimho}
\end{equation}

\noindent
This value compares very well with that derived by \citet{F01} from
SNe~Ia ($71\pm2$~\dimho), and reasonable well with that found by
\citet{San06} ($62.3\pm1.3$~\dimho) with a similar Type~Ia
sample. With only two calibrating SNe the SCM still has plenty of room
to deliver a more precise value for $H_0$. \vspace{1cm}
\begin{table}[h]
\begin{center}
\small
{\scshape \caption{$H_0$ calculations}\label{TbHo}}
\vspace{2mm}
\begin{tabular}{l|ccc|r@{$\,\pm\,$}lr@{$\,\pm\,$}lr@{$\,\pm\,$}l}
\hline\hline
          &\multicolumn{3}{c|}{$M+\alpha\log(\upsilon/5000)-\beta(V-I)$} &\multicolumn{2}{c}{ }  &\multicolumn{2}{c}{ }  &\multicolumn{2}{c}{ } \\
\cline{2-4}
SN         &$B$                   &$V$                    &$I$           &\multicolumn{2}{c}{$H_0(B)$}  &\multicolumn{2}{c}{$H_0(V)$}  &\multicolumn{2}{c}{$H_0(I)$} \\
\hline
1999em  &--17.94$\,\pm\,$0.20  &--18.41$\,\pm\,$0.20  &--18.24$\,\pm\,$0.19         &64.7 &6.8                    &62.2  &6.1                   &62.9  &6.0    \\
2004dj  &--17.08$\,\pm\,$0.30  &--17.28$\,\pm\,$0.27  &--17.13$\,\pm\,$0.26         &95.7 &14.0                   &104.5 &13.7                  &104.8 &12.9   \\
\hline
Average &--17.67$\,\pm\,$0.39  &--18.02$\,\pm\,$0.53  &--17.84$\,\pm\,$0.53    &{\bf 71} &{\bf 12}            &{\bf 69} &{\bf 16}            &{\bf 70} &{\bf 16} \\
\hline
\end{tabular}

\tablecomments{The uncertainties in $\langle H_0\rangle$ and $\langle
  M_{corr}\rangle$ (the corrected absolute magnitude for $BVI$ in the
  second column) correspond to the dispersion in each pair of values,
  which are about $\times$2--3 greater than the formal errors.}

\vspace{-4mm}
\end{center}
\end{table}

\clearpage
\section{Distances}\label{DIST}

Armed with standardized absolute magnitudes for \sneiip, we are in a
position to calculate distances to all the SNe of our sample. This can
be done with the following expression

\begin{equation}
 \mu=m+\alpha\,\log{(\upsilon_{\mbox{\scriptsize FeII}}/5000)}-\beta
 (\mbox{\vi})-\langle M_{corr}\rangle
\end{equation}

\noindent
where $\langle M_{corr}\rangle$ is the weighted mean of the corrected
absolute magnitudes of the two calibrating SNe given in
Table~\ref{TbHo}. The resulting values are given in Table~\ref{TbMu}
for the 37~\sneiip.  The last column of the table shows the weighted
averages of their distance moduli.

\begin{figure}[p]
\begin{center}
\includegraphics[angle=0,scale=0.8]{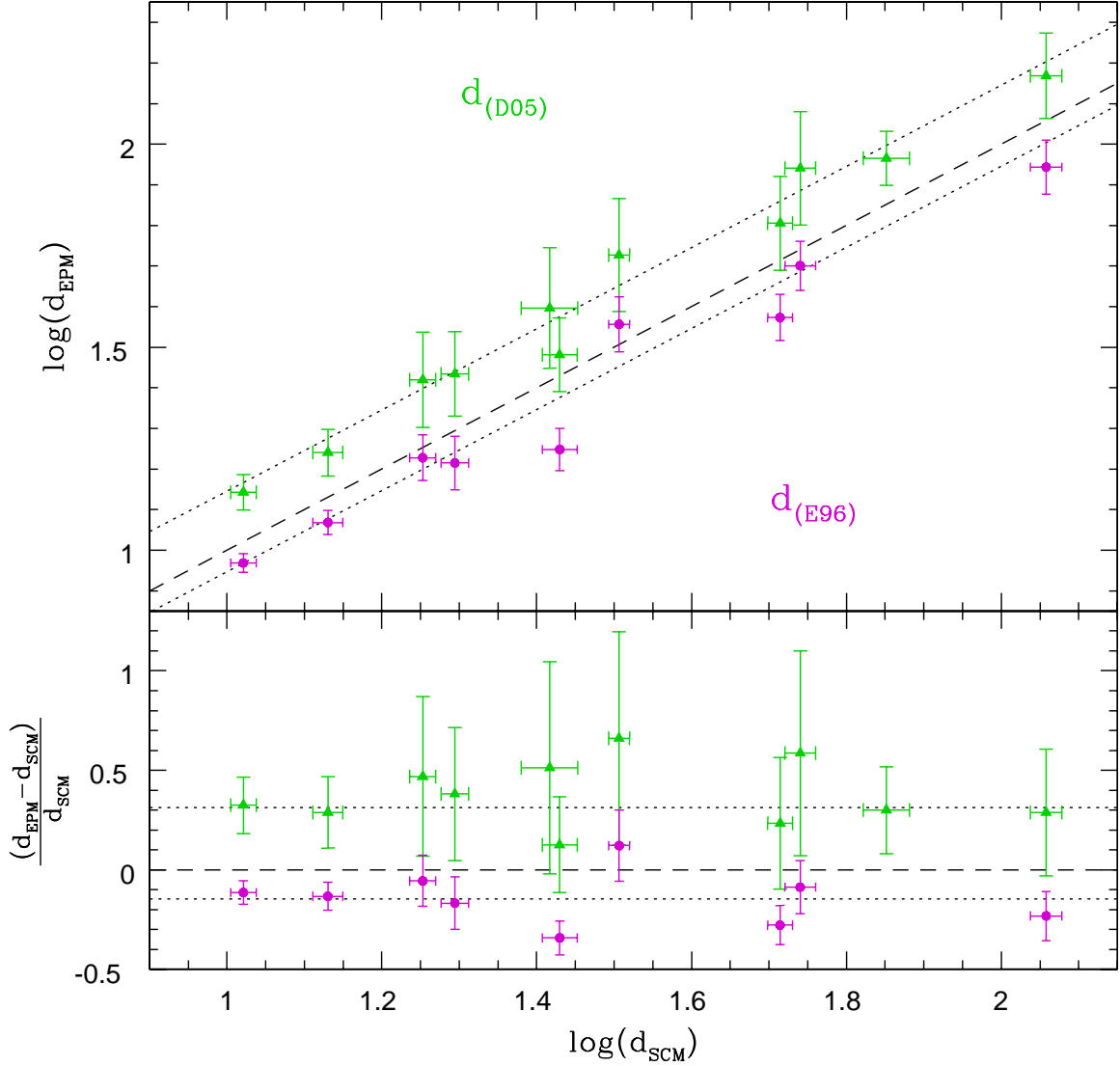}
\vspace{-5mm}
\caption[Comparison between SCM and EPM distances]{Direct comparison
  between SCM and EPM distances calculated by \citet{JH08}. In magenta
  are shown the EPM distances computed from the atmosphere models of
  E96, while the green triangles are EPM distances obtained from the
  D05 models. The bottom panel shows the fractional differences
  between both techniques against $\log(d_{SCM})$. In both panels the
  dotted lines trace the systematic shifts of the EPM distances with
  respect to the SCM distances, $\sim$~40\% and $\sim$~12\% using D05
  and E96 atmosphere models, respectively.
\label{EPSC}}
\end{center}
\end{figure}

As mentioned in the introduction, we can evaluate the precision of the
SCM from a comparison with EPM distances. For this purpose we employ
EPM distances recently calculated by \citet{JH08} which are summarized
in Table~\ref{TbEPM} along with our distances for the 11~objects in
common between SCM and EPM. \citet{JH08} determine EPM distances with
two different sets of atmosphere models. The EPM distances determined
using the models of \citet{Ea96} (E96; column~2) are 12\%$\,\pm\,$5\%
shorter than our SCM distances. On the other hand, the EPM distances
determined using the models of \citet{DH05} (D05; column~3) are
40\%$\,\pm\,$10\% greater than our SCM distances. These shifts are
calculated weighting by the error in the EPM and SCM distances.  These
systematic differences can be clearly seen in the upper panel of
Figure~\ref{EPSC}, and more clearly in the fractional differences
$(d_{EPM}-d_{SCM})/d_{SCM}$ plotted in the bottom panel.

\begin{figure}[p]
\begin{center}
\includegraphics[angle=0,scale=0.8]{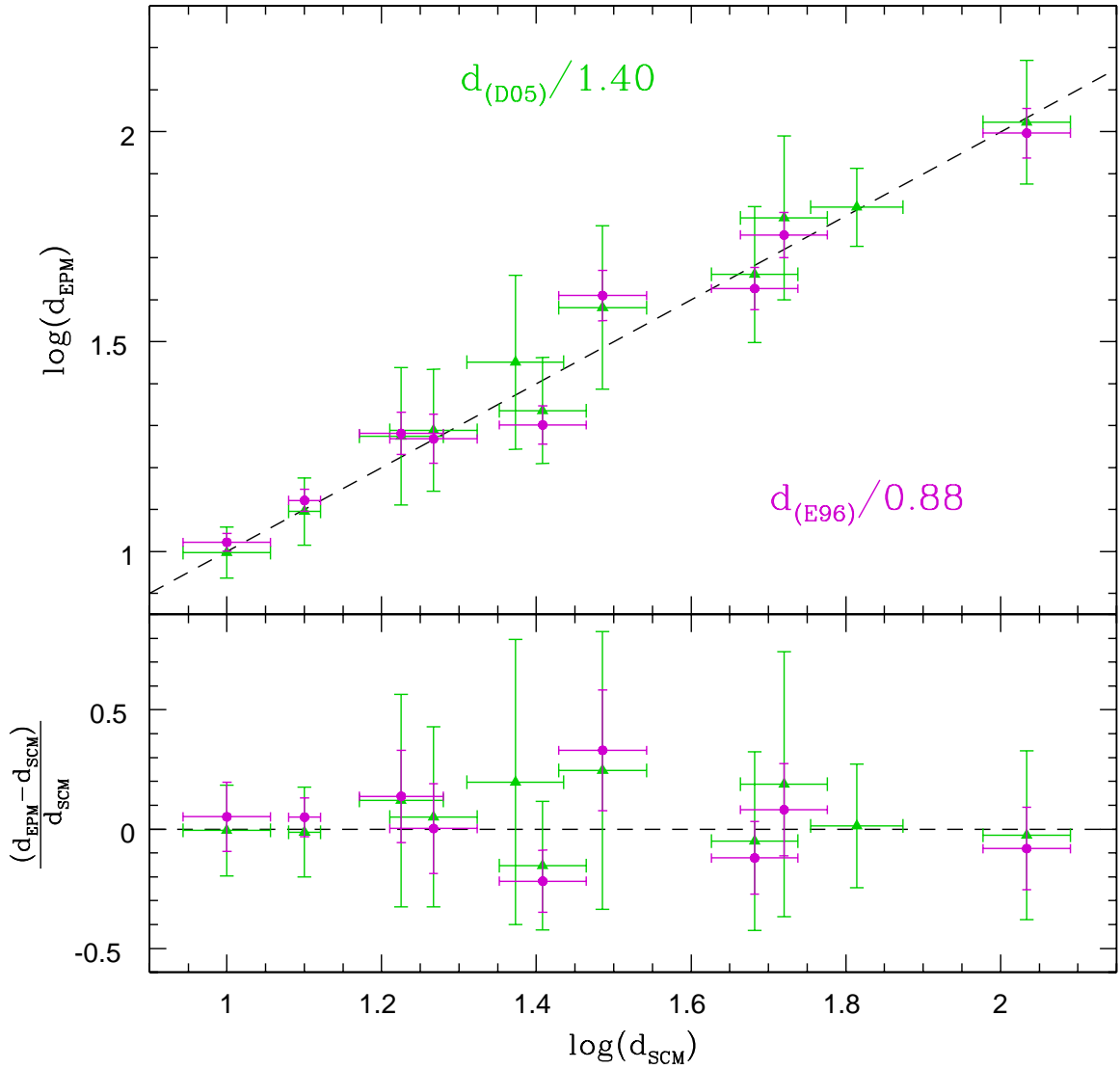}
\vspace{-5mm}
\caption[Comparison between SCM and scaled-EPM distances]{Similar to
  Figure~\ref{EPSC} after bringing the EPM distances to the SCM
  scale. The removal of the systematic differences leads to random
  differences between EPM and SCM distances. For the E96 case the
  discrepancies have a spread of 16\%, while for the D05 case the
  scatter is 13\%.
\label{EPSC_c}}
\end{center}
\end{figure}

The systematic differences among the two sets of EPM distances can be
solely attributed to the atmosphere models of E96 and D05. New
radiative transfer models of \sneiip\ are necessary to understand the
origin of this discrepancy. Besides the systematic errors in both EPM
implementations, we can get an understanding of the internal precision
of EPM and SCM after removing the systematic differences and bringing
all distances to a common scale. For this purpose we correct the EPM
distances to the SCM distance scale by removing the percentage shifts
of 1.40 and 0.88. The comparison is shown in Figure~\ref{EPSC_c}.  The
distance differences have dispersions of $\sim$~13\% and $\sim$~16\%
using D05 and E96 respectively. This implies that either SCM or EPM
produce relative distances with an internal precision between
13--16\%. This agrees with the dispersions of 0.25--0.30~mag seen in
the HDs restricted to SNe with $cz_{CMB}>300$~\kmpsec.

\clearpage
\begin{table}[p]
\begin{center}
\small
{\scshape \caption{Distance Moduli}\label{TbMu}}
\vspace{3mm}
\begin{tabular}{lc|ccc|c}
\hline\hline
SN name  &Host Galaxy     &$\mu_B$       &$\mu_V$      &$\mu_I$     &$\langle\mu\rangle$ \\
\hline
1991al   &LEDA 140858     &34.14(41)  &34.04(54)  &33.89(53)  &34.05(28) \\
1992af  &ESO 340--G038    &34.18(41)  &34.13(54)  &34.13(54)  &34.15(28) \\
1992am  &MCG--01--04--039 &\nodata    &\nodata    &\nodata    &\nodata   \\
1992ba   &NGC 2082        &31.34(40)  &31.31(54)  &31.34(53)  &31.33(28) \\
1993A    &[MH93a]073838.4 &\nodata    &\nodata    &\nodata    &\nodata   \\
1999br   &NGC 4900        &31.67(48)  &31.89(59)  &32.10(57)  &31.86(31) \\
1999ca   &NGC 3120        &\nodata    &\nodata    &\nodata    &\nodata   \\
1999cr   &ESO 576--G034	  &34.50(42)  &34.63(55)  &34.59(54)  &34.56(28) \\
1999em   &NGC 1637 	  &30.07(40)  &29.96(54)  &29.94(53)  &30.00(28) \\
1999gi   &NGC 3184 	  &30.48(40)  &30.51(54)  &30.52(53)  &30.50(28) \\
0210   &MCG +00--03--054  &37.52(50)  &37.27(57)  &37.09(54)  &37.31(31) \\
2002fa   &NEAT J205221.51 &37.23(44)  &37.16(55)  &37.03(54)  &37.16(29) \\
2002gw   &NGC 922 	  &33.35(41)  &33.46(54)  &33.44(53)  &33.41(28) \\
2002hj   &NPM1G+04.0097   &34.90(41)  &34.93(54)  &34.97(53)  &34.93(28) \\
2002hx   &PGC 23727 	  &35.49(42)  &35.53(54)  &35.45(54)  &35.49(28) \\
2003B    &NGC 1097 	  &32.21(40)  &32.18(54)  &32.15(53)  &32.18(27) \\
2003E  &MCG--4--12--004   &33.91(42)  &34.04(54)  &34.10(54)  &34.01(28) \\
2003T    &UGC 4864 	  &35.21(42)  &35.15(54)  &35.13(54)  &35.17(28) \\
2003bl   &NGC 5374 	  &33.95(45)  &34.09(57)  &34.23(55)  &34.07(30) \\
2003bn   &2MASX J10023529 &33.67(42)  &33.54(55)  &33.53(54)  &33.60(28) \\
2003ci   &UGC 6212 	  &\nodata    &35.31(54)  &35.28(53)  &35.30(38) \\
2003cn   &IC 849 	  &34.77(42)  &34.86(55)  &34.85(54)  &34.81(28) \\
2003cx   &NEAT J135706.53 &36.17(44)  &36.23(55)  &36.23(54)  &36.20(29) \\
2003ef   &NGC 4708        &\nodata    &\nodata    &\nodata    &\nodata   \\
2003fb   &UGC 11522 	  &34.41(44)  &34.56(55)  &34.55(54)  &34.49(29) \\
2003gd   &M74 		  &29.99(42)  &29.98(55)  &29.94(54)  &29.98(28) \\
2003hd &MCG--04--05--010  &35.93(40)  &35.85(54)  &35.77(53)  &35.86(28) \\
2003hg   &NGC 7771 	  &33.50(42)  &33.18(54)  &33.14(54)  &33.31(28) \\
2003hk   &NGC 1085 	  &34.45(40)  &34.41(54)  &34.35(53)  &34.41(28) \\
2003hl   &NGC 772 	  &32.08(41)  &32.01(54)  &31.10(53)  &32.04(28) \\
2003hn   &NGC 1448 	  &31.14(40)  &31.15(54)  &31.11(53)  &31.13(27) \\
2003ho   &ESO 235--G58 	  &34.10(41)  &34.18(54)  &34.13(53)  &34.13(28) \\
2003ip   &UGC 327         &33.81(41)  &33.75(54)  &33.67(54)  &33.76(28) \\
2003iq   &NGC 772 	  &32.50(40)  &32.40(54)  &32.34(53)  &32.43(28) \\
2004dj   &NGC 2403        &28.06(43)  &28.22(55)  &28.19(54)  &28.14(29) \\
2004et   &NGC 6946        &28.64(50)  &28.21(58)  &28.17(55)  &28.37(31) \\
2005cs   &NGC 5194	  &\nodata    &\nodata    &\nodata    &\nodata   \\
\hline
\end{tabular}
\tablecomments{These values were calculated according to the approach
  exposed in \S~\ref{DIST}.  The mean value is an average of the
  values for each filter.}
\end{center}
\end{table}

\begin{table}[h]
\begin{center}
{\scshape \caption{EPM distances \citep{JH08} and SCM
    distances}\label{TbEPM}}
\vspace{4mm}
\begin{tabular}{lr@{~}lr@{~}lr@{~}l}
\hline\hline
SN name  &\multicolumn{2}{c}{$d_{\mbox{\scriptsize E96}}$\tablenotemark{a}}  &\multicolumn{2}{c}{$d_{\mbox{\scriptsize D05}}$\tablenotemark{b}}   &\multicolumn{2}{c}{$d_{\mbox{\scriptsize SCM}}$} \\  
         &\multicolumn{2}{c}{(Mpc)} &\multicolumn{2}{c}{(Mpc)} &\multicolumn{2}{c}{(Mpc)}  \\
\hline
1992ba    &16.4 &(2.5)                                  &27.2  &(6.5)                  &18.5  &(2.4)                    \\  
1999br    &\multicolumn{2}{c}{\nodata}                  &39.5  &(13.5)                 &23.6  &(3.4)                    \\  
1999em    &9.3  &(0.5)                                  &13.9  &(1.4)                  &10.0  &(1.3)                    \\  
1999gi    &11.7 &(0.8)                                  &17.4  &(2.3)                  &12.6  &(0.6)                    \\  
2002gw    &37.4 &(4.9)                                  &63.9  &(17.0)                 &48.1  &(6.2)                    \\  
2003T     &87.8 &(13.5)                                 &147.3 &(35.7)                 &108.1 &(14.0)                   \\  
2003bl    &\multicolumn{2}{c}{\nodata}                  &92.4  &(14.2)                 &65.2  &(9.0)                    \\  
2003bn    &50.2 &(7.0)                                  &87.2  &(28.0)                 &52.5  &(6.8)                    \\  
2003hl    &17.7 &(2.1)                                  &30.3  &(6.3)                  &25.6  &(3.3)                    \\  
2003hn    &16.9 &(2.2)                                  &26.3  &(7.1)                  &16.8  &(2.1)                    \\  
2003iq    &36.0 &(5.6)                                  &53.3  &(17.1)                 &30.6  &(4.0)                    \\  
\hline
\end{tabular}
\vspace{-3mm}

\tablenotetext{a}{\,EPM distances from atmosphere models by \citet{Ea96}.}
\tablenotetext{b}{\,EPM distances from atmosphere models by \citet{DH05}.}

\end{center}
\end{table}

\clearpage

\chapter{Discussion\label{DISC}}

\section{Variations of the extinction law}
The reddening law that minimizes the dispersion in the $BVI$ Hubble
diagrams ($R_V=1.4\pm0.1$) turns out to be very different than the
standard Galactic extinction law \citep[$R_V=3.1$;][]{Car89}. To our
knowledge this is the first study of the interstellar extinction in
external galaxies based on \sneiip. On the other hand, there have been
several studies on this subject using Type~Ia SNe.
Most recently, \citet{Fol08} solved for $R_V$ in a similar manner to
our approach, i.e., by minimizing the dispersion in the Hubble diagram
and obtained a value of $R_V\simeq1.5$, remarkably consistent with our
value. \citet{Alt04} and \citet{Rei05} applied a similar procedure by
minimizing the dispersion in the $M_B-\Delta m_{15}(B)$ relation for
SNe~Ia and found $R_B$ ranging between~3.5 and~3.7, significantly
smaller than the standard value of~4.3.  An even smaller value of
$R_B=1.7$ was found by \citet{Cap90} based on the same kind of
analysis for SNe~Ia. Further evidence for low $R_{\lambda}$ values
were reported by \citet{Phi99} and \citet{Kno03}.
Studies of individual Type~Ia SNe, such as SN~1999cl \citep{Kri07},
SN~2003cg \citep{Eli06}, SN~2002cv \citep{Eli08}, and SN~2006X
\citep{WX08} also yielded low values around $R_V\sim1.5$.
Other studies of dust reddening based on diverse methods were
performed to nearby galaxies obtaining values for $R_B$ ranging
between 2.4 and 4.3 \citep{Rif90,dVF92,BL91}. The ratio of total to
selective absorption varies significantly between $R_V\simeq2.6-5.5$
even within our own Galaxy \citep{CC88}.
A lower value of $R_V$ could be due to dust grains smaller than those
in our Galaxy, since for a given value of $A_V$ the $E(B-V)$ reddening
decreases if the size of the grains grows. On the other hand,
\citet{WL05} suggests the idea that scattering by dust clouds located
in the circumstellar medium of the SN tends to reduce the effective
$R_{\lambda}$ in the optical. This effect should be opposite in the
ultraviolet, hence it could be further tested with photometry at these
wavelenghts.



\section{$H_0$ comparison with other methods}

The HDs constructed from the \vi\ color at day~--30 as the extinction
estimator (\S~\ref{HDAcs}) give values of $H_0$ between 70--73~\dimho,
which turn out to be very similar to those derived from the HDs where
we leave $R_V$ as a free parameter ($70\pm8$~\dimho). This shows that
if we use \vi\ color based extinctions, the value of the Hubble
constant is not be too sensitive to the adopted $R_V$.

\begin{table}[h]
\begin{center}
{\scshape \caption{$H_0$ values from the literature}\label{Tb_H0}}
\vspace{4mm}
\begin{tabular}{lcl}
\hline\hline
Method                &$H_0$           &Reference       \\
                      &(\dimho)        &                \\
\hline
SNe Ia                &71$\;\pm\;$2    &\citet{F01}       \\ 
SNe Ia                &71$\;\pm\;$7    &\citet{Alt04}     \\ 
SNe Ia                &73$\;\pm\;$4    &\citet{Rie05}     \\ 
SNe Ia                &62$\;\pm\;$1    &\citet{San06}     \\ 
SNe Ia                &72$\;\pm\;$4    &\citet{Wan06}     \\
\hline
\sneiip~(EPM)         &73$\;\pm\;$6    &\citet{Sch94}     \\ 
\sneiip~(EPM\tablenotemark{*})   &52$\;\pm\;$4    &\citet{JH08}      \\ 
\sneiip~(EPM\tablenotemark{**})  &92$\;\pm\;$7    &\citet{JH08}      \\ 
\sneiip~(SCM)         &55$\;\pm\;$12   &\citet{HP02}      \\ 
\sneiip~(SCM)         &72$\;\pm\;$6    &\citet{H03}       \\ 
\sneiip~(EPM)         &72$\;\pm\;$9    &\citet{F01}       \\ 
\sneiip~(SCM)         &70$\;\pm\;$8   &this study        \\ 
\hline 
Tully-Fisher          &71$\;\pm\;$3    &\citet{F01}       \\ 
Tully-Fisher          &59$\;\pm\;$2    &\citet{San06}     \\ 
\hline
TRGB\tablenotemark{a} &62$\;\pm\;$2    &\citet{San06}     \\ 
SBF\tablenotemark{b}  &70$\;\pm\;$5    &\citet{F01}       \\ 
FP\tablenotemark{c}   &82$\;\pm\;$6    &\citet{F01}       \\
\hline
\end{tabular}

\tablenotetext{a}{\,TRGB = tip of the red giant branch \citep{Sak04}}

\tablenotetext{b}{\,SBF = surface brightness fluctuations \citep{TS88}}

\tablenotetext{c}{\,FP = ``fundamental plane''}

\tablenotetext{*}{\,Using dilution factors of \citet{DH05}}

\tablenotetext{**}{\,Using dilution factors of \citet{Ea96}}

\end{center}
\end{table}
 Table~\ref{Tb_H0} summarizes several modern measurements
of the expansion rate of the Universe, $H_0$, derived from different
methods: SNe~Ia, EPM and SCM for \sneiip, the Tully-Fisher relation,
{\it tip of the red giant branch} (TRGB) distances, the {\it surface
  brightness fluctuations} (SBF) method, and the ``fundamental plane''
method for early type galaxies. According to Table~\ref{Tb_H0} the
value of the Hubble constant ranges between 52--82~\dimho, and the
most accepted value today is 70~\dimho, which is in very good
agreement with our SCM value of $70\pm8$~\dimho.

This study shows that the SCM can deliver accurate distances. The
$BVI$ Hubble diagrams yield scatter of 0.3~mag which implies a
precision in individual distances $\sim$~15\%. Part of this scatter
could be due to the peculiar velocities of the SN host galaxies and
the intrinsic precision of SCM could be even lower than this. In fact,
when we perform a comparison for SNe in common between SCM and EPM,
the distance differences range between 13--16\% (after removing the
systematic differencce among the SCM and EPM). This comparison is
independent of the SN host-galaxy redshifts and implies that the
internal precision in any of these two techniques must be lower than
13--16\%, since these differences comprise the combined uncertainties
of both methods. This is an encouraging result, since it implies that
both EPM and SCM can produce high precision relative distances, thus
offering a new route to cosmological parameters.

\section{Final remarks}

We note that the dispersion in the HDs increases with wavelength. This
seems contrary to expectations given that 1) the extinction effects
decrease with wavelength, and 2) metallicity affects the
$B$-band. Perhaps the luminosity is not only a function of velocity
but also of metallicity, i.e., $M_V=f(\upsilon,\mbox{[Fe/H]})$. If so,
the velocity term in eqs.~\ref{HDeq1}--\ref{HDeq2} might be removing
metallicity effects, more efficientlyat shorter wavelengths. This
could be tested with [Fe/H] measurements of the Type~II-P SNe. We plan
to address this issue in the near future. Given the evidence we have,
we can only claim a dependence of $L$ with $\upsilon$. The obvious
physical explanation is that the internal energy is correlated with
the kinetic energy. The implication is that the ratio between the
internal energy and kinetic energy is approximately independent of the
explosion energy.

\chapter{Conclusions\label{CONC}}

We established a library of 196 SN~II optical spectra and developed a
code which allows us to correct the observed photometry for Galactic
extinction, $K$-terms and host-galaxy extinction. We applied our code
to the $BVRI$ photometry of 37~\sneiip\ (\S~\ref{cor}).  We developed
fitting procedures to the light curves, color curves and velocity
curves which allow us to precisely determine the transition time
between the plateau and the tail phases. The use of this parameter as
the time origin permited us to line up the SNe to a common phase. The
additional benefit of these fits is the interpolation of magnitudes,
colors and velocities over a wide range of epochs. The methodology
explained above yields the following conclusions:


\begin{enumerate}
\item The comparison between our color-based dereddening technique and
  the spectroscopic reddenings of \citet{D08} is satisfactory within
  the errors of both techniques. This is particularly encouraging
  since our method uses late-time photometric information, while the
  other method uses early-time spectroscopic data, i.e. completely
  independent information.

\item Using our new sample of SNe we recover the luminosity-velocity
  trend (\lumvel\ relation) previously reported by \citet{HP02}.

\item We construct $BVI$ HDs using two sets of host-galaxy
  reddenings. We demonstrated that the (\vi)-based extinctions do a
  much better job than the spectroscopic determinations by
  \citet{D08}, reaching dispersions of $\sim$ 0.4~mag in the Hubble
  diagrams. This scatter is somewhat higher than that found previously
  by \citet{H03} of 0.35~mag. A much smaller dispersion of 0.3~mag was
  achieved when we used \vi\ colors to estimate reddening and allowed
  $R_V$ to vary. We obtain $R_V=1.4\pm0.1$, much smaller than the
  Galactic value of 3.1. The low value of $R_V$ can be explained by a
  different nature of the dust grains in host-galaxies along the line
  of sight to Type II-P SNe.

\item We derive a Hubble constant of $70\pm8$~\dimho\ from $BVI$
  photometry, calibrating our HDs with Cepheid distances to SN~1999em
  and SN~2004dj, which agrees very well with the value obtained by the
  HST Key Project \citep{F01}.

\item Finally, we calculate the distance moduli to our SN sample, and
  make a comparison against EPM distances from \citet{JH08}. The
  11~SNe in common show a systematic difference in distance between
  EPM and SCM, depending on the atmosphere models employed by
  EPM. Correcting for these shifts we bring the EPM distances to the
  SCM distance scale, from which we measure a dispersion of
  13--16\%. This spread reflects the combined internal precision of
  EPM and SCM. Therefore the internal precision in any of these two
  techniques must be $<$~13--16\%.
\end{enumerate}

This analysis reconfirms the usefulness of \sneiip\ as cosmological
probes, providing strong encouragement to future high-$z$ studies. We
found that one can determine relative distances from \sneiip\ with a
precision of 15\% or better. This uncertainty could be further reduced
by including more SNe in the Hubble flow.  In its current form the SCM
requires both photometric and spectroscopic data. Since the latter are
expensive to obtain (especially at high-$z$) it would be desirable to
look for a photometric observable as a luminosity indicator instead of
the expansion velocities.


\appendix
\chapter{\scshape The Computation of Synthetic Magnitudes\label{appA}}

The implementation of SCM requires the implementation of AKA
corrections to the observed SN magnitudes (\S~\ref{cor}). This process
involves synthesizing broadband magnitudes from the library of
\sneiip\ spectra. It is crucial, therefore, to place the synthetic
magnitudes on the same photometric system employed in the observations
of the SN.

Since the SN magnitudes are measured with photon detectors, a
synthetic magnitude is the convolution of the observed photon number
distribution, $N_\lambda~$, with the filter transmission function
$S(\lambda)$, i.e.,

\begin{equation}
 mag = -2.5~log_{10}~\int N_{\lambda}~S(\lambda)~d\lambda ~+~ZP,
\label{mageqn}
\end{equation}

\noindent where $ZP$ is the zero point for the magnitude scale
and $\lambda$ is the wavelength in the observer's frame.

\begin{figure}[p]
\begin{center}
\includegraphics[scale=0.7,angle=0]{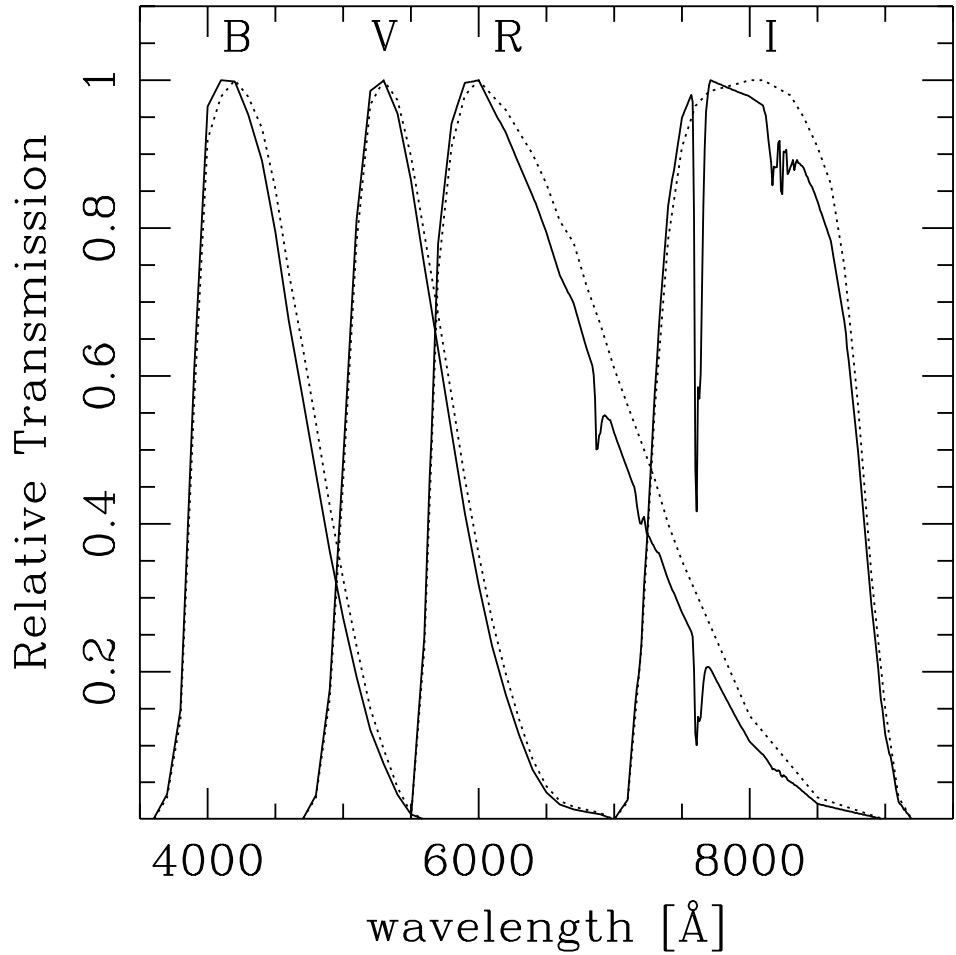}
\caption[$BVRI$ band-passes]{$BVRI$ filters functions of \citet{Bsl90}
  meant for use with energy distributions (dotted curves).  With solid
  lines are shown the curves modified for use with photon
  distributions, to which I added the telluric lines.
\label{filters_bvri}}
\end{center}
\end{figure}

For an adequate use of equation~\ref{mageqn}, S($\lambda$) must
include the transparency of the Earth's atmosphere, the filter
transmission, and the detector quantum efficiency (QE).  For $BVRI$ I
adopt the filter functions $B_{90}$, $V_{90}$, $R_{90}$, $I_{90}$
published by \citet{Bsl90}. However, since these curves are meant for
use with energy and not photon distributions \citep[see Appendix
in~][]{Bsl83}, I must divide them by $\lambda$ before employing them
in equation \ref{mageqn}. Also, since these filters do not include the
atmospheric telluric lines, I add these features to the $R$ and $I$
filters (in $B$ and $V$ there are no prominent telluric features)
using my own atmospheric transmission spectrum. Figure
\ref{filters_bvri} shows the resulting curves.
\begin{figure}[p]
\begin{center}
\includegraphics[scale=.7, angle=0]{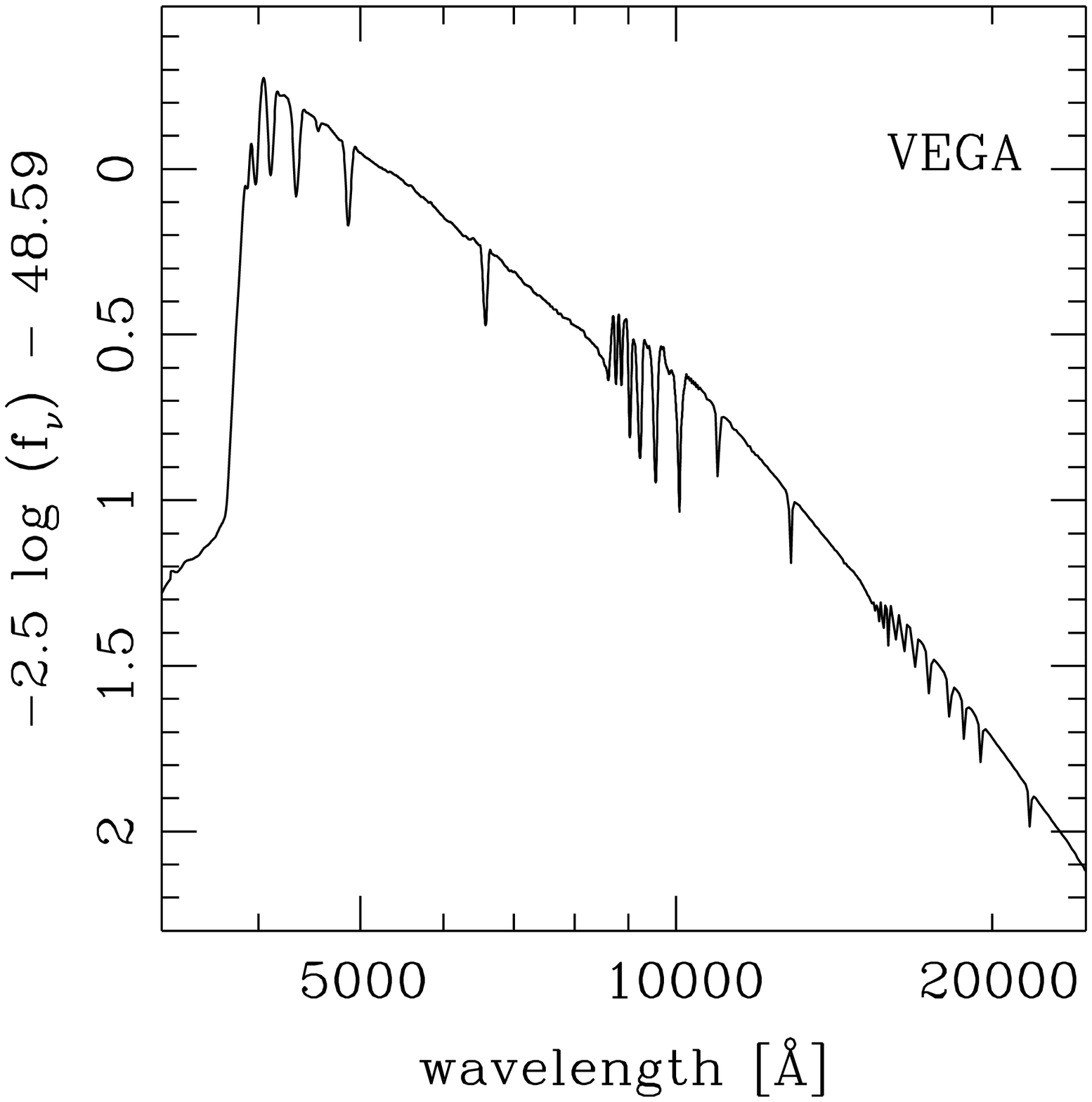}
\caption[Adopted spectrophotometric calibration for Vega]{Adopted
  spectrophotometric calibration for Vega. In the optical ($\lambda$
  $\leq$ 10,500~\AA) the calibration is from \citet{Ha85}, and at
  longer wavelengths I adopted the Kurucz spectrum with parameters
  $T_{eff}$=9,400 K, log $g$=3.9, [Fe/H]=--0.5, $V_{microturb}$=0.
\label{vega}}
\end{center}
\end{figure}

The $ZP$ in eq.~\ref{mageqn} must be determined by forcing the
synthetic magnitude of a star to match its observed magnitude. I use
the spectrophotometric calibration of Vega published by \citet{Ha85}
in the range 3300-10405~\AA~and the $V$ magnitude of 0.03~mag measured
by \citet{John66}, from which I solve for the $ZP$ in the $V$-band.
In principle, I can use the same procedure for $BRI$, but Vega's
photometry in these bands is not very reliable as it was obtained in
the old Johnson standard system.  To avoid these problems I employ ten
stars with excellent spectrophotometry \citep{H94} and photometry in
the modern Kron-Cousins system \citep{Cou71,Cou80,Cou84}. Before using
these standards I remove the telluric lines from the spectra since the
filter functions already include these features. With this approach I
obtain an average and more reliable zero point for the synthetic
magnitude scale with rms uncertainty of $\sim$0.01 mag. With these
$ZP$s I find that the synthetic magnitudes of Vega are brighter than
the observed magnitudes \citep{John66} by 0.016~mag in $B$, 0.025 in
$R$, and 0.023 in $I$ (see Table \ref{vega_zp}), which is not so
surprising considering that this comparison requires transforming the
Johnson $RI$ magnitudes to the Kron-Cousins system \citep{Tay86}.
Figure \ref{vega} shows the adopted spectrophotometric calibration for
Vega. \begin{table}[h]
\begin{center}
{\scshape \caption{Photometric Zero points and Synthetic Magnitudes
    for Vega}\label{vega_zp}}
\vspace{3mm}
\begin{tabular}{crrrr}
\hline \hline 
\multicolumn{1}{c}{}  & 
\multicolumn{1}{c}{$B$} & 
\multicolumn{1}{c}{$V$} &
\multicolumn{1}{c}{$R$} &
\multicolumn{1}{c}{$I$} \\
\multicolumn{5}{c}{} \\
\hline
Zero point & 35.287  & 34.855  & 35.060  & 34.563  \\
Vega & 0.014 & 0.030 & 0.042 & 0.052 \\
\hline
\end{tabular}
\end{center}
\end{table}

Table \ref{vega_zp} summarizes the zero points computed with
eq.~\ref{mageqn}, and the corresponding magnitudes for Vega in such
system.  For the proper use of these $ZP$s it is necessary to express
$N_\lambda$ in sec$^{-1}$~cm$^{-2}$~cm$^{-1}$ and $\lambda$ in~\AA.
From the ten secondary standards I estimate that the uncertainty in
the zero points is $\sim$~0.01~mag in $BVRI$.

\chapter{Finding $\mathcal{F}(\vec v)$ using the DSM in Multidimensions\label{appB}}

\begin{figure}[p]
\begin{center}
\includegraphics[angle=0,scale=1.0]{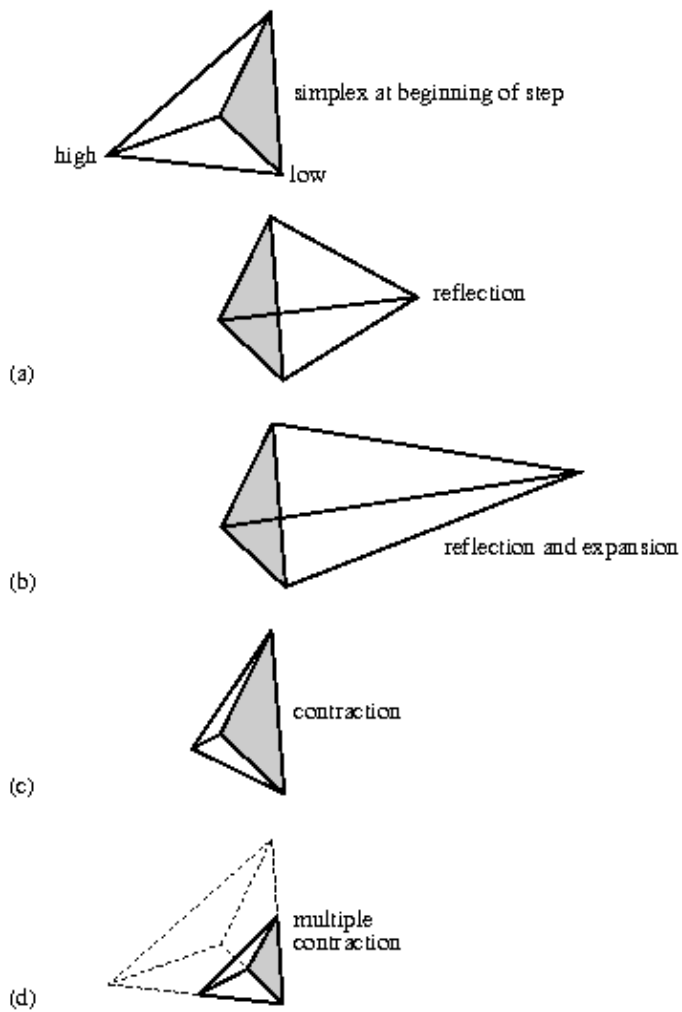}
\caption[Steps of the {\tt amoeba} algorithm]{Steps of the {\tt
    amoeba} algorithm trying to ``swallow'' the minimum using the
  DSM. This example draws the simplex in 3-D (tetrahedron). The
  possible outcomes, that move away from the highest point, are (a)
  the vertice is reflected through its opposite face, (b) a reflection
  plus a 1-D expansion, and (c) a 1-D contraction. A more unusual step
  performs (d) a contraction along all dimensions towards the lowest
  vertice.
\label{FgAmo}}
\end{center}
\end{figure}

Given the 8-parameter function we have to deal with (refer to
$\mathcal{F}$ in eq.~\ref{eqF}), we had to explore different methods
of multidimensional minimization. The {\it Downhill Simplex Method}
(DSM)\footnote{Consult for further details Numerical Recipes in
Fortran \citep[chapter~10.4;][]{Pr92} as the main bibliographic
reference of this appendix.}, due to \citet{NM65}, does not use
one-dimensional minimization as a part of their computational
strategy. Derivatives are not required, only function evaluations. Its
funcionality is based on the simplex, a geometrical figure consisting
in $N$ dimensions, of $N+1$ vertices, all their interconecting line
segments, and polygonal faces. For example, in 2-dimensional space a
simplex is a triangle. The starting guess should not be just a single
point, but $N+1$, defining an initial simplex. The DSM now takes a
series of steps (shown in Figure~\ref{FgAmo}), most steps just moving
the vertice of the simplex where the function is largest through the
opposite face of the simplex to a vertice with a lower value of the
function. The routine {\tt amoeba}, called after the unicellular
organism, tries to ``swallow'' the minimum being intended to be
descriptive of this kind of behavior. The hungry amoeba is the simplex
which is ``fed'' with minima. The purpose after a few step is the
reduction of the $N$-dimensional volume of the simplex as shown in
Fig.~\ref{FgAmo} for the 3D case. Hence the minimum gets enclosed and
finally found with an accuracy given by the size of the simplex.

The input parameters for the Fortran version of this routine are the
following
\begin{center}
{\tt {\bf amoeba}(isimplex[mp,np],y[mp],mp,np,ndim,ftol,funk,iter)}
\end{center}

\noindent
where
\begin{description}
\item {\tt isimplex[mp,np]} : the matrix containing the coordinates of
  the initial simplex, {\tt np+1} vertices defined by a {\tt
    np}-dimensional vector
\item {\tt y[mp]} : vector composed by the value function evaluated in
  each of the {\tt np+1} initial vertices (rows) of {\tt isimplex}
\item {\tt mp = np + 1}
\item {\tt np = ndim}
\item {\tt ndim} : number of dimensions or variables of the function
  (parameters to be fitted)
\item {\tt ftol} : the fractional convergence tolerance to be achieved
  in the value of the function
\item {\tt funk} : function of the form {\tt funk(x)} where {\tt
  x[mp]} is a {\tt np+1}-dimensional vector
\item {\tt iter} : total number of function evaluations after the
  convergence
\end{description}

The output consists of the {\tt isimplex} matrix and the {\tt y}
vector overwritten by {\tt np+1} new points all within a tolerance
{\tt ftol} of the minimum function value, whereas the number of
function evaluations is saved in {\tt iter}. Thus, the result is a new
and smaller simplex containing the minimum. This procedure guarantees
a succesful search of at least a local minimum value of the function.

In our specific case we want to minimize $\chi^2(\vec v)$ as a
function of the parameter vector $\vec v$, which is written as
\begin{equation}
 \chi^2(\vec v) = \sum_{i=1}^{nobs}
 \left(\frac{mag[i]-\mathcal{F}(\vec
   v;t[i])}{\sigma_{mag}[i]}\right)^2
\label{chi_eq}
\end{equation}

\noindent
where $\mathcal{F}$ is the function defined in eq.~\ref{fnc_eq} whose
role is to model the \sneiip\ light
curves. ($mag[i]\,\pm\,\sigma_{mag}[i]$) is the observed magnitude of
the SN with its associated uncertainty for a given Julian date $t[i]$,
all together completing a total of $nobs$ measurements that,
basically, draw the observed SN light curve.

Although simple and user-friendly, the DSM finds problems working on
an 8-dimensional space but not as most algorithm would do. In most
cases the initial simplex has to be accurately defined according to
the desired results. In order to make a semi-automatic search for the
initial simplex, we introduce new parameters in the minimization
routine. From the photometric observations we extract the date of the
first data point $t_i$, and using a simple algorithm we calculate the
approximate end of the plateau $t_{end}$ (refer to \S~\ref{LCF}). By
means of these calculations we constrain the $\mathcal{F}$-parameters
having temporal components. In order to contract the magnitude
component of the $\mathcal{F}$-parameters we extract from the
photometry the brightest and the dimmest magnitude, $m_{bri}$ and
$m_{dim}$ respectively. For example, below is shown the initial
simplex we feed into {\tt amoeba} to fit the V~light curve of
SN~1999em,
\begin{equation}
 \mbox{\tt isimplex} = {\small\bordermatrix{
 &a_0  &t_{PT}      &w_0  &p_0    &m_0      &P     &Q     &R    \cr
 &4    &t_{end}-5   &4.5  &0.08   &m_{bri}  &-1.2  &60    &t_i+9   \cr
 &1.6  &t_{end}+25  &4 5  &0.08   &m_{bri}  &-1.2  &60    &t_{end}-10\cr
 &1.6  &t_{end}-5   &8    &0.08   &m_{bri}  &-1.2  &60    &t_{end}-10\cr
 &1.6  &t_{end}-5   &4.5  &0.012  &m_{bri}  &-1.2  &60    &t_{end}-10\cr
 &1.6  &t_{end}-5   &4.5  &0.08   &25       &-1.2  &60    &t_{end}-10\cr
 &1.6  &t_{end}-5   &4.5  &0.08   &m_{bri}  &1.7   &60    &t_{end}-10\cr
 &1.6  &t_{end}-5   &4.5  &0.08   &m_{bri}  &-1.2  &21    &t_{end}-10\cr
 &1.6  &t_{end}-5   &4.5  &0.08   &m_{bri}  &-1.2  &60    &t_{end}-10\cr
 &4    &t_{end}+25  &8    &0.012  &25       &1.7   &60    &t_i+9   \cr
 }},
\end{equation}

\noindent
where the rows are the 9~initial vertices of the simplex in the
8-parameter coordinate system. This is one structure of the many one
can choose to build the {\tt isimplex} that gives us the best
convergence. Each filter requires slightly different parameter
ranges. By averaging the vertices of the final converging simplex, we
obtain a unique parameter vector with its corresponding uncertainty
vector calculated from the dispersions of the 9~vertices around the
mean. The resulting fits for SN~1999em are shown in the upper-right
panel of Fig.~\ref{FgLCs} and the corresponding parameters are
summarized in Table~\ref{TbFpar}.

\begin{table}[t]
\begin{center}
{\scshape \caption{$\mathcal{F}$-parameters for the V light curve of SN~1999em}\label{TbFpar}}\vspace{4mm}
\begin{tabular}{c|r@{$\,\pm\,$}ll}
\hline\hline
$a_0$      &1.978            &0.015  &mag     \\
$t_{PT}$   &JD\ 2451590.1    &1.1    &        \\
$w_0$      &4.52             &0.85   &day     \\
$p_0$      &9.0              &0.4    &10$^{-3}\,$mag/day  \\
$m_0$      &16.297           &0.014  &mag     \\
$P$        &--0.403          &0.019  &mag     \\
$Q$        &JD\ 2451475.3    &3.7    &      \\
$R$        &64.6             &5.7    &day     \\
\hline
\end{tabular}
\end{center}
\end{table}


\end{document}